\documentclass[a4paper,11pt]{article}
\usepackage{amsmath,amssymb}
\usepackage{a4wide}
\usepackage{xspace}
\usepackage{dsfont}
\usepackage{lscape}
\usepackage{booktabs}
\usepackage{multirow}
\usepackage{amsthm}
\usepackage{graphicx}
\usepackage{xcolor,cite}
\usepackage{algorithm}
\usepackage{algpseudocode}

\def\gsim{\mathrel{\rlap{\lower4pt\hbox{\hskip1pt$\sim$}}
    \raise1pt\hbox{$>$}}}
\def\lsim{\mathrel{\rlap{\lower4pt\hbox{\hskip1pt$\sim$}}
    \raise1pt\hbox{$<$}}}

\newcommand{\be}{\begin{equation}}
\newcommand{\ee}{\end{equation}}
\newcommand{\bea}{\begin{eqnarray}}
\newcommand{\eea}{\end{eqnarray}}
\newcommand{\bi}{\begin{itemize}}
\newcommand{\ei}{\end{itemize}}
\newcommand{\ben}{\begin{enumerate}}
\newcommand{\een}{\end{enumerate}}

\newcommand{\lc}{\left[}
\newcommand{\rc}{\right]}
\newcommand{\lp}{\left(}
\newcommand{\rp}{\right)}

\usepackage[colorlinks=true, linkcolor=black!50!blue, urlcolor=blue, citecolor=blue, anchorcolor=blue]{hyperref}

\usepackage[font=small,labelfont=bf,margin=0mm,labelsep=period,tableposition=top]{caption}

\usepackage[a4paper,top=3cm,bottom=2.5cm,left=2.5cm,right=2.5cm,bindingoffset=0mm]{geometry}

\numberwithin{equation}{section}
\numberwithin{figure}{section}
\theoremstyle{plain}

\theoremstyle{definition}

\theoremstyle{remark}

\newcommand{\bzero}{{\ensuremath{\mathbf{0}}}\xspace}
\newcommand{\bx}{\ensuremath{\boldsymbol{x}}\xspace}
\newcommand{\bc}{\ensuremath{\boldsymbol{c}}\xspace}

\newcommand{\bv}{\ensuremath{\boldsymbol{v}}\xspace}

\newcommand{\TeV}{\ensuremath{\,\text{Te\hspace{-.08em}V}}\xspace}
\newcommand{\GeV}{\ensuremath{\,\text{Ge\hspace{-.08em}V}}\xspace}

\newcommand{\POWHEG} {{\textsc{powheg}}\xspace}

\newcommand{\Delphes} {\textsc{Delphes}\xspace}
\newcommand{\cPqb}{\ensuremath{\mathrm{b}}}
\newcommand{\cPqc}{\ensuremath{\mathrm{c}}}

\newcommand{\pt}{\ensuremath{p_{\mathrm{T}}}\xspace}

\providecommand{\PZ}{\ensuremath{\mathrm{Z}}\xspace}
\DeclareRobustCommand{\bone}{\text{\usefont{U}{bbold}{m}{n}1}}

\newcommand{\bz}{\ensuremath{\boldsymbol{z}}\xspace}
\newcommand{\bn}{\ensuremath{\boldsymbol{\nu}}\xspace}

\newcommand{\dd}{\ensuremath{{\textrm{d}}}\xspace}

\newcommand{\ttbar}{{\ensuremath{{\text{t}\overline{\text{t}}}}}\xspace}

\newcommand{\E}{{\ensuremath{{\mathbb{E}}}}\xspace}

\DeclareMathOperator*{\argmin}{arg\,min}

\begin{document}
\newgeometry{top=1.5cm,bottom=1.5cm,left=1.5cm,right=1.5cm,bindingoffset=0mm}

\vspace{-2.0cm}
\begin{flushright}
MBI-ML-26-03\;\;MBI-CMS-26-01\\
\end{flushright}
\vspace{0.3cm}

\begin{center}
  {\Large \bf Proton Structure from Neural Simulation-Based Inference at the LHC}\\
  \vspace{1.1cm}
  {
 Ricardo~Barru\'e$^{1}$,
 Lisa~Benato$^{1}$,
 Ali Kaan Güven$^{1}$,
 Elie~Hammou$^{2}$,
 Jaco~ter~Hoeve$^{3}$,\\[0.05cm]
Claudius~Krause$^{1}$,
 Ang~Li$^{1}$,
 Luca~Mantani$^{4}$,
 Juan~Rojo$^{2,5}$,
Sergio~S\'anchez~Cruz$^{6}$,
 Robert~Sch\"ofbeck$^{1}$,\\[0.05cm]
 Maria~Ubiali$^{7}$, and
 Daohan Wang$^{1}$
  }\\

\vspace{1.0cm}

{\it
~$^{1}$Marietta Blau Institute for Particle Physics of the \"OAW, Dominikanerbastei 16, 1010 Vienna, Austria\\[0.1cm]
~$^{2}$Nikhef Theory Group, Science Park 105, 1098 XG Amsterdam, The Netherlands\\[0.1cm]

~$^{3}$The Higgs Centre for Theoretical Physics, University of Edinburgh,\\[0.1cm]
JCMB, KB, Mayfield Rd, Edinburgh EH9 3FD, Scotland\\[0.1cm]

~$^{4}$Instituto de F\'isica Corpuscular (IFIC), Universidad de Valencia-CSIC, E-46980 Valencia, Spain\\[0.1cm]

~$^{5}$Department of Physics and Astronomy, Vrije Universiteit Amsterdam, \\NL-1081 HV Amsterdam, The Netherlands\\
~$^{6}$Department of Physics and ICTEA, Universidad de Oviedo, C/ San Francisco, 3, 33003 Oviedo, Asturias, Spain\\[0.1cm]
~$^{7}$DAMTP, University of Cambridge, Wilberforce Road, Cambridge CB3 0WA, United Kingdom
}

\vspace{0.7cm}

{\bf Abstract}
\end{center}

The precise determination of the parton distribution functions (PDFs) of the proton is an essential ingredient for LHC analyses, including for those at the upcoming High-Luminosity LHC.
So far, PDFs are determined from global fits to binned low-dimensional data obtained from unfolded hard-scattering cross section measurements.
In this work we demonstrate for the first time the feasibility of neural simulation-based inference (NSBI) for constraining the proton PDFs using a high-dimensional unbinned data set. Exploiting the full statistical power of unbinned data removes the loss of information inherited by the binning procedure.
As a proof-of-concept, we determine the gluon PDF from simulated data of top quark pair production at the LHC with $\sqrt{s}=13$ TeV.
Taking into account both experimental and theoretical systematic uncertainties in the detector-level features, we demonstrate how the NSBI pipeline achieves significant improvements in precision compared to low-dimensional binned analyses.
Our results illustrate the potential of unbinned inference to reduce the reliance on coarse approximations of uncertainties and their correlations entering PDF determinations, hence contributing to a new paradigm of unbinned detector-level ML-assisted measurements at the LHC.

\clearpage

\tableofcontents

\clearpage

\section{Introduction}
\label{sec:intro}

The determination of the collinear substructure of the proton, encoded in its parton distribution functions (PDFs)~\cite{Gao:2017yyd,Ethier:2020way,Kovarik:2019xvh,Amoroso:2022eow}, is one of the key targets of the LHC, its High-Luminosity upgrade (HL-LHC)~\cite{Cepeda:2019klc}, as well as of other ongoing and planned experiments, from the Electron Ion Collider (EIC)~\cite{AbdulKhalek:2021gbh} to FASER~\cite{FASER:2025myb} and the Forward Physics Facility (FPF)~\cite{Anchordoqui:2021ghd,Feng:2022inv}.
Pinning down the proton PDFs has a two-fold motivation.
On the one hand, to address open questions in Quantum Chromodynamics (QCD) such as the behaviour of the gluon in the small-$x$ region where novel dynamical regimes may arise~\cite{xFitterDevelopersTeam:2018hym,Ball:2017otu,Morreale:2021pnn}, the pattern of matter/antimatter asymmetries in the nucleon~\cite{SeaQuest:2021zxb,Faura:2020oom}, and the heavy quark content of the proton~\cite{Ball:2022qks,Guzzi:2022rca,Brodsky:2015fna}.
On the other hand, to improve the precision and accuracy of hard cross sections at the LHC~\cite{Chiefa:2025loi} and other experiments for processes from Higgs and Drell-Yan production~\cite{AbdulKhalek:2018rok} to the extraction of $\alpha_s(m_Z)$~\cite{Ball:2025xgq,dEnterria:2022hzv,Ablat:2025gbp,Cridge:2024exf}, $m_t$~\cite{Ball:2026qno} and  $m_W$~\cite{CMS:2024lrd,ATLAS:2024erm}. Additionally, searches for new phenomena beyond the Standard Model~(SM)~\cite{Carrazza:2019sec,Ball:2022qtp,Kassabov:2023hbm,Greljo:2021kvv,Hammou:2023heg,Costantini:2024xae,Hammou:2024xuj,Cole:2026eex} are, or will be soon, limited by PDF uncertainties.

Given their non-perturbative nature, and despite encouraging progress from lattice QCD~\cite{Constantinou:2020hdm,Lin:2017snn}, PDFs need to be extracted from experimental data through a phenomenological global analysis.
Recent PDF determinations~\cite{Alekhin:2017kpj,NNPDF:2021njg,Hou:2019efy,Bailey:2020ooq,ATLAS:2021vod} differ in the choices of input data, theoretical calculations, and methodological framework to parameterise the PDFs and to estimate the associated uncertainties, see also~\cite{Risse:2025qlo,Derakhshanian:2026zkx}.
Furthermore, for many phenomenological applications, it is advisable to consolidate PDF fits into combined sets, such as PDF4LHC15/21~\cite{PDF4LHCWorkingGroup:2022cjn,Butterworth:2015oua} or the combination of approximate N$^3$LO PDFs presented in~Ref.~\cite{Cridge:2024icl}.
Most PDF analyses rely on fixed functional forms chosen from a combination of theoretical insight and the requirements of describing a broad range of observables.
The use of feed-forward neural networks as universal unbiased interpolants for PDF parameterisation has been adopted by the NNPDF Collaboration~\cite{Ball:2008by,NNPDF:2014otw} as well as by other analyses beyond collinear PDFs~\cite{Bacchetta:2025ara,Dutrieux:2021wll}.
A third option has been presented recently in Ref.~\cite{Costantini:2025wxp}, built upon the {\sc\small Colibri} framework~\cite{Costantini:2025agd}, and is based on proper orthogonal decomposition (POD) to construct a linear model for the PDFs whose parameters can then be inferred or fitted from the data.

While available PDF determinations thus differ in several technical aspects, they all share a common denominator: the use of binned low-dimensional observables as input to their fits.
Indeed, experimental collaborations release their PDF-sensitive measurements in terms of central values, statistical and systematic uncertainties, and a coarse approximation of the inter-bin correlations which are assumed to follow a multivariate Gaussian distribution and, typically, conflate several, if not all, systematic uncertainties.
The comparison between theoretical predictions and experimental data, which underlies all PDF determinations, is then carried out by means of the standard Gaussian likelihood or $\chi^2$.

While binned multivariate Gaussian measurements have represented the foundation of PDF determinations for decades, they are affected by a number of bottlenecks.
First, uncertainties in experimental data in general do not follow a multivariate Gaussian distribution, especially close to the edges of kinematic phase space.
Second, the dimensionality of binned measurements is drastically limited so that relevant information, which could additionally constrain the PDFs, is integrated out.
Third, precision measurements that serve as inputs to PDF fits are often limited by systematic uncertainties, where even small variations of the underlying correlation model~\cite{Kassabov:2022pps} can significantly affect the interpretation of the results.

A promising pathway towards bypassing these limitations is provided by recent developments in Neural Simulation Based Inference (NSBI)~\cite{Cranmer:2019eaq} and, more broadly, in machine learning~(ML) techniques, where unbinned measurements~\cite{Benato:2025rgo,ATLAS:2024xxl,Arratia:2021otl,Schofbeck:2024zjo,H1:2023fzk,Electron-PositronAlliance:2025hze} are emerging as a powerful alternative to the traditional binned approach.
Unbinned measurements can display enhanced sensitivity as compared to their binned counterparts, in particular in situations where statistical uncertainties dominate.
This powerful feature is illustrated by the recent ATLAS measurement of the Higgs width from the combination of on-shell and off-shell cross sections~\cite{ATLAS:2024jry} based on unbinned observables~\cite{ATLAS:2025clx}. It achieves a 30\% increase in precision using the same data set as compared to a binned analysis.
Unbinned approaches have also been extensively explored in the context of determinations of Wilson coefficients~\cite{GomezAmbrosio:2022mpm,Brehmer:2019xox,DAgnolo:2019vbw,Brehmer:2018kdj,Brehmer:2018eca,Silva:2025hzo} in the Standard Model Effective Field Theory (SMEFT) framework~\cite{Isidori:2023pyp}.
At the same time, the widespread application of unbinned methods at the LHC has so far been hindered by the lack of a realistic treatment of systematic uncertainties.
This limitation has only recently begun to be addressed in uncertainty-aware NSBI frameworks~\cite{Schofbeck:2024zjo,Benato:2025rgo,Valsecchi:2026kpp}, which provide a detailed treatment of systematic effects~\cite{Schofbeck:2024zjo,Benato:2025rgo}.

In this work, we demonstrate how NSBI combined with unbinned multivariate measurements can be used to carry out a direct determination of the proton structure at the LHC.
As a proof of concept, we consider the extraction of the gluon PDF from unbinned top-quark pair production from simulated detector-level data, accounting for both experimental and theoretical systematic uncertainties.
Our approach is based on constructing a linear model for the gluon PDF~\cite{Costantini:2025wxp}, satisfying all theoretical constraints, and implementing it within the uncertainty-aware NSBI pipeline developed in~Refs.~\cite{Schofbeck:2024zjo,Benato:2025rgo} for SMEFT applications.
The comparison of the expected precision in this unbinned projection with a reference binned analysis highlights major sensitivity improvements that can be achieved with this novel methodology.
In addition, our approach potentially offers the LHC experiments the option of an internal calibration of proton structure with reduced reliance from external data sets, as we demonstrate with the application to differential Higgs production in gluon fusion.

The outline of this paper is as follows.
In Sec.~\ref{sec:linear_model} we summarise the main features of the linear model for the gluon PDF and its validation.
Section~\ref{sec:nsbi_for_pdfs} provides a self-contained discussion of NSBI and its application to PDF determination.
The simulated unbinned data set, the detector-level features, our account of systematic uncertainties, and the surrogate training are provided in Sec.~\ref{sec:unbinned-obs}.
Results for the expected precision of the gluon PDF, based on simulated data, are presented in Sec.~\ref{sec:results}, including an initial study of the implications for Higgs production.
Finally, in Sec.~\ref{sec:summary} we summarise and discuss the outlook for future work, in particular the integration of unbinned measurements alongside the binned ones in global PDF fits.

Technical details are collected in appendices.
Appendix~\ref{app:linear_models_PDF} provides additional information on the linear model for the gluon PDF.
Appendix~\ref{app:benchmarking_binned} provides a benchmarking of the binned gluon PDF determination baseline with global PDF fits, as well as with dedicated variants of NNPDF4.0 based on synthetic data.
Finally App.~\ref{app:mse_bpt} describes the multi-parameter extension of the Boosted Information Tree (BIT) algorithm, originally introduced in Refs.~\cite{Chatterjee:2021nms,Chatterjee:2022oco}.

\section{A linear model for the gluon PDF}
\label{sec:linear_model}

In this work we adopt the formalism presented in Ref.~\cite{Costantini:2025wxp} to construct a linear model for the NSBI determination of the gluon PDF.
We refer the reader to the original work for all details of the formalism.
Such a linear model is particularly well suited to our approach, because the PDF parameterisation enters the parton-level description of hard proton-proton scattering at most quadratically.
The resulting predictions are therefore at most quadratic functions of the PDF coefficients, which are our parameters of interest~(POIs).
ML algorithms fully exploiting this structure are already available in the SMEFT context~\cite{Chen:2020mev,GomezAmbrosio:2022mpm,Schofbeck:2024zjo}, where the analytic dependence on the Wilson coefficients is similarly restricted.

In Sec.~\ref{subsec:formalism} we summarize the main features of the linear model formalism and describe how it is adapted to match the needs of the present analysis.
Section~\ref{sec:dglap_evolution} considers the role of DGLAP evolution, and finally Sec.~\ref{subsec:validation} validates the reconstruction accuracy of the model.

\subsection{Linear model formalism}
\label{subsec:formalism}

Let us denote by $\mathcal{H}$ the space of all physically admissible functions for the gluon $f_g(x,Q_0)$ and quark singlet $f_\Sigma(x,Q_0)$ PDFs at a given reference scale $Q_0$.
The elements of $\mathcal{H}$ should satisfy all relevant theory constraints at $Q_0$, including the endpoint behavior $f_i(x,Q_0)\to 0$ as $x\to1$, integrability, positivity constraints~\cite{Altarelli:1998gn,Candido:2020yat,Candido:2023ujx}, and the momentum and valence sum rules.
While $\mathcal{H}$ can include any combination of parton flavors, only the gluon $f_g(x,Q_0)$ and the total quark singlet $f_\Sigma(x,Q_0)$ are relevant
to our purposes and other flavors are thus neglected in the following.
Within a Bayesian perspective, $\mathcal{H}$ corresponds to the prior space of all physically acceptable possible functional forms at the chosen reference scale $Q_0$.

The starting point of the linear model formalism is the construction of this functional space $\mathcal{H}$.
Following~\cite{Costantini:2025wxp}, this space is spanned by a large sample of PDF candidates, each of them parameterised by a randomly initialised deep neural network.
This choice is motivated by the well-known property of neural networks to be universal function approximators.
Specifically, we adopt the same neural network architecture as in NNPDF4.0~\cite{NNPDF:2021njg,NNPDF:2021uiq} and draw $M=2\times 10^4$ samples,
each corresponding to a different random initialization of its weights and thresholds.

At variance with the procedure in Ref.~\cite{Costantini:2025wxp}, in which all flavours are constrained by the data, here we only constrain the gluon PDF, and set the quark PDFs
to match a given reference PDF set ${\boldsymbol f}^{\rm ref}(x,Q_0)$.
If we denote by $\widetilde{f}_g^{(m)}$ and $\widetilde{f}_\Sigma^{(m)}$ the $m$-th member of the space $\mathcal{H}$, consisting of a total of $M$ elements, we have
\bea
x\widetilde{f}_g^{(m)}(x,Q_0) &=&A_mx^{1-\alpha_m}(1-x)^{\beta_m}{\rm NN}_m(x;\boldsymbol{\theta}) \, , \nonumber\\
x\widetilde{f}_\Sigma^{(m)}(x,Q_0)&=&xf_{\Sigma}^{\rm ref}(x,Q_0) \, ,
\label{eq:H_candidate_space}
\eea
where ${\rm NN}_m(x;\boldsymbol{\theta})$ stands for the output of a fully connected neural network, randomly initialized, with the same architecture as NNPDF4.0.
In Eq.~(\ref{eq:H_candidate_space}), the preprocessing exponents $\alpha_m$ and $\beta_m$ impose the physically motivated asymptotic behavior of the gluon PDF at small-$x$ and large-$x$ respectively.
For each element $m$, these exponents are sampled at random from a flat distribution in the interval
\be
\label{eq:preprocessing_range_variations}
\alpha_m \in \lc 0.5,1.8 \rc \, ,\qquad
\beta_m \in \lc 2, 6 \rc \, ,
\ee
which covers the corresponding asymptotic behaviour of a broad range of different PDF fits~\cite{Ball:2016spl} and is consistent with the integrability and positivity constraints~\cite{NNPDF:2021njg}.
We have verified that our results are stable upon variations of the ranges defined in Eq.~(\ref{eq:preprocessing_range_variations}).
The normalisation constant $A_m$ is fixed by requesting the momentum sum rule which implies
\be
A_m = \frac{1-\int_0^1 dx\,xf^{\rm ref}_{\Sigma}(x,Q_0)}{\int_0^1 dx\,x^{1-\alpha_m}(1-x)^{\beta_m}{\rm NN}_m(x;\boldsymbol{\theta})} \, , \qquad m=1,\ldots,M \, ,
\ee
where the denominator depends on the preprocessing exponents and the initialisation of the PDFs.
By sampling Eq.~(\ref{eq:H_candidate_space}), we construct the Hilbert space $\mathcal{H}$ which contains all {\it a priori} acceptable functional forms for
the gluon PDF, for a given assumption on the reference quark singlet PDF.
Elements with excessive arc length may arise due to spurious fluctuations and are removed using the outlier pruning procedure of~\cite{Costantini:2025wxp}.

As discussed in~\cite{Costantini:2025wxp}, it is advantageous to consider a shifted version of the PDF space $\mathcal{H}$ defined as
\be
\label{eq:hilbert-space-redefinition}
\widehat{\mathcal{H}} = \mathcal{H} - \varphi^{(0)} \, ,
\ee
where $\varphi^{(0)}$ is the mean of the $M$ generated samples.
This way, $\widehat{\mathcal{H}}$ spans the space of the allowed deviations with respect to the central element $\boldsymbol{\varphi}^{(0)}$.
Following this redefinition, the elements of $\widehat{\mathcal{H}}$ are now indicated by $\widehat{f}$ and are given by
\bea
x\widehat{f}_g^{(m)}(x,Q_0) &=&\lp A_m\,x^{1-\alpha_m}(1-x)^{\beta_m}{\rm NN}_m(x;\boldsymbol{\theta})\rp  - x\varphi_{g}^{(0)}(x,Q_0)   \, , \nonumber\\
x\widehat{f}_\Sigma^{(m)}(x,Q_0)&=& 0 \, ,
\label{eq:H_candidate_space_v2}
\eea
Since all the singlet functions were set to the same
reference, as presented in Eq.~\eqref{eq:H_candidate_space}, subtracting their
mean sets them all to zero.
From Eq.~\eqref{eq:hilbert-space-redefinition} it follows that the elements of $\widehat{\mathcal{H}}$ satisfy
\be
\label{eq:msr_Htilde}
\int_0^1dx\,x \lp \widehat{f}_g^{(m)}(x,Q_0) + \widehat{f}^{(m)}_\Sigma(x,Q_0) \rp = \int_0^1\,dx\,x  \widehat{f}_g^{(m)}(x,Q_0)= 0 \, ,\quad \, m =1,\ldots,M \, ,
\ee
which is the homogeneous variant of the original momentum sum rule.

Following the procedure described in Ref.~\cite{Costantini:2025wxp}, we can now construct a linear representation of $\widehat{\mathcal{H}}$, defined in terms of $N\ll M$ basis functions, satisfying the theoretical requirements whose numerical coefficients will be constrained from experimental data. This is done via the Proper Orthogonal Decomposition (POD) described in the next paragraph.
Such a linear model for the gluon PDF is given by
\begin{equation}
\label{eq:linear_model_definition}
f_g(x, Q_0,\boldsymbol{c}) \equiv \varphi_{g}^{(0)}(x, Q_0)  + \sum_{a=1}^N c_a \varphi_{g}^{(a)}(x,Q_0) \, ,
\end{equation}
whose first term is the same average of the initially generated functions, used for the construction of the Hilbert space $\widehat{\mathcal{H}}$
as described in Eq.~\eqref{eq:hilbert-space-redefinition}.
In Eq.~(\ref{eq:linear_model_definition}), $\varphi_a^{(g)}(x,Q_0)$ are the basis functions which parametrise the deviations of the linear model $f_g(x, Q_0,\boldsymbol{c})$ with
respect to the mean function $\varphi_{g}^{(0)}$, the coefficients \bc, individually denoted by $c_a$, are to be constrained from the data, and $N$ specifies the dimensionality of the model.
With an appropriate choice of basis functions $\{\varphi_a(x,Q_0)\}$ at any given reference scale $Q_0$
and for sufficiently large values of $N$, the linear model should be able to describe all functional forms present in $\widehat{\mathcal{H}}$.

The elements of the linear model satisfy the momentum sum rule
\begin{equation}
\label{eq:momentum_sum_rule}
    \int_0^1 dx\,x\,\left[ f_g(x, Q_0,\boldsymbol{c})+f_{\Sigma}^{\rm ref}(x,Q_0) \right] = 1
\end{equation}
for any value of the reference scale $Q_0$ and the coefficients $\boldsymbol{c}$.
It can be checked by inserting Eq.~(\ref{eq:linear_model_definition}) that this momentum sum rule demands
\be
\label{eq:MRS_basis_function_level}
\sum_{a=1}^N c_a \int_0^1 dx\,x\,  \varphi_g^{(a)}(x,Q_0)  = 0 \, .
\ee
The basis functions $\varphi_a^{(g)}$ are the output of the application of the POD method to $\widehat{\mathcal{H}}$, and any
linear, homogeneous property of $\widehat{\mathcal{H}}$ is preserved in the resulting basis
vectors.
Hence, Eq.~(\ref{eq:msr_Htilde}) implies
\be
\label{eq:MRS_basis_function_level_2}
\int_0^1 dx\,x\,  \varphi_g^{(a)}(x,Q_0)  = 0, \, \qquad a=1,\ldots,N \, ,
\ee
and the momentum sum rule Eq.~(\ref{eq:MRS_basis_function_level}) is satisfied irrespective of \bc.

In addition to the momentum sum rule, we need to impose integrability of the gluon PDF, else the momentum integral in Eq.~(\ref{eq:momentum_sum_rule}) becomes undefined.
Integrability can be translated to the requirement that the small-$x$ PDFs do not rise too steeply,
\be
\label{eq:integrability_constraints}
\lim_{x\to 0} x^2 f_g(x, Q_0,\boldsymbol{c})  =0 \, ,
\ee
which at the level of the elements of the space of candidate PDFs $\mathcal{H}$ implies
\be
\lim_{x\to 0} \lp A_m x^{2-\alpha_m}(1-x)^{\beta_m} {\rm NN}_m(x;\boldsymbol{\theta})\rp  = 0,
\ee
which imposes $\alpha_m <2$.
If this value of the small-$x$ preprocessing exponent is enforced, as done in Eq.~(\ref{eq:preprocessing_range_variations}), then all elements of $\mathcal{H}$ satisfy integrability.
Additionally, since the reference PDF also satisfies integrability, it follows that also the elements of the shifted space of candidate PDFs $\widehat{\mathcal{H}}$ satisfy integrability,
and, since the POD method described below ensures that any
linear, homogeneous property of the original sample is maintained by the basis vectors, we have
\be
\lim_{x\to 0} x^2\varphi_g^{(a)}(x,Q_0)=0 \,.
\ee
Thus our associated linear model Eq.~(\ref{eq:linear_model_definition}) will also satisfy integrability for any \bc.

In addition to integrability, the linear model for the gluon PDF must also satisfy positivity~\cite{Candido:2023ujx}.
We account for this by introducing a penalty term which disfavors samples from Eq.~(\ref{eq:H_candidate_space}) leading to negative gluons at the reference scale $Q_0$.

Below, we give a more thorough description of the Proper Orthogonal Decomposition (POD) method that is used in Ref.~\cite{Costantini:2025wxp} to reduce the
dimensionality of the initial basis $M$.
The POD method enables modal decomposition of a given ensemble of functions $\{ \widehat{f}_g^{(m)}(x,Q_0)\}_{m=1}^M$ spanning the infinite-dimensional Hilbert space $\widehat{\mathcal{H}}$,
yielding a basis of functions $\{ \varphi_g^{(a)}(x)\}_{a=1}^N$ that provides a finite-dimensional representation of $\widehat{\mathcal{H}}$ such that each member of the large $M$-dimensional
basis
can be approximated by a linear combination of the elements of a smaller $N$-dimensional orthogonal basis, namely
\be
\widehat{f}_g^{(m)}(x,Q_0)\approx
\sum_{a=1}^N c_a \varphi_g^{(a)}(x,Q_0) \, ,
\ee
from which the linear model for the gluon PDF is constructed by Eq.~(\ref{eq:linear_model_definition}).
POD provides the most efficient means of capturing the key features of $\widehat{\mathcal{H}}$ using only a finite number $N$ of modes.
Once discretised, the POD becomes the Singular Value Decomposition (SVD), which has been used for related PDF applications in~\cite{Carrazza:2015aoa,Carrazza:2016htc}.
The POD procedure starts with tabulating the ensemble $\{ \widehat{f}_g^{(m)}(x,Q_0)\}_{m=1}^M$ in a finite $x$ grid $\{ x_\alpha\}$, with $\alpha=1,\ldots, n$:
\be
{\boldsymbol{\widehat{f}}_g^{(m)}} =\lp \begin{array}{c} \widehat{f}_g^{(m)}(x_1,Q_0) \\
\cdots
\\
\widehat{f}_g^{(m)}(x_n,Q_0)
\end{array}
\rp \, ,
\ee
where $x_1$ and $x_n$ denote the end-points of the $x$-grid.
The same tabulation is done for the basis functions
\be
{\boldsymbol{\varphi}_{g}^{(a)}} =\lp
\begin{array}{c}
\varphi_g^{(a)}(x_1,Q_0) \\
\cdots
\\
\varphi^{(a)}_g(x_n,Q_0)
\end{array}
\rp.
\ee
To determine the orthogonal basis vectors $\{ \varphi_g^{(a)}(x,Q_0)\}_{a=1}^N$ at the $x$-grid nodes, we evaluate the square projection into the space spanned by the elements of $\widehat{\mathcal{H}}$,
namely
\be
\label{eq:inner_product}
\phi\lc \boldsymbol{\varphi}_g \rc = \sum_{m=1}^M \frac{\left|\boldsymbol{\varphi}_g^T \boldsymbol{\widehat{f}}_g^{(m)}\right|^2}{||\boldsymbol{\varphi}_g||^2}
= \frac{1}{||\boldsymbol{\varphi}_g||^2}
\boldsymbol{\varphi}_g^{T} \lp \sum_{m=1}^M \boldsymbol{\widehat{f}}_g^{(m)}\boldsymbol{\widehat{f}}_g^{(m)T}  \rp \boldsymbol{\varphi}_g
\, ,
\ee
where the denominator corresponds to the norm of the basis vector.
We can then diagonalise the autocorrelation matrix of the original ensemble
\be
\label{eq:autocorrelation_matrix}
A = \sum_{m=1}^M \boldsymbol{\widehat{f}}_g^{(m)}\boldsymbol{\widehat{f}}_g^{(m)T} \, ,
\ee
with dimensions $(n \times n)$ dictated by the size of the $x$ grid.
The eigenvectors of the autocorrelation matrix Eq.~(\ref{eq:autocorrelation_matrix}) are the POD modes which maximize the inner product Eq.~(\ref{eq:inner_product}) and, equivalently, the overlap between $\widehat{\mathcal{H}}$ and its linear representation.
By diagonalising Eq.~(\ref{eq:autocorrelation_matrix}) we obtain the sought-for eigenvectors
\be
{\boldsymbol{\varphi}_g^{(a)}} =\lp \begin{array}{c} \varphi_g^{(a)}(x_1,Q_0) \\
\cdots
\\
\varphi_g^{(a)}(x_n,Q_0)
\end{array}
\rp \, , \qquad a=1,\ldots,N \, ,
\ee
ordered by the magnitude of the associated eigenvector $\lambda_a$.
From this finite-dimensional representation, the continuous basis functions $\{\varphi_g^{(a)}(x,Q_0)\}_{a=1}^N$ over the full $x$ range are reconstructed using the {\sc\small LHAPDF} interpolation procedure~\cite{Buckley:2014ana}, with extrapolation applied outside the tabulated $x$ grid.

\begin{figure}[htbp]
    \centering
\includegraphics[width=0.99\linewidth]{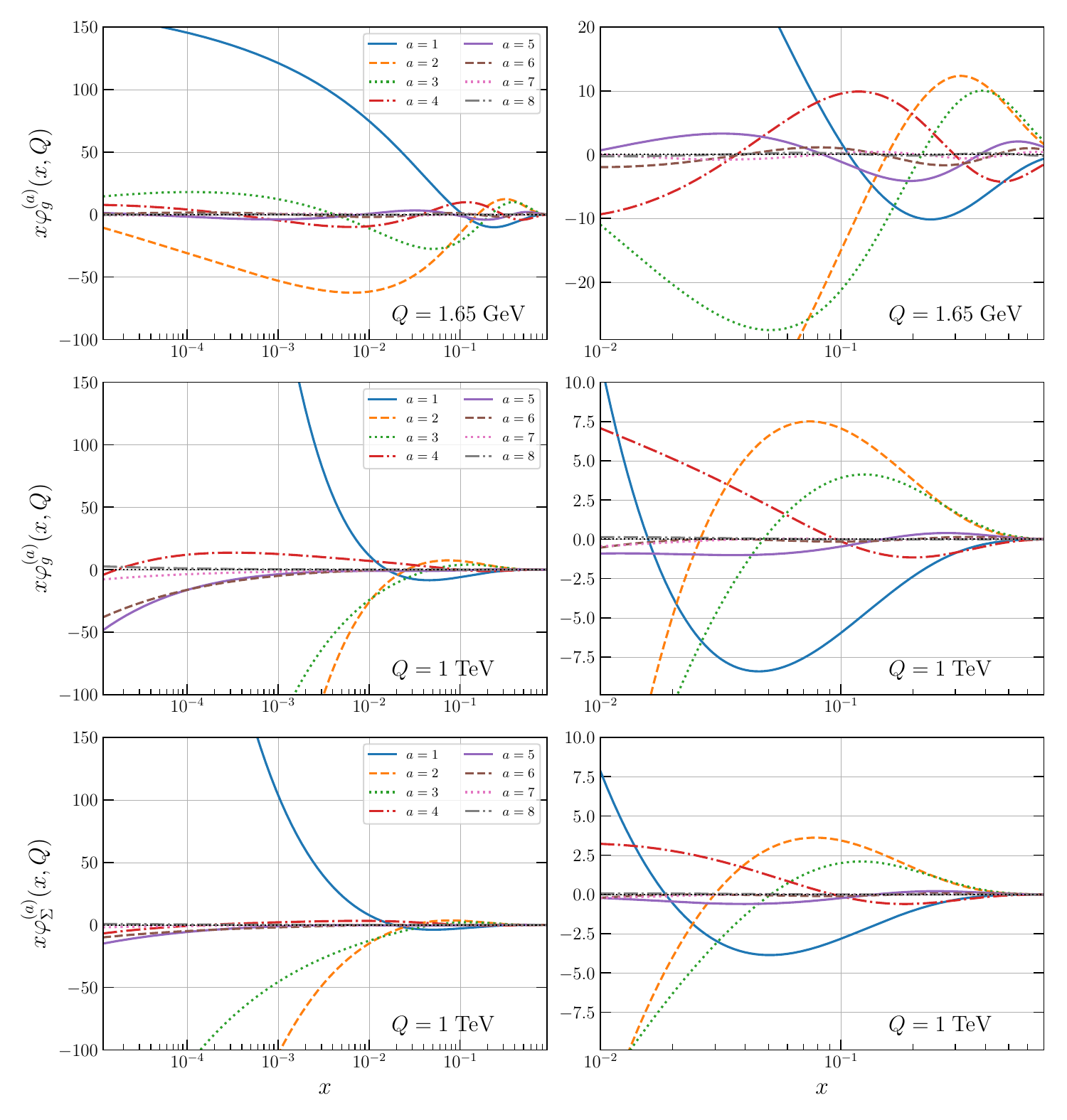}
    \caption{The first $N=8$ eigenvectors $x\varphi^{(a)}_g(x,Q_0)$ for the linear model of the gluon PDF~(top) obtained with the POD procedure for the reference scale $Q_0=1.65$ GeV for the ranges $2\cdot 10^{-5}\leq x\leq 1$ (left) and $10^{-2}\leq x\leq 1$ (right).
    Quark PDFs are set to be equal to the central value of PDF4LHC21.
   The same eigenvectors are also shown for $Q=1$ TeV~(middle), highlighting the effects of DGLAP evolution on the basis vectors.
   The bottom panel shows the same eigenvectors for the singlet quark PDF basis vectors $x\varphi^{(a)}_\Sigma(x,Q)$, generated by DGLAP evolution for $Q > Q_0$ in Eq.~(\ref{eq:singlet_basis_vectors2}).
   In all panels, the linear model basis elements are normalized to the associated eigenvector.
    }
    \label{fig:gluon-basis-elements}
\end{figure}

In the top panels of Fig.~\ref{fig:gluon-basis-elements}, we display the first $N=8$ eigenvectors for the resulting linear model of the gluon PDF. The quark PDFs were set to the central value of PDF4LHC21 NNLO~\cite{PDF4LHCWorkingGroup:2022cjn}.
We adopt $Q_0=1.65$ GeV as reference scale and normalize each eigenvector to its corresponding eigenvalue.
As we will show in Sec.~\ref{subsec:validation}, between $N=6$ and 9 eigenvectors are necessary to achieve the target reconstruction accuracy.

\begin{figure}[t]
    \centering
\includegraphics[width=0.70\linewidth]{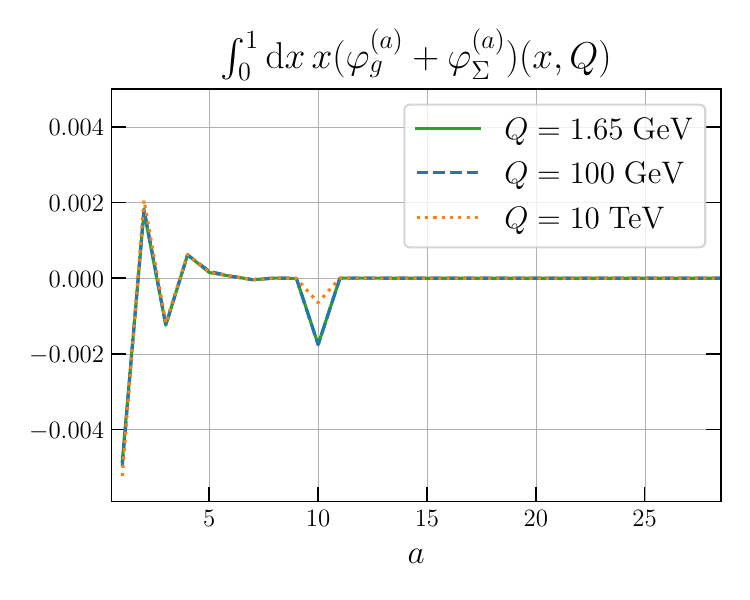}
\vspace{-0.5cm}
    \caption{The numerical value of the homogeneous momentum integral, Eq.~\eqref{eq:MRS_basis_function_level_2}, Eq.~(\ref{eq:homogeneous_momentum_integral}),
    which should be satisfied by the basis functions of our linear model for the gluon PDF at all scales $Q\ge Q_0$.
    We evaluate it for the first $N=30$ eigenvectors of the model at three different scales: $Q=1.65$ GeV, $Q=100$ GeV, and $Q=10$ TeV.
    Note that for $Q=Q_0 = 1.65$~GeV $\varphi_\Sigma^{(a)}(x)=0$ for all $a$.
    The quark PDFs have been set to the central element of PDF4LHC21.
    }
    \label{fig:msr_check}
\end{figure}

\subsection{DGLAP evolution}
\label{sec:dglap_evolution}

So far, we considered the linear model only at the input scale $Q_0$.
To evaluate the model for $Q \ge Q_0$ in the perturbative region, we must account for DGLAP evolution.
Starting from the linear model for the gluon at $Q_0$ in Eq.~(\ref{eq:linear_model_definition}), we write the singlet-sector solution of the DGLAP equations as
\bea
 f_g(x,Q) &=& \Gamma_{gg}(x,Q,Q_0)\otimes f_g(x,Q_0) + \Gamma_{g\Sigma}(x,Q,Q_0)\otimes f_\Sigma(x,Q_0) \, , \\\nonumber
 f_\Sigma(x,Q) &=& \Gamma_{\Sigma g}(x,Q,Q_0)\otimes f_g(x,Q_0) + \Gamma_{\Sigma\Sigma}(x,Q,Q_0)\otimes f_\Sigma(x,Q_0) \, ,
\eea
 where $\otimes$ stands for the convolution operator\footnote{The QCD convolution operator is defined here as
 \begin{equation}
 \nonumber
\left( \Gamma_{ij} \otimes f_j \right)(x,Q,Q_0)
\equiv
\int_x^1 \frac{dy}{y} \,
\Gamma_{ij}\!\left( \frac{x}{y}, Q, Q_0 \right) f_j(y,Q_0) \,,
\qquad i,j = g,\Sigma \,.
\end{equation}
 } and the singlet evolution kernels $\Gamma_{ij}$ can be evaluated in perturbation theory up to approximate N$^3$LO by means of packages such as {\sc\small APFEL}~\cite{Bertone:2013vaa}, {\sc\small HOPPET}~\cite{Salam:2008qg}, or {\sc\small EKO}~\cite{Candido:2022tld,NNPDF:2024nan}.
One can verify that for $Q \ge Q_0$ the linear model is given by
\bea
\label{eq:linearmodel_highq}
 f_g(x,Q,{\boldsymbol{c}}) &=& \varphi_g^{(0)}(x,Q) + \sum_{a=1}^N c_a \varphi_g^{(a)}(x,Q) \, ,  \\
 f_\Sigma(x,Q,{\boldsymbol{c}}) &=&
f_\Sigma^{\rm ref}(x,Q) + \sum_{a=1}^N c_a \varphi_\Sigma^{(a)}(x,Q) \, ,
\label{eq:linearmodel_highq2}
\eea
where we have defined the elements of the PDF linear model for $Q\ne Q_0$ as
\bea
\label{eq:singlet_basis_vectors1}
\varphi_g^{(a)}(x,Q)&\equiv & \Gamma_{gg}(x,Q,Q_0)\otimes \varphi_g^{(a)}(x,Q_0) \, ,\\
\varphi_\Sigma^{(a)}(x,Q)&\equiv & \Gamma_{\Sigma g}(x,Q,Q_0)\otimes \varphi_g^{(a)}(x,Q_0) \, .
\label{eq:singlet_basis_vectors2}
\eea
For the quark singlet PDF, these elements are not independent basis functions, but instead those of the gluon convoluted with the DGLAP kernels.\footnote{Note that the terms in Eqns.~(\ref{eq:singlet_basis_vectors1})--(\ref{eq:singlet_basis_vectors2}) proportional to $\Gamma_{g\Sigma}$ and $\Gamma_{\Sigma\Sigma}$ vanish since $\varphi_{\Sigma}^{(0)}(x,Q_0)=0$ exactly.}
Consequently, evolution does not mix basis elements, such that each element evolves separately under DGLAP.
Eqs.~(\ref{eq:linearmodel_highq})--(\ref{eq:linearmodel_highq2}) indicate that the linear model preserves its additive structure for $Q\ge Q_0$, provided the basis
elements are evolved with corresponding DGLAP kernels.

The middle and bottom panels of Fig.~\ref{fig:gluon-basis-elements} display the evolved basis functions Eqs.~(\ref{eq:linearmodel_highq})--(\ref{eq:linearmodel_highq2}) at $Q=1$\TeV,
both for the gluon and the quark singlet PDFs.
We use NNLO evolution kernels computed with {\sc\small EKO}, with the exact solution of the DGLAP equations, and the same values of $m_c$ and $m_b$ as in NNPDF4.0.
For the gluon PDF, the impact of DGLAP evolution is sizable at both large-$x$ and at small-$x$.
In the latter case, a steep gradient in $Q$ is observed for $a=1,2,3$.
The basis elements of the singlet PDF for $Q \ge Q_0$ follow closely their gluon counterparts, consistent with the solution of the DGLAP evolution equations Eqs.~(\ref{eq:singlet_basis_vectors1})--(\ref{eq:singlet_basis_vectors2}).
It can be shown that the resulting basis functions are identical up to a small overall rescaling if PDF4LHC21 is replaced with NNPDF4.0 for the central value of the quark and antiquark PDFs, as shown in App.~\ref{app:linear_models_PDF}.

The momentum sum rule is conserved upon DGLAP evolution, and the linear model must satisfy
Eq.~\eqref{eq:momentum_sum_rule} at all scales $Q\ge Q_0$.
Using Eq.~(\ref{eq:msr_Htilde}), one can show that for $Q \geq Q_0$ the homogeneous momentum integral is satisfied by the sum of individual basis functions
\be
\label{eq:homogeneous_momentum_integral}
\int_0^1\,dx\,x \lp  \varphi_g^{(a)}(x,Q) + \varphi_\Sigma^{(a)}(x,Q)  \rp =0, \qquad a=1,\ldots,N ,
\ee
and that the momentum sum rule is respected for all allowed values of $Q$.
Figure~\ref{fig:msr_check} displays the homogeneous momentum integral, Eq.~(\ref{eq:homogeneous_momentum_integral}),
for the first $N=30$ eigenvectors of the linear model at three different scales: $Q=1.65$ GeV, $Q=100$ GeV, and $Q=10$ TeV.
The homogeneous sum rules are satisfied with a residual accuracy of $\le 0.5\%$ for all eigenvectors, and its numerical stability with respect to
variations of $Q$ further confirms the correct implementation of DGLAP evolution effects in the linear PDF model.

\subsection{Validation}
\label{subsec:validation}

To determine how many elements $N$ of the linear gluon model are needed to surpass a given reconstruction accuracy, we adopt the following strategy.
First, we apply the linear model Eq.~(\ref{eq:linear_model_definition}) to directly reconstruct
a chosen PDF target $f_T$ composed of $N_T$ individual members.
Note that this set of replicas can be constructed freely.
This reconstruction provides a sample of best-fit coefficients
$\hat c^{(j)}_{a}$,
for $j=1,\ldots,N_T$, which can be used to estimate the distribution
of the theory parameters $\hat c_{a}$ for $a=1,\ldots,N$.
For a specific target $f_T$, and for each member $j$ of the target PDF set, we estimate the best fit values at $Q_0$ by
minimizing the distance estimate
\begin{equation}
\label{eq:MSE_modified_uncertainty_individual}
    d^{(j)}_{{\rm rec},T}(N,Q_0) \equiv \frac{1}{n_x}\sum_{i=1}^{n_x}
    \Big|
    f_{g,T}^{(j)}(x_i,Q_0) - \left( \varphi_g^{(0)}(x,Q_0)
    + \sum_{a=1}^N \hat c_{a}^{(j)}\varphi_g^{(a)}(x,Q_0) \right)
    \Big|\Big/ \sigma_{g,{\rm T}}(x_i,Q_0) \, ,
\end{equation}
with $j=1,\ldots,N_T$, and where $n_x$ denotes the number of points in the $x$ grid on which the distance is computed.
This distance estimate incorporates the uncertainty $\sigma_{\rm T}(x_i,Q_0)$ in the target PDF, which we take to be the size of the 68\% CL interval evaluated at $x_i$ and for a scale $Q_0$.
A similar expression holds for
$Q>Q_0$, in this case we also include the singlet contribution which has, however, no independent degrees of freedom.
Our figure of merit is the arithmetic mean of these distances over the $N_T$ members,
\begin{align}
D_{T}(Q)\equiv\frac{1}{N_T} \sum_{j=1}^{N_T} d^{(j)}_{\rm rec,T}(N,Q),\label{eq:MSE_modified_uncertainty}
\end{align}
together with their associated standard deviation.
Eq.~(\ref{eq:MSE_modified_uncertainty}) has a straightforward interpretation:
a value $D_T$ implies that for $N$ basis elements, the reconstruction accuracy is a factor $D_T$ of the underlying PDF uncertainty averaged over the $N_T$ target gluon replicas and $n_x$ grid points.

In the following, we consider the $N_T=100$ replicas of PDF4LHC21 NNLO set~\cite{PDF4LHCWorkingGroup:2022cjn} in its Monte Carlo representation~\cite{Carrazza:2015hva} as the target gluon PDFs.
The rationale for this choice is that PDF4LHC21 is an unweighted combination of the CT18~\cite{Hou:2019efy}, MSHT20~\cite{Bailey:2020ooq}, and NNPDF3.1~\cite{NNPDF:2017mvq} global PDF determinations, and hence the spread of its replicas for the gluon covers a broad range of results determined from recent phenomenological analyses.
Nevertheless, we have verified that our qualitative findings concerning the reconstruction accuracy of the linear model are unchanged if one uses instead different targets such as ABMP16~\cite{Alekhin:2017kpj} or ATLASpdf21~\cite{ATLAS:2021vod}, with rather different gluon PDFs compared to PDF4LHC21.
Furthermore, we restrict the evaluation of Eq.~(\ref{eq:MSE_modified_uncertainty_individual}) to a region of $x$ defined by $x\in \lc 3\times 10^{-3},0.6\rc$ and take $Q=70$ GeV, which maps to the kinematic region relevant for top-quark pair production at the LHC, as we show in Sec.~\ref{sec:unbinned-obs}.

\begin{figure}[p]
    \centering
\includegraphics[width=0.8\linewidth]{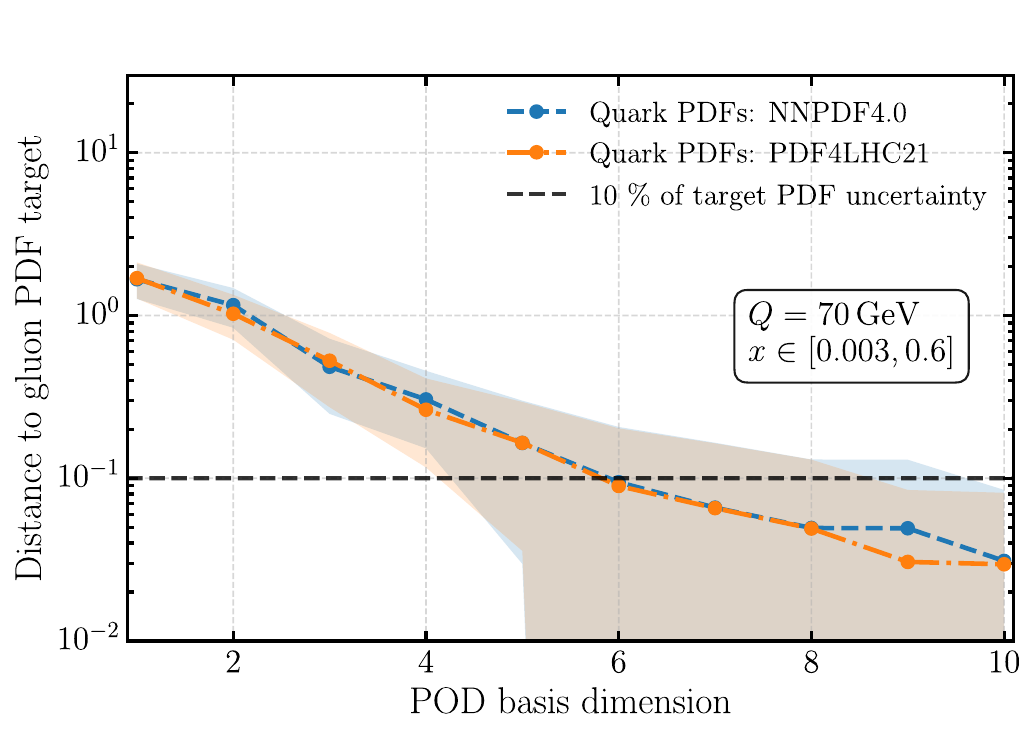}
    \caption{The reconstruction quality of PDF4LHC21 gluon replicas is quantified as the distance to the gluon PDF targets $d^{(T)}_{{\rm rec},j}(N,Q)$, Eq.~(\ref{eq:MSE_modified_uncertainty}), evaluated for the $N_T=100$ original and reconstructed
    replicas of the PDF4LHC21 NNLO set and as a function of the POD basis dimension $N$.
    The central values (dashed curves) indicate the median distance $D_T(Q)$, and the associated band indicates the corresponding standard deviation evaluated over the $N_T$ replicas.
    The dashed horizontal line corresponds to $D=0.1$, for which the inaccuracy of the linear model becomes smaller than 10\% of the native $68\%$ CL PDF uncertainty of PDF4LHC21.
    We show results for two linear models in which the quark PDFs are set to match those of either NNPDF4.0 or PDF4LHC21: the performance is identical in both cases.
    }
\label{fig:pod_validation}
\end{figure}

Figure~\ref{fig:pod_validation} displays the distance to the gluon PDF targets evaluated between the $N_{T}=100$ original and reconstructed replicas of PDF4LHC21 as a function of the POD basis dimension $N$.
The central curves indicate the median distance $D_T$, and the associated band the corresponding standard deviation evaluated over the replicas.
The dashed horizontal line corresponds to $D_T=0.1$, for which the median inaccuracy of the linear model becomes smaller than 10\% of the native $68\%$ CL PDF uncertainty of PDF4LHC21.
Qualitatively similar results are obtained for other PDF sets, as shown explicitly in the case of NNPDF4.0.

\begin{figure}[p]
    \centering
\includegraphics[width=0.8\linewidth]{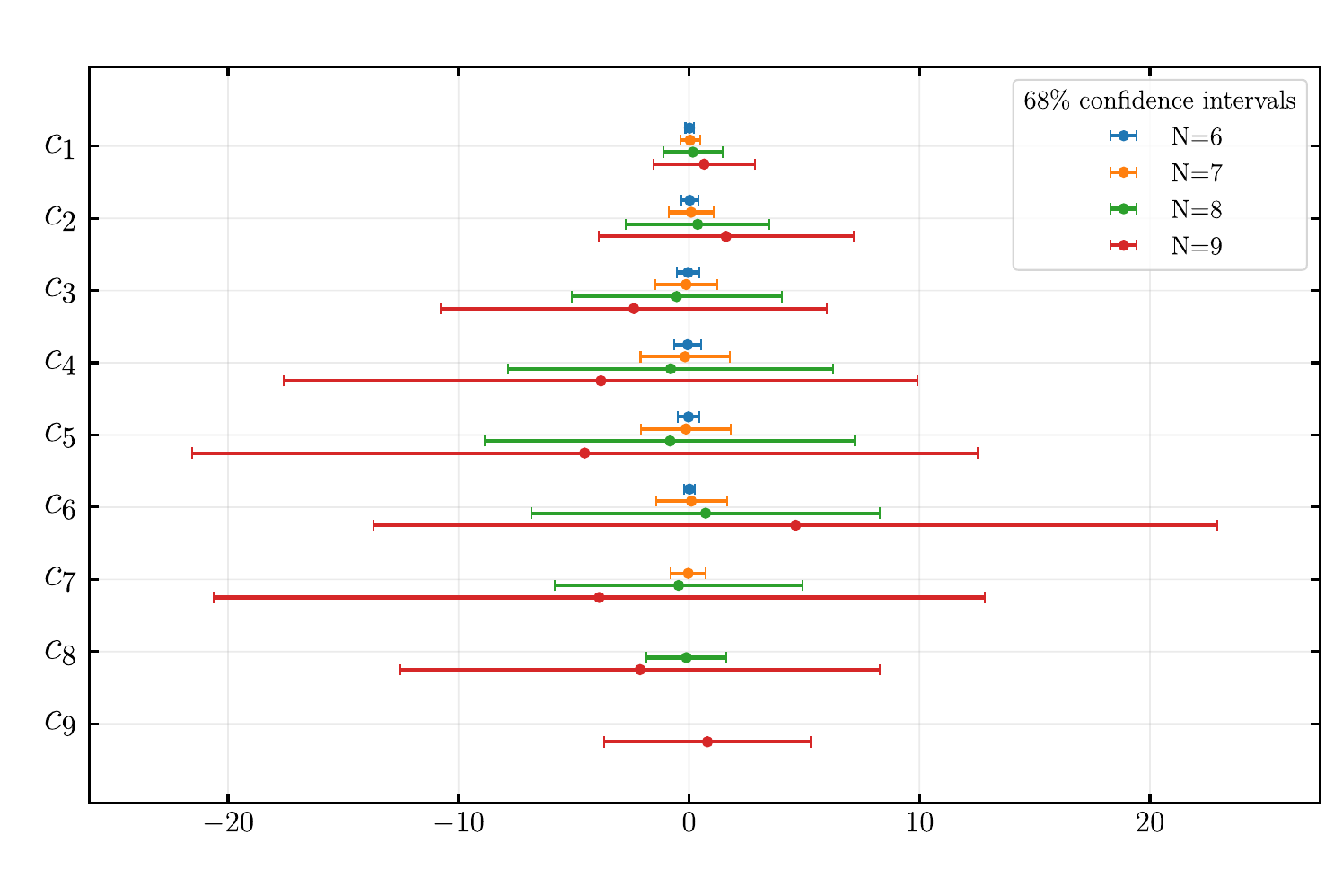}
    \caption{The 68\% CL intervals on the coefficients $c_a$ with $a=1,\ldots,9$ for the linear PDF model fitted to the same PDF4LHC21 replicas as in Fig.~\ref{fig:pod_validation}.
    We show results for linear models with $N=6,\ldots,9$ basis elements.
    }
\label{fig:weight_bounds_PDF4LHC21}
\end{figure}

From Fig.~\ref{fig:pod_validation} one can determine the dimension $N$ of the linear model basis required to reconstruct the original gluon PDF with a mismatch which is on average less than 10\% of the corresponding PDF uncertainty.
We find that $N=6$ is sufficient to achieve $D_T \le 0.1$, while for $N=9$ the same condition is also achieved for the upper range of the standard deviation.
Our analysis indicates that if one aims to a target reconstruction accuracy of 10\% for the gluon PDF (in units of the PDF uncertainty), a range $6 \le N \le 9$ for the linear model is sufficient.\footnote{Within a Monte Carlo PDF determination such as NNPDF or PDF4LHC21\_mc, the ``uncertainty on the PDF uncertainty'' has been estimated to be at the 10\% level~\cite{Demartin:2010er}, motivating our choice of threshold.}
Adding more elements is both unnecessary and would introduce numerical instabilities in the fit.
Furthermore, we have verified that the local (fine-grained) reconstruction accuracy (in narrower intervals of $x$) is similar to the global (coarse-grained) one shown in Fig.~\ref{fig:pod_validation}.

Finally, Fig.~\ref{fig:weight_bounds_PDF4LHC21} displays the 68\% CL intervals associated to the coefficients $c_a$ with $a=1$ to 9 for the linear PDF model fitted to the same PDF4LHC21 replicas as in Fig.~\ref{fig:pod_validation}.
We show results for linear models with a maximum of $N=6,7,8$ and $9$ basis elements.
We find that the spread in the distribution of $\bc$ increases rapidly with $N$, both for the low-$a$ and high-$a$ coefficients.
This result indicates that when $N$ becomes too large (in terms of the complexity of the underlying PDF to be reconstructed), the linear model still works but only through large cancellations between the basis elements, which may (and actually does) lead to numerical instabilities when fitting the linear model to the data.

Motivated by these findings, for the results shown in the following of the paper we adopt $N=7$ as the baseline dimension of the linear model.
We further demonstrate in Sect.~\ref{subsec:stability} the stability of our results with respect to this choice  both for the binned and unbinned determinations of the gluon PDF.

\section{NSBI for PDF determinations}
\label{sec:nsbi_for_pdfs}

In this section we describe the NSBI strategy used to constrain the coefficients \(\bc\) of the linear gluon model through unbinned observables at the LHC.
A key consideration is the dependence of hadronic cross sections on the parameters of the linear PDF model.
Since proton--proton scattering involves the product of the two PDFs of the incoming partons, and since each PDF depends linearly on the coefficients \(\bc\), both parton-level and detector-level predictions are at most quadratic functions\footnote{We neglect the small PDF dependence entering the simulation of initial-state radiation. This dependence involves PDF ratios and is therefore very mild in the perturbative region.} of \(\bc\).
This configuration is thus formally analogous to the quadratic dependence on Wilson coefficients encountered in SMEFT analyses, and it simplifies the training of machine-learned surrogates that exploit this analytic structure, e.g. Refs.~\cite{Chen:2020mev,GomezAmbrosio:2022mpm,Schofbeck:2024zjo,Brehmer:2019xox}.

We denote observed detector-level features by \(\bx\), not to be confused with the partonic momentum fraction \(x\).
We collectively denote the latent variables (not directly accessible by the experiment) by
\begin{equation}
\bz=(m,n,x_1,x_2,\Phi,\bz_{\rm rest})\,,
\end{equation}
where \(m,n\) are the incoming parton flavours, \(x_1,x_2\) their momentum fractions, \(\Phi\) the hard-process phase space, and \(\bz_{\rm rest}\) the remaining shower, hadronisation, and detector degrees of freedom. The renormalisation and factorisation scales are denoted by \(\mu_R\) and \(\mu_F\), respectively.
The dependence on the parameters of interest \(\bc\) thus enters only through the PDFs \(f_m(x,\mu_F,\bc)\), primarily affecting the parton-level description. The dilution of PDF effects at the subsequent stages, in particular at the level of the detector, poses a challenging inverse problem that we solve with NSBI. This approach is firmly rooted in a principled statistical model that, in turn, relies on the analytic structure of the theoretical predictions and accurate simulation of intractable components. Hence, we provide a detailed account of binned and unbinned predictions and how these can be used to extract information on PDFs in a nearly optimal way.

\subsection{Simulating linear PDF effects with systematic uncertainties}
\label{sec:parton-vs-detector}

At the parton level, the fiducial cross section of a hard-scattering reaction can be written as an integral over the hard partonic differential cross section
\begin{equation}
  \bar\sigma(\bc)
  =
  \sum_{m,n}\int \dd x_1\,\dd x_2 \int \dd\Phi\;
  f_m(x_1;\mu_F,\bc)\,f_n(x_2;\mu_F,\bc)\,
  \frac{\dd\hat\sigma^{mn}(\Phi;\mu_F,\mu_R)}{\dd\Phi}\,\label{eq:partonl-level-xsec}
\end{equation}
Because our PDF model is linear in \(\bc\), any parton-level inclusive or differential observable is an exact quadratic function of the coefficients \(\bc\).
To obtain detector-level predictions, the partonic final state must be propagated through showering, hadronisation, the underlying event, detector simulation, and event reconstruction.
We absorb these effects formally into a simulator-implicit transfer density \(p_{X|\Phi}(\bx|\Phi)\), which is only available through sampling.
The resulting detector-level fiducial differential cross section is
\begin{equation}
  \frac{\dd\sigma(\bx|\bc)}{\dd\bx}
  =
  \sum_{m,n}\int \dd x_1\,\dd x_2 \int \dd\Phi\;
  f_m(x_1;\mu_F,\bc)\,f_n(x_2;\mu_F,\bc)\,
  \frac{\dd\hat\sigma^{mn}(\Phi;\mu_F,\mu_R)}{\dd\Phi}\,
  p_{X|\Phi}(\bx|\Phi)\, .
  \label{eq:detector-xsec}
\end{equation}
Since \(p_{X|\Phi}\) is independent of the parameters \(\bc\), the quadratic dependence on the PDF coefficients is preserved at detector level.
The integral in Eq.~(\ref{eq:detector-xsec}) is approximated using Monte Carlo generators together with subsequent software for simulating particle decays, the parton shower, hadronisation, the interaction with the detector material, and event reconstruction.

Rather than sampling Eq.~(\ref{eq:detector-xsec}) with the \(\bc\)-dependent PDFs, we can generate an event sample with a PDF set \(f_m^{\rm(gen)}(x;\mu_F)\) and then apply a weighting defined by
\begin{align}
w_i(\bc)=w_{i,0}\,\omega(\bz_i,\bc),\label{eq:reweight-c}
\end{align}
where $w_{i,0}$ is the nominal weight of the $i$-th event, provided by the generator with this PDF, and the factor \(\omega(\bz_i,\bc)\) is the ratio of the desired PDF to the generator PDF, evaluated for the corresponding parton-level configuration.
Specifically, we use
\begin{equation}
  \omega(\bz_i,\bc)
  =
  \frac{f_{m_i}(x_{1i};Q_i,\bc)\,f_{n_i}(x_{2i};Q_i,\bc)}
       {f_{m_i}^{\rm(gen)}(x_{1i};Q_i)\,f_{n_i}^{\rm(gen)}(x_{2i};Q_i)},
  \label{eq:event-ratio_2}
\end{equation}
where \(Q_i\) is the value of the dynamic (factorization) scale associated to the event \(i\).
We have verified (see Sec.~\ref{sec:unbinned-obs}) that this approximation accurately reproduces the exact PDF reweighting provided by the {\sc\small POWHEG} internal algorithm.
We note that the intractable transfer density cancels in the event-wise ratio in Eq.~(\ref{eq:event-ratio_2}).

Next, we associate systematic uncertainties with a vector of nuisance parameters, denoted by \(\bn\).
Systematic uncertainties can be grouped coarsely into theoretical, modeling, and experimental effects, and more finely into individual sources such as uncertainties in jet energy scale, resolution, etc.
The value \(\bn=\bzero\) corresponds to the nominal simulation, including the best available calibrations of the parton-, particle-, and detector-level data. Non-zero nuisance parameters then correspond to deviations from the nominal simulation and are normalised such that values of \(\pm1\) correspond to the Gaussian approximation of the likelihood of auxiliary measurements constraining these effects.

Accounting for systematic effects that modify the detector-level observations, such as variations of the jet energy scale, is computationally expensive, effectively preventing the generation of simulated data that can be varied continuously with \(\bn\).
We assume, however, that simulated data sets are available for a sufficient number of fixed non-zero \(\bn\) values such that ML surrogates can learn an interpolation.
When a systematic variation can be described by reweighting an existing data set, the computational load is typically milder, and it is sufficient to extend Eq.~(\ref{eq:reweight-c}) by a nuisance-parameter-dependent factor as
\begin{align}
w_i(\bc,\bn)=w_i(\bz_i,\bc)\,\omega_i(\bz_i,\bx_i,\bn).\label{eq:reweight-c-nu}
\end{align}
The factor \(\omega_i(\bz_i,\bx_i,\bn)\) adjusts the probabilistic weight of a given event and can depend on latent or observable quantities\footnote{For a pedagogical discussion, see Ref.~\cite{Schofbeck:2024zjo}.}. We use it, for example, to reweight the simulation so that it reflects changes in the b-tagging efficiency and similar effects.
Combining these systematic variations with the simulation tool chain ultimately produces $N_{\rm sim}$ weighted event samples
\begin{align}
\mathcal{D}_{\bc,\bn}=\{\bz_i,\bx_i,w_{i}(\bc,\bn)\}_{i=1}^{N_\text{sim}},\label{eq:def-variied-sim-data}
\end{align}
where for each event we have the latent variables $\bz_i$, the observed detector-level features $\bx_i$, and the weights $w_i(\bc,\bn)$ which encode the dependence on both the PDF and the nuisance parameters.

It is convenient to set the overall scale of the generator weights \(w_{i,0}\) to a given integrated luminosity.
This is achieved by normalising the inclusive sample for \(\bc=\bn=\bzero\), before any selection requirements, according to
\begin{equation}
\sum_{\mathcal{D}_{\bzero,\bzero}}w_{i,0}=\mathcal{L}_0\,\bar\sigma(\bzero),
\label{eq:norm-total-sample}
\end{equation}
that is, we demand that the overall sum over all nominal event weights equals the product of the inclusive cross section evaluated at \(\bc=\bzero\) and a chosen integrated luminosity, denoted by \(\mathcal{L}_0\), which in this work we set to \(137\)~fb\(^{-1}\) (total Run II luminosity).

In this setup, it is now straightforward to compute detector-level yields.
If we consider an arbitrary detector-level phase-space region \(\Delta \bx\), for example a histogram bin defined by a set of potentially complicated selection requirements, the predicted Poisson yield is given by
\begin{align}
\lambda_{\Delta \bx}(\bc,\bn)\approx\sum_{\bx_i\in\Delta\bx\,\cap\,\mathcal{D}_{\bc,\bn}}w_{i}(\bc,\bn).\label{eq:lumi-yield-norm}
\end{align}

From Eqs.~(\ref{eq:event-ratio_2}--\ref{eq:reweight-c-nu}) it follows that Eq.~(\ref{eq:lumi-yield-norm}), viewed as a function of \(\bc\), is a sum of per-event quadratic polynomials.
Hence, any Poisson yield is also a quadratic polynomial in \(\bc\). This holds, in particular, for the inclusive detector-level cross section, which differs from Eq.~(\ref{eq:partonl-level-xsec}) by the effects of the transfer density function \(p_{X|\Phi}\).
Denoting our full detector-level acceptance region by \(\mathcal{X}\), we have
\begin{align}
\mathcal{L}_0\,\sigma(\bc,\bn)=\sum_{\bx_i\in\mathcal{X}\,\cap\,\mathcal{D}_{\bc,\bn}}w_i(\bc,\bn).
\end{align}

Simulated data sets provide detector-level predictions sufficient for evaluating the likelihood at the sampled values of \(\bn\), but these predictions need not be continuous in \(\bn\). In the binned case, the yields from Eq.~(\ref{eq:lumi-yield-norm}) are therefore interpolated with simple parameterisations, such as those implemented, for example, in the \textsc{Combine} package~\cite{CMS:2024onh}, which then allow one to evaluate the likelihood continuously in \(\bc\) and \(\bn\). In the unbinned case, one must additionally model the continuous dependence on the reconstructed features \(\bx\), which is typically encoded in an ML surrogate. In our approach, these surrogates are constructed with the same log-polynomial dependence on \(\bn\) as in the binned case, together with a quadratic dependence on \(\bc\), reflecting the exact dependence on the parameters of the linear PDF model. In the next section, we construct these surrogates from the unbinned profiled likelihood-ratio test statistic.

\subsection{The extended likelihood}
\label{sec:ext-like}

Consider an observed unbinned data set \(\mathcal D=\{\bx_j\}_{j=1}^{N_{\rm obs}}\) corresponding to an integrated luminosity \(\mathcal L\).
The extended likelihood for a point \((\bc,\bn)\) in model parameter space is
\begin{equation}
L(\mathcal D|\bc,\bn)
=
{\rm Pois}\!\left(N_{\rm obs}\,\middle|\,\mathcal L(\bn)\,\sigma(\bc,\bn)\right)\mathcal{N}(\bn|\bzero,\bone)
\prod_{j=1}^{N_{\rm obs}}
p(\bx_j|\bc,\bn)\, ,\label{eq:ext-likelihood}
\end{equation}
where the probability density function for observing the detector-level features $\bx$ for this point in model parameter space is given by
\begin{equation}
p(\bx|\bc,\bn)=\frac{1}{\sigma(\bc,\bn)}\frac{\dd\sigma(\bx|\bc,\bn)}{\dd\bx}\, .
\end{equation}
The constraint term $\mathcal{N}(\bn|\bzero,\bone)$ is an approximation of the likelihood of real or hypothetical auxiliary measurements.
Relative to the reference point \((\bc,\bn)=(\bzero,\bzero)\), the corresponding log-likelihood ratio is
\begin{equation}
  \log\frac{L(\mathcal D|\bc,\bn)}{L(\mathcal D|\bzero,\bzero)}
  =
  -\Big[\mathcal L(\bn)\sigma(\bc,\bn)-\mathcal L(\bzero)\sigma(\bzero,\bzero)\Big]
  +\sum_{j=1}^{N_{\rm obs}}
  \log\!\left[
  \frac{\mathcal L(\bn)}{\mathcal L(\bzero)}
  \frac{\dd\sigma(\bx_j|\bc,\bn)}{\dd\sigma(\bx_j|\bzero,\bzero)}
  \right]
  -\frac{1}{2}|\bn|^2\, .
  \label{eq:ext-like-mse-2}
\end{equation}
Here we used the fact that the powers of \(\sigma(\bc,\bn)\) from the Poisson term combine with the normalised event density \(p(\bx|\bc,\bn)\) to yield the un-normalized detector-level differential cross section ratio~\cite{Chen:2020mev,Schofbeck:2024zjo}. The reference hypothesis corresponds to the reference gluon PDF and \(\bn=\bzero\), that is, to the nominal choice of nuisance parameters. For the sake of brevity, we do not write the Gaussian penalty term $-\tfrac{1}{2}\bn^2$ from now on.

For the applications considered in this work, a single-process description is sufficient and we therefore suppress process labels. The extension to several contributing processes is straightforward and discussed in detail in Ref.~\cite{Schofbeck:2024zjo}.
We assume that the dependence on the PDF coefficients \(\bc\) factorizes approximately from the dependence on the nuisance parameters \(\bn\).
To make this explicit, we write
\begin{equation}
\frac{\dd\sigma(\bx|\bc,\bn)}{\dd\sigma(\bx|\bzero,\bzero)}
=
\frac{\dd\sigma(\bx|\bc,\bn)}{\dd\sigma(\bx|\bc,\bzero)}
\frac{\dd\sigma(\bx|\bc,\bzero)}{\dd\sigma(\bx|\bzero,\bzero)}
\approx
S(\bx,\bn)\,R(\bx,\bc)\, ,
\label{eq:systematic_PoI_factorisation}
\end{equation}
where we have defined
\begin{equation}
R(\bx,\bc)
\equiv
\frac{\dd\sigma(\bx|\bc,\bzero)}{\dd\sigma(\bx|\bzero,\bzero)}\, ,
\qquad
S(\bx,\bn)
\equiv
\frac{\dd\sigma(\bx|\bzero,\bn)}{\dd\sigma(\bx|\bzero,\bzero)}\, .\label{eq:true-S-R}
\end{equation}
Here \(R(\bx,\bc)\) describes the dependence on the PDF coefficients, while \(S(\bx,\bn)\) accounts for systematic effects.
\subsection{Surrogates for machine learning}
The factorization assumption in Eq.~(\ref{eq:systematic_PoI_factorisation}) is parallel to the same assumption in binned analyses, but in the unbinned case it is typically milder. The reason is that it is imposed at fixed reconstructed event features \(\bx\), rather than only after integrating over a coarse analysis bin. In a binned description, any interplay between the PDF parameters \(\bc\) and the nuisance parameters \(\bn\) within the same bin is averaged over, and can therefore induce violations of the factorization assumed in Eq.~(\ref{eq:systematic_PoI_factorisation}).
In the unbinned case instead, one conditions on a much more informative representation of the event, so that a smaller fraction of the latent-event variation is left unresolved.
As a result, the residual non-factorizing effects are expected to be reduced. This argument becomes stronger the richer the feature vector \(\bx\) is, that is, the more completely it captures the information relevant for the detector-level response.

The quadratic \(\bc\)-dependence of \(R(\bx,\bc)\) can be used to curtail the set of functional forms considered for machine-learned surrogates.
For fixed \(\bx\), we make an ansatz for the surrogate ratio \(\hat R(\bx,\bc)\) as an exact quadratic polynomial in \(\bc\) of the form
\begin{equation}
  \hat R(\bx,\bc)
  =1+\sum_a c_a\,\hat R_a(\bx) + \sum_{a,\,b\leq a}c_a c_b \,\hat R_{ab}(\bx) \equiv
  1+c_A\,\hat R_A(\bx)\,,
  \label{eq:rhat-ansatz}
\end{equation}
where \(c_A=\{c_a,c_a c_b\}\) collects all linear monomials \(c_a\) and symmetric quadratic monomials \(c_a c_b\), and the functions \(\hat R_A(\bx)\), where \(A\in\{a,ab\}\), are represented by neural networks taking only the reconstructed features \(\bx\) as input. Repeated indices $A,B,\ldots$ are summed over. For \(N\) basis functions, \(A\) thus indexes the \(N\) linear term as well as the \(\tfrac{1}{2}N(N+1)\) quadratic terms, so that there are \(N+\tfrac{1}{2}N(N+1)\) different values for \(A\).
Any machine-learned surrogate quantity is denoted with a hat, to differentiate it from the intractable quantities in Eq.~(\ref{eq:true-S-R}).

Similarly, we parameterise the dependence on systematic effects as
\begin{equation}
  \hat S(\bx,\bn)
  =
  \exp\!\left(\nu_A\,\hat\Delta_A(\bx)\right)\, ,\label{eq:S-training-task}
\end{equation}
where, again, the sum over \(A\) can include linear and, in principle, arbitrary quadratic monomials of nuisance parameters.
The functions \(\hat \Delta_A(\bx)\) encode the coefficients of a polynomial expansion of the log-dependence of the differential cross section ratio on the nuisance parameters.

In the case where mixed quadratic terms among groups of nuisances can be neglected, we can factorize \(\hat S(\bx,\bn)\) further as
\begin{equation}
\hat S(\bx,\bn)
=\prod_{p}\hat S_p(\bx,\bn_p)\label{eq:S-fact}
\end{equation}
where \(p\) labels groups of effects, each with its vector \(\bn_p\) of nuisance parameters.
Moreover, many effects are well described by a single (exponentiated) linear term, and the need for quadratic or bilinear terms is the exception.
We will learn the functions \(\hat S_p(\bx,\bn)\) in Eq.~(\ref{eq:S-training-task}--\ref{eq:S-fact}) using the neural-network based approach from Ref.~\cite{Benato:2025rgo} with procedures for obtaining the training data similar to Ref.~\cite{Schofbeck:2024zjo}. Details are provided in Sec.~\ref{sec:unbinned-obs}.
The uncertainty in the luminosity is incorporated through the explicit factor \(\mathcal L(\bn)/\mathcal L(\bzero)\).

\subsection{Binned and unbinned profiled likelihood-ratio test statistic}\label{sec:likelihood-eval}
For a numerically stable evaluation of the likelihood in terms of the surrogate functions, it is convenient to define
\begin{equation}
  \hat T(\bx;\bc,\bn)
  \equiv
  \frac{\mathcal L(\bn)}{\mathcal L(\bzero)}\,\hat S(\bx,\bn)\,\hat R(\bx,\bc)-1
  \label{eq:dcr-in-T}
\end{equation}
which approaches zero when $\bc$ and $\bn$ are small.
Using Eq.~(\ref{eq:dcr-in-T}), the extended log-likelihood ratio relative to the nominal point \((\bc,\bn)=(\bzero,\bzero)\) can then be approximated by
\begin{equation}
\log\frac{L(\mathcal D|\bc,\bn)}{L(\mathcal D|\bzero,\bzero)}
=
-\mathcal L(\bzero)\int \dd\sigma(\bx|\bzero,\bzero)\,
\hat T(\bx;\bc,\bn)
+\sum_{j=1}^{N_{\rm obs}}
\log\!\Big(1+\hat T(\bx_j;\bc,\bn)\Big).\label{eq:likelihood-final}
\end{equation}
where the integral in the first term is evaluated with simulated data as
\begin{align}
  \mathcal{L}(\bzero) \int \dd\sigma(\bx|\bzero,\bzero)\,T(\bx;\bc,\bn) \; \simeq
\sum_{\{w_{i,0},\bx_i\}\in\mathcal D^{\rm sim}_{\bzero,\bzero}}w_{i,0}\hat T(\bx_i;\bc,\bn),
\end{align}
which is the computational bottleneck because, typically, $N_\text{sim}\gg N_\text{obs}$.
The profiled likelihood-ratio test statistic is then
\begin{equation}
q_{\bc}(\mathcal D)
=
-2\left[
\max_{\bn}\log\frac{L(\mathcal D|\bc,\bn)}{L(\mathcal D|\bzero,\bzero)}
-
\max_{\bc',\bn'}\log\frac{L(\mathcal D|\bc',\bn')}{L(\mathcal D|\bzero,\bzero)}
\right].
\label{eq:test-stat}
\end{equation}

For Asimov studies~\cite{Cowan:2010js}, the sum over observed events is replaced by its expectation value under a chosen injected point $(\bc',\bn')$.
For an Asimov data set corresponding to \((\bc,\bn)=(\bzero,\bzero)\) one finds from Eq.~(\ref{eq:likelihood-final}) the expectation
\begin{equation}
\E_{\bc',\bn'}\left[
\log\frac{L(\mathcal D|\bc,\bn)}{L(\mathcal D|\bzero,\bzero)}
\right]
\simeq
\sum_{\{w_{i,0},\bx_i\}\in\mathcal D^{\rm sim}_{\bzero,\bzero}}
w_{i,0}\,
\left[
-\hat T(\bx_i;\bc,\bn)
+
\big(1+\hat T(\bx_i;\bc',\bn')\big)\,
\log\big(1+\hat T(\bx_i;\bc,\bn)\big)
\right].
\label{eq:asimov-test-stat}
\end{equation}

For comparison, the corresponding binned log-likelihood ratio is
\begin{equation}
\log\frac{L_{\rm binned}(\mathcal D|\bc,\bn)}{L_{\rm binned}(\mathcal D|\bzero,\bzero)}
=
\sum_{i=1}^{N_{\rm bins}}
\left[
-\big(\lambda_i(\bc,\bn)-\lambda_i(\bzero,\bzero)\big)
+
N_{i,{\rm obs}}
\log\frac{\lambda_i(\bc,\bn)}{\lambda_i(\bzero,\bzero)}
\right],
\end{equation}
with Asimov expectation under an injected point \((\bc',\bn')\) obtained by replacing
\(N_{i,{\rm obs}}\to\lambda_i(\bc',\bn')\).

\section{Unbinned observables in \texorpdfstring{$t\bar{t}$}{ttbar} production}
\label{sec:unbinned-obs}

In this work we determine the sensitivity of the NSBI analysis to the gluon PDF using top quark pair production at the LHC in a simulated event sample of $pp$ collisions at $\sqrt{s}=13~\TeV$. We consider the dileptonic decay channel
\be
\textrm{pp}\rightarrow\ttbar\rightarrow\textrm{b}\ell^+\nu_\ell\overline{\textrm{b}}\ell^-\overline{\nu}_\ell \, ,
\ee
which we abbreviate as $\ttbar(2\ell)$ in the following.
First, in Sec.~\ref{sec:tt2l-event-simulation}, we describe the event generators that provide all necessary quantities at the parton, particle, and detector levels.
Section~\ref{subsec:reco_level_features} presents the event features reconstructed with \Delphes~\cite{deFavereau:2013fsa} that enter the NSBI likelihood.
Section~\ref{sec:learn-logratio} describes how to machine-learn the PDF dependence in this event sample, and Sec.~\ref{sec:PCA} how to carry out a Principal Component Analysis (PCA) on the PDF dependence.
Section~\ref{sec:learning-systematis} discusses the treatment of experimental and theoretical systematic uncertainties entering our analysis and how to machine-learn the associated surrogates.
Finally, Sec.~\ref{sec:binned-reference} presents the construction of the reference binned measurement.

Many uncertainties in reconstructed objects (jets, missing transverse momentum, and leptons) can be reasonably estimated by the recipes provided by the ATLAS and CMS open-data projects~\cite{CMS-Open-Data,ATLAS-Open-Data}.
A fully realistic detector simulation including all data-dependent systematic effects is neither feasible nor necessary here. Instead, we adopt a heuristic treatment of the dominant systematic uncertainties.
A detailed binned measurement of $\ttbar(2\ell)$, including a full treatment of systematic uncertainties, is available from ATLAS~\cite{ATLAS:2023gsl} and CMS~\cite{CMS:2024ybg}.

\subsection{Event generation and selection requirements}
\label{sec:tt2l-event-simulation}

We generate the $\ttbar(2\ell)$ signal process at next-to-leading order (NLO) in QCD using the \POWHEG-v2 Monte Carlo generator~\cite{Nason:2004rx,Frixione:2007vw,Alioli:2010xd,Frixione:2007nw} at $\sqrt{s}=13\TeV$.
The nominal PDF set is taken to be {\tt NNPDF31\_nnlo\_as\_0118}~\cite{NNPDF:2017mvq}, and the (pole) top-quark mass is set to $m_t=172.5$ GeV.
The strong coupling is set to $\alpha_s(m_Z)=0.118$, with variations of $\delta \alpha_s=\pm 0.001$ provided by the same PDF set.
The renormalisation ($\mu_R$) and factorisation ($\mu_F$) scales are chosen dynamically~\cite{Czakon:2016dgf} on an event-by-event basis using
\be
\label{eq:scale_choice}
\mu_F=\mu_R = H_T/4= \lp \sqrt{m_t^2+p_{T,t}} + \sqrt{m_t^2+p_{T,\bar{t}}}\rp /4\, ,
\ee
with $p_{T,t}$ and $p_{T,\bar{t}}$ being the transverse momentum of the top quark and antiquark, respectively, in the $\ttbar$ rest frame.
The NLO matrix-element calculation is matched to {\textsc{PYTHIA}}~v8.226~\cite{Skands:2014pea} using the CP5 tune~\cite{CMS:2015wcf,CMS:2019csb} for parton showering, fragmentation, hadronisation, multiparton interactions, and the modeling of the underlying event.
We normalise the \Delphes simulation of a total of $10^7$ events to the inclusive prediction evaluated with the \texttt{Top++2.0} program~\cite{Czakon:2011xx}, which includes next-to-next-to-leading-order (NNLO) QCD corrections and the resummation of next-to-next-to-leading logarithmic (NNLL) soft-gluon terms~\cite{Cacciari:2011hy,Beneke:2011mq,Barnreuther:2012wtj,Czakon:2012zr,Czakon:2012pz}.
This cross section calculation yields
\be
\bar\sigma(\bzero)
=
\sigma(\ttbar)\times \mathrm{BR}(t\to b\ell\nu)^2
=
831.8\,\mathrm{pb}\times (3\times 0.1086)^2
=
88.3\,\mathrm{pb}.
\ee

The generated particle-level events are subsequently processed with a \Delphes-based simulation model of the CMS detector.
Kinematic requirements are placed on jets, electrons, and muons.
Jets are reconstructed with the anti-$k_T$ algorithm\cite{Cacciari:2008gp} using a distance parameter of $R=0.4$ in the \textsc{FastJet} software package\cite{Cacciari:2011ma}.
The nominal \cPqb-tagging of jets in \Delphes is based on parton-matching and a parameterisation of the default CMS \cPqb-tagging efficiency.
Electrons and muons must be isolated from jets, satisfy $\pt^\ell>20$\GeV, and be reconstructed within absolute pseudorapidity $|\eta_\ell|<2.5$.
If there are two same-flavor lepton candidates of opposite electric charge within a $10$~\GeV window around the \PZ~boson mass, ${|m_{\ell \bar{\ell}}-m_\PZ|<10\GeV}$, the event is rejected to reduce electroweak $Z$-production backgrounds.
According to the analysis of Ref.~\cite{CMS:2024ybg}, the purity after the \PZ~boson mass veto is 95\%, with small background from the Drell-Yan process and semi-leptonic top-quark pair production with a fake or non-prompt lepton.
We ignore these contributions in the following.
Reconstructed jets must satisfy $\pt^{\rm jet}>30$\GeV and $|\eta_{\rm jet}|<2.4$, and there must be more than two jets in the event, among which at least two must be \cPqb~tagged.

Using the \Delphes objects, we reconstruct the top quark kinematic quantities following the procedure described in Ref.~\cite{CMS:2024ybg}.
We ignore events where the top quark reconstruction algorithm does not find a viable solution.
Subsequently, we construct the observables that will be used as input for the PDF determination, including the top quarks' invariant masses, angles, and transverse momenta.

\begin{figure}[t]
    \centering
\includegraphics[width=0.49\linewidth]{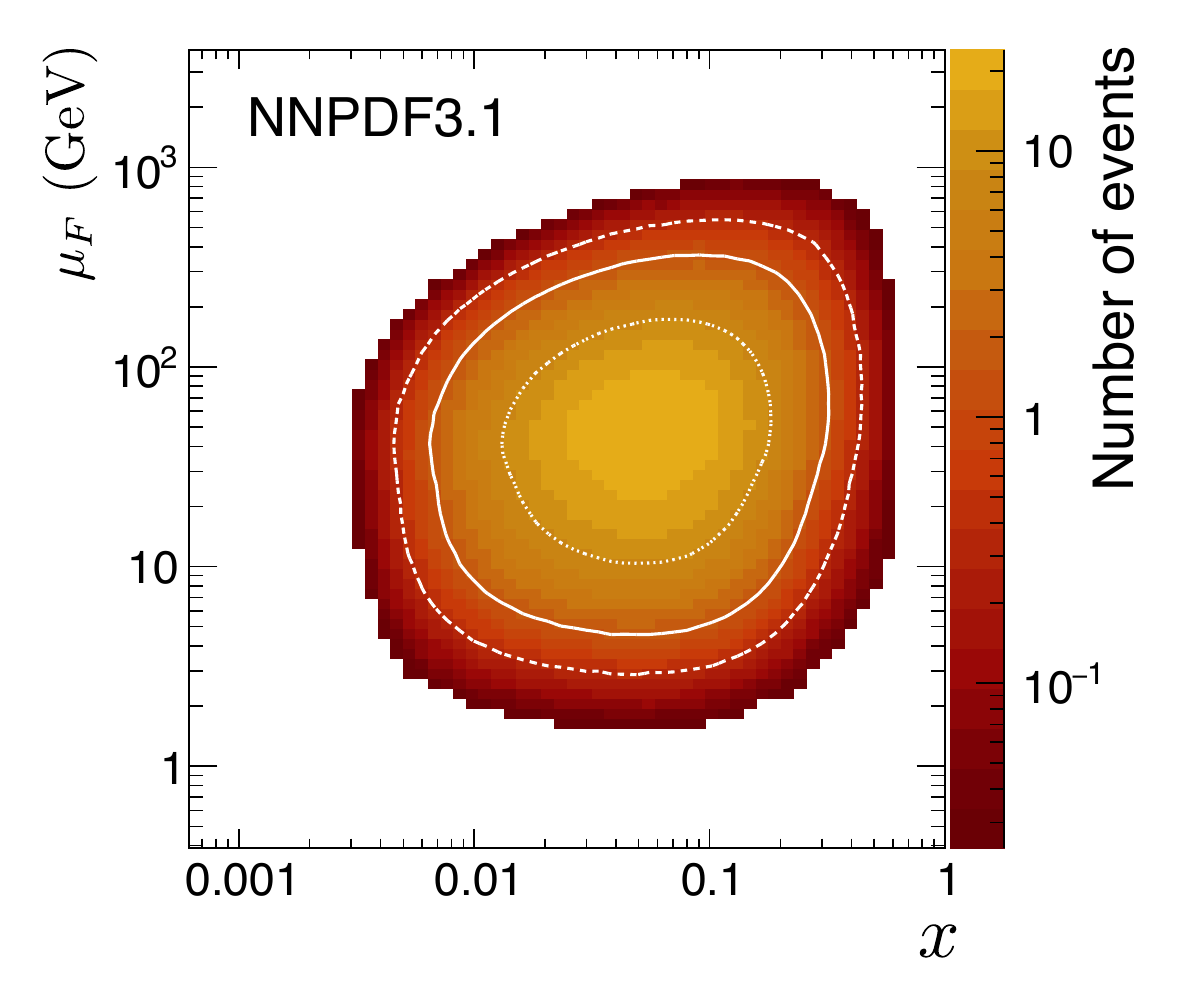}
\includegraphics[width=0.49\linewidth]{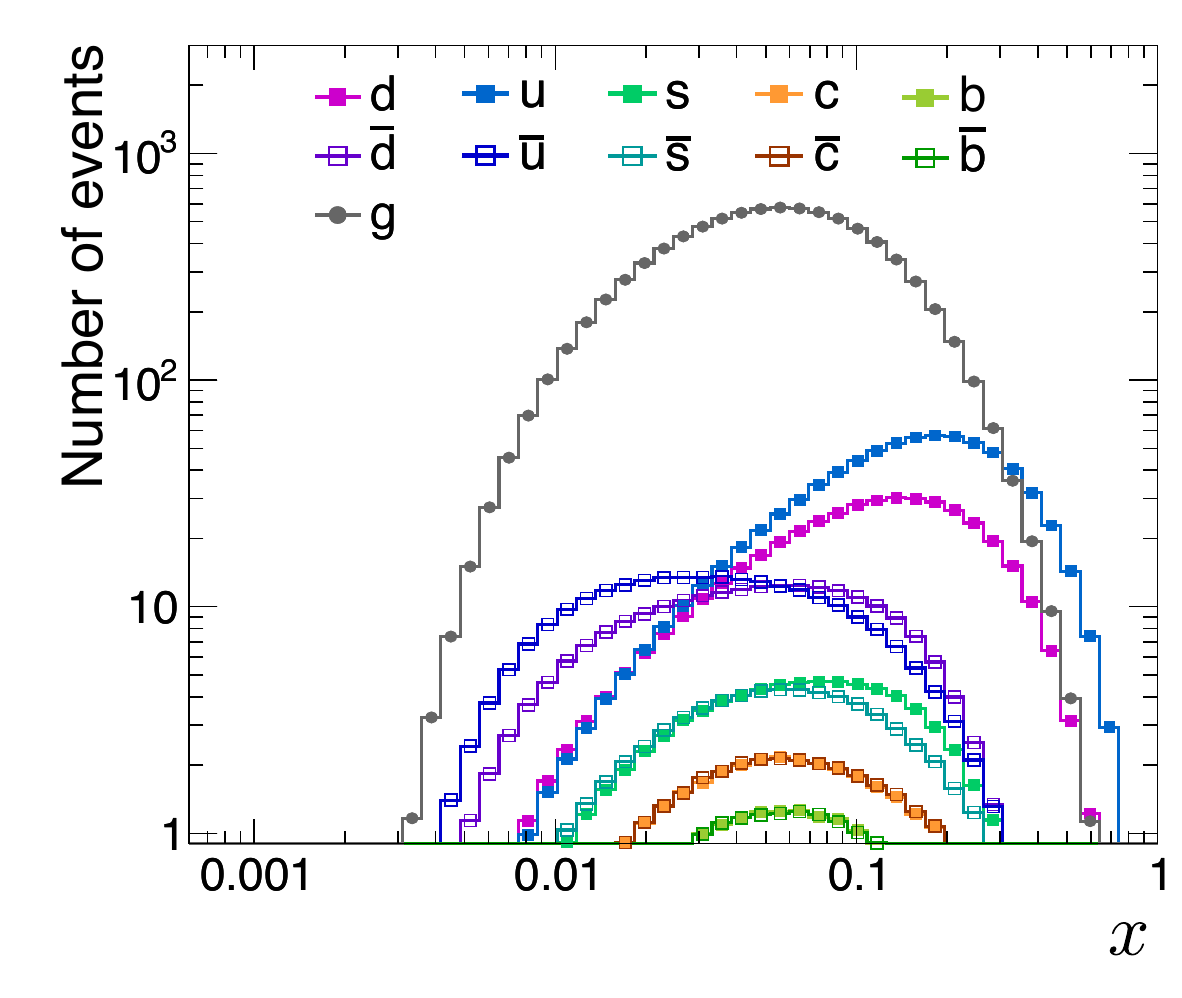}
    \caption{The distribution of $x$ and $\mu_F$ of gluon-initiated sub-processes of the $\ttbar(2\ell)$ signal sample, evaluated with {\sc\small POWHEG} and NNPDF3.1 in the acceptance region~(left).
    The contours indicate the region containing 68\%, 95\%, and 99\% of the events in the sample.
One-dimensional distribution of $x$ for all partonic sub-processes~(right).
    }
    \label{fig:x_muF_density_id_21}
\end{figure}

\paragraph{Partonic coverage.}
From the generated events and the described selection and acceptance cuts, we can determine the kinematic coverage of the $\ttbar(2\ell)$ signal on the partonic kinematics $(x,\mu_F)$.
Figure~\ref{fig:x_muF_density_id_21} displays the distribution of $x$ and $\mu_F$ for events involving gluon scattering ({\tt pid=21}).
The bulk of the events are concentrated in the kinematic region $10^{-2}\le x \le 0.3$ and $30~{\rm GeV}\le \mu_F \le 300$ GeV, with some events reaching up to $x\approx 0.6$ and $\mu_F \approx 1$ TeV.
Therefore, the coverage of the gluon PDF in the region $x\le 5\times 10^{-3}$ can be safely neglected.
The left panel of Fig.~\ref{fig:x_muF_density_id_21} indicates that a determination of the gluon PDF from this LHC top quark pair production data set will be sensitive to the kinematic region $10^{-2}\le x \le 0.5$, and hence the validation of the linear model presented in Sec.~\ref{subsec:validation} can be restricted to this region.
We note that the $\mu_F$ distribution peaks at $Q\approx 70$ GeV, justifying the choice to validate the model at this scale in Sec.~\ref{subsec:validation}.

The right panel of Fig.~\ref{fig:x_muF_density_id_21} presents information similar to that in the left one now integrated over $\mu_F$ and extended to the contributions of the remaining partonic sub-channels.
For $x\ge 0.4$, the contribution from scatterings involving the up quark becomes dominant, and that of down quarks comparable to the gluon contribution.
The contributions from strange quarks and from antiquarks can be neglected over the whole range of $x$.
We recall that the large-$x$ valence quark PDFs, $u_V(x)$ and $d_V(x)$, are already well constrained from fixed-target deep-inelastic scattering measurements~\cite{Gao:2017yyd}.

\begin{figure}[t]
    \centering
\includegraphics[width=0.325\linewidth]{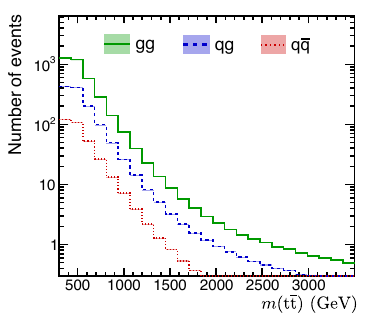}
\includegraphics[width=0.325\linewidth]{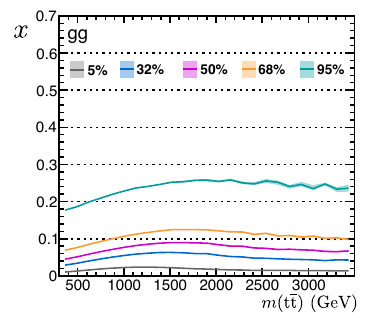}
\includegraphics[width=0.325\linewidth]{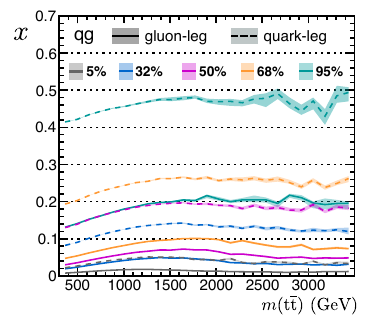}

    \caption{One-dimensional distribution of $m(\ttbar)$ for the partonic channels $gg$, $qg$, and $q\bar{q}$~(left).
    The quantiles of the distribution of $x$ in bins of $m(\ttbar
    )$, separately for the $gg$~(middle) and $qg$~(right) partonic channels.
    For $qg$, we separate the gluon leg and the quark leg.
    Statistical uncertainties are shown as shaded bands.
    }
\label{fig:tr_ttbar_mass_partonic_channels}
\end{figure}

The left panel of Fig.~\ref{fig:tr_ttbar_mass_partonic_channels} displays the decomposition of event weights in top-quark pair production as a function of the invariant mass $m(t\bar{t})$ for the three partonic channels: $gg$, $qg$, and $q\bar{q}$.
Over the full range of $m(t\bar{t})$, the gluon-gluon scattering channel dominates, with the gluon-quark channel subdominant by around a factor of 2.
As indicated by the right panel of Fig.~\ref{fig:x_muF_density_id_21}, scattering between gluons and up quarks represents the main contribution to the quark-gluon channel, which arises at the level of the NLO correction to the Born process.
We also see that the quark-antiquark scattering channel only contributes at the few-percent level, and hence can be safely neglected in the context of a PDF determination.

The middle and right panels of Fig.~\ref{fig:tr_ttbar_mass_partonic_channels} display the quantiles in the momentum fraction $x$ as a function of $m(t\bar{t})$ for the $gg$ and $qg$ partonic channels, respectively.
For the latter, we separate the quantiles from the gluon leg and those from the quark leg.
For gluon-gluon scattering, we see that the $x$ range covered extends to $\approx0.25$ at the 95\% CL, while for quark-gluon scattering the quark leg reaches $x\approx 0.5$ for the 95\% CL quantile, consistent with the information presented in the right panel of Fig.~\ref{fig:x_muF_density_id_21}.

The results of Figs.~\ref{fig:x_muF_density_id_21} and~\ref{fig:tr_ttbar_mass_partonic_channels} confirm that the top-quark pair production data set that we consider in this work has excellent sensitivity to the gluon PDF over a broad range of $x$, and that the contribution from quark PDFs is subleading and restricted to the large-$x$ up quark, which is well determined from inclusive DIS structure function measurements.

\paragraph{PDF reweighting validation.}
In a {\sc\small POWHEG} NLO simulation, the treatment of PDF uncertainties is exact at the level of inclusive quantities at fixed underlying Born.
The treatment of PDFs entering QCD radiation is, instead, approximate.
If one considers, for example, $\ttbar$ production, the PDF error for the $m_{t\bar{t}}$ or $y_{t\bar{t}}$ distributions at NLO, which is inclusive over the radiation, obtained by means of PDF reweighting is exact, whereas the PDF error on the $p_T$ of the leading jet in the event is an approximation.

As discussed in Sec.~\ref{sec:nsbi_for_pdfs}, the NSBI procedure requires the calculation of event weights for each of the basis elements of the PDF linear model.
For each event, this weight can be computed using
\begin{equation}
  w_i^{(j)}
  \;=\;
  \frac{f^{(j)}_{m_i}(x_{1,i};\mu_{F,i})\,f^{(j)}_{n_i}(x_{2,i};\mu_{F,i})}
       {f^{\mathrm{(0)}}_{m_i}(x_{1,i};\mu_{F,i})\,f^{\mathrm{(0)}}_{n_i}(x_{2,i};\mu_{F,i})}\,,
  \label{eq:event-ratio}
\end{equation}
with $i$ labeling the event, and $x_{1,2}$, $\mu_F$, and $(m,n)$ indicating the associated momentum fractions, factorisation scale, and partonic flavor indices, respectively.
In Eq.~(\ref{eq:event-ratio}), $(j)$ labels the element of the linear model, while $(0)$ stands for the central value of the PDF set used in the nominal {\sc\small POWHEG} event generation.
Note that the latter does not need to coincide with the reference PDF set entering the construction of the linear model in Eq.~(\ref{eq:linear_model_definition}).

\begin{figure}[t]
    \centering
\includegraphics[width=0.7\linewidth]{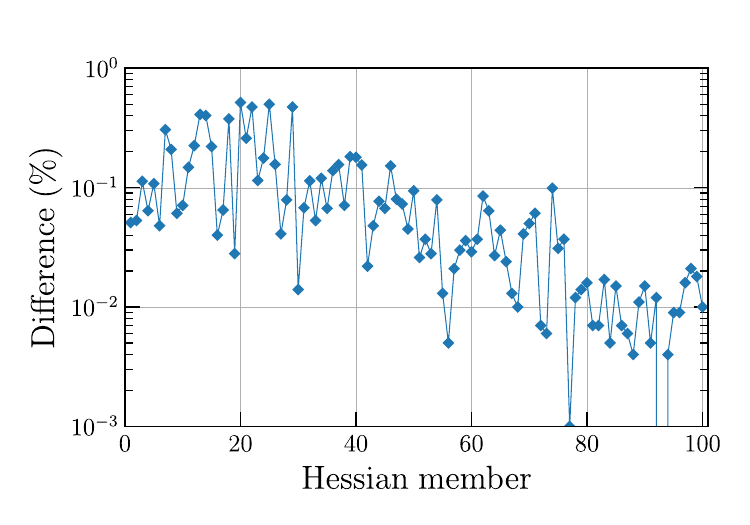}
    \caption{The relative difference between the PDF weight evaluated by {\sc\small POWHEG}'s native reweighting algorithm and the offline evaluation based on Eq.~(\ref{eq:event-ratio}) for each of the $N_{\rm eig}=100$ eigenvectors of NNPDF3.1 (Hessian variant) for a specific Monte Carlo event from our sample.
    Qualitatively similar results are obtained for other events.
        }
    \label{fig:gluon-PDFRW-check}
\end{figure}

In order to verify the validity of Eq.~(\ref{eq:event-ratio}) for the sample used in this study, we compare it with the built-in PDF reweighting feature of {\sc\small POWHEG}, which is provided in the event record, in the case of NNPDF3.1 NNLO.
Figure~\ref{fig:gluon-PDFRW-check} displays the percentage difference between the PDF weight evaluated by {\sc\small POWHEG} and the corresponding evaluation based on Eq.~(\ref{eq:event-ratio}) for each of the $N_{\rm eig}=100$ eigenvectors of the NNPDF3.1 Hessian set and for a specific Monte Carlo event.
For all eigenvectors, Eq.~(\ref{eq:event-ratio}) agrees with the {\sc\small POWHEG} evaluation to better than 1\%.
We conclude that we can reliably adopt Eq.~(\ref{eq:event-ratio}) to construct ML surrogates based on the basis elements of the linear PDF model, as required by the procedure of Sec.~\ref{sec:nsbi_for_pdfs}.
The same qualitative picture arises if we consider other events in the {\sc\small POWHEG} sample.

\subsection{Reconstructed event features}
\label{subsec:reco_level_features}

Measurements of top-quark pair production at the LHC are typically presented in terms of single-, double- or triple-differential distributions of certain kinematic variables~\cite{ATLAS:2022xfj,CMS:2019esx,CMS:2023qyl}, either at the parton level (in terms of reconstructed top quarks) or at the particle level (in terms of final-state leptons and jets).

In the case of parton-level measurements, results are often obtained as a function of the rapidities, $y(t),y(\bar{t}),y(\ttbar)$, transverse momenta, $p_T(\ttbar),p_T(t),p_T(\bar{t})$, and invariant mass $m(\ttbar)$.
In the context of applications of top-quark pair production to constrain the gluon PDF~\cite{Czakon:2016olj,Czakon:2019yrx,Bailey:2019yze}, the rapidity and invariant mass distributions are typically preferred due to their more direct connection with the kinematics of the underlying hadronic collision, while instead transverse momentum distributions are more sensitive to the pattern of QCD radiation.
The invariant mass distribution $m(\ttbar)$ and the rapidity distributions at high values are particularly sensitive to the gluon PDF at large-$x$, where PDF uncertainties are the largest.

The unbinned observables that we use in this work to constrain the gluon PDF are defined in terms of 16 event-level kinematic features $\bx$.
From the reconstructed top quark four-momenta, we compute the invariant mass $m(\ttbar)$, the transverse momentum $p_\textrm{T}(\ttbar)$, and the rapidity $y(\ttbar)$ of the $\ttbar$ system.
In addition, we consider the rapidity difference
$\Delta\eta(\ttbar)=\eta(\textrm{t})-\eta(\overline{\textrm{t}})$ and the difference of absolute rapidities of the top and anti-top quark,
$\Delta|\eta|(\ttbar)=|\eta(\textrm{t})|-|\eta(\overline{\textrm{t}})|$.
The quantities $m(\ttbar)$ and $p_\textrm{T}(\ttbar)$ are sensitive to PDF effects at high Bjorken $x$, while $\Delta|\eta|(\ttbar)$ is sensitive to the effects of the charge asymmetry~\cite{CMS:2022ged,ATLAS:2022waa}, with potential benefit to discriminate against modifications of the light-quark PDFs.
Furthermore, as discussed above, we include the transverse momenta and rapidities of the individual top and anti-top quarks, $p_\textrm{T}(\textrm{t})$, $p_\textrm{T}(\overline{\textrm{t}})$, $y(\textrm{t})$, and $y(\overline{\textrm{t}})$.

Lepton-level observables in the $\ttbar(2\ell)$ channel provide clean probes of the gluon PDF which are largely independent of the hadronic activity and the associated experimental uncertainties in jet momenta and missing transverse momentum.
Here we consider the transverse momenta of the leading and subleading leptons, $p_\textrm{T}(\ell_0)$ and $p_\textrm{T}(\ell_1)$, as well as kinematic properties of the dilepton system: the invariant mass $m(\ell^+\ell^-)$, the transverse momentum $p_\textrm{T}(\ell^+\ell^-)$, the pseudo-rapidity $\eta(\ell^+\ell^-)$, the rapidity difference $\Delta\eta(\ell^+\ell^-)$, and the difference of absolute rapidities $\Delta|\eta|(\ell^+\ell^-)$, defined analogously to their top-level counterparts.

Table~\ref{tab:features} summarizes the 16 event-level observables that constitute the detector-level observation $\bx$ used in this analysis to constrain the gluon PDF both at the particle level (in terms of the reconstructed leptons) and at the parton level (in terms of top quark kinematics).
Since we evaluate event-wise probabilities from the simulation, no double counting of information is incurred by considering e.g., both top-level and lepton-level quantities when constructing the likelihood.
This property represents a crucial advantage of unbinned measurements as compared to the binned counterparts.
For instance, in the latter case one cannot include in the PDF fit at the same time top-quark-level and particle-level measurements corresponding to the same data set, since the statistical correlations are not provided, and the same holds frequently for the binned measurements presented in terms of different top-quark-level kinematic variables.

{\renewcommand{\arraystretch}{1.6}
\begin{table}[t]
\centering
\begin{tabular}{|l|l||l|l|}
\hline
Observable & Description & Observable & Description \\
\hline
$m(\ttbar)$                & invariant mass of the top quark pair          & $p_T(\ttbar)$              & $\pt$ of the top quark pair \\
$y(\ttbar)$                & rapidity of the top quark pair            & $\Delta\eta(\ttbar)$       & difference of $\eta$ of the $\ttbar$ system\\
$\Delta|\eta|(\ttbar)$     & absolute rapidity difference of the $\ttbar$ system     & $p_T(t)$                   & $\pt$ of the top quark \\
$p_T(\bar t)$              & $\pt$ of the anti-top quark              & $y(t)$                     & rapidity of the top quark\\
$y(\bar t)$                & rapidity of the anti-top quark          & $p_T(\ell_0)$              & $\pt$ of the leading lepton \\
$p_T(\ell_1)$              & $\pt$ of the subleading lepton         & $p_T(\ell\overline{\ell})$            & $\pt$ of the $\ell\overline{\ell}$ system \\
$m(\ell\overline{\ell})$              & invariant mass of the dilepton system                   & $\eta(\ell\overline{\ell})$           & pseudorapidity of the $\ell\overline{\ell}$ system\\
$\Delta\eta(\ell\overline{\ell})$     & difference of $\eta$ of the $\ell\overline{\ell}$ system   & $\Delta|\eta|(\ell\overline{\ell})$   & difference of $|\eta|$ of the $\ell\overline{\ell}$ system \\
\hline
\end{tabular}
\vspace{0.4cm}
\caption{Description of the 16 event-level observables that constitute the detector-level observables $\bx$ used in this analysis to constrain the gluon PDF.
}\label{tab:features}
\end{table}
}

\subsection{Machine-learning the PDF dependence}
\label{sec:learn-logratio}

As discussed in Sec.~\ref{sec:nsbi_for_pdfs}, the ML surrogate for \(R(\bx,\bc)\) is obtained by minimizing the weighted mean-squared-error (MSE) loss function,
\begin{equation}
  \mathcal L_{\rm MSE}
  =
  \sum_{\bc\in\mathcal V}\sum_i
  w_{i,0}\,
  \Big(\hat R(\bx_i,\bc)-\omega(\bz_i,\bc)\Big)^2,
  \label{eq:mse-loss}
\end{equation}
where the per-event generator weights \(w_{i,0}\) are normalised according to Eq.~(\ref{eq:norm-total-sample}) and the training points $\mathcal V$ are defined below. The sum over $i$ extends over all events in the training data set
$\{\bx_i,\bz_i,w_{i,0}\}\in\mathcal{D}^{\text{train}}$. The reconstructed per-event features \(\bx_i\) enter the surrogate of the differential cross section ratio, \(\hat R(\bx,\bc)\), while the latent variables \(\bz_i\) determine the corresponding PDF-varied weight at parameter point \(\bc\) through Eq.~(\ref{eq:event-ratio_2}).
Because the nominal sample is distributed according to \(\dd\sigma(\bx|\bzero)\), the minimizer satisfies
\begin{equation}
  \hat R^\star(\bx,\bc)
  =
  \mathbb E\!\left[\omega(\bz,\bc)\mid \bx,\bc=\bzero\right]
  =
  \frac{\dd\sigma(\bx|\bc)}{\dd\sigma(\bx|\bzero)}
  =
  R(\bx,\bc)\, ,
  \label{eq:cond-exp}
\end{equation}
where \(\mathbb E[\cdots\mid \bx,\bc=\bzero]\) denotes the conditional expectation with respect to the nominal simulation measure.
Thus, the MSE objective learns the detector-level cross section ratio itself, without requiring an explicit model for the transfer density function~\cite{Brehmer:2018kdj,Brehmer:2018eca,Brehmer:2019xox}.

The tree-based ML implementation is an extension of the ``Boosted Information Tree''~(BIT) algorithm~\cite{Chatterjee:2021nms,Chatterjee:2022oco}, see App.~\ref{app:mse_bpt} for more details.
Rather than fitting one coefficient at a time, this variant learns the complete quadratic dependence on \(\bc\) in a single boosting sequence as derived in App.~\ref{app:mse_boosting}.
The dependence on the chosen set of training points \(\mathcal V\) enters only through the Gram matrix \(V_{AB}=\sum_{\bc\in\mathcal{V}}c_A c_B\). Note that this matrix contains fourth powers of the coordinates of the training points, because each factor $c_A$ comprises linear and quadratic terms in the coordinates.
This matrix therefore specifies how the loss in Eq.~(\ref{eq:mse-loss}) weights the linear and quadratic components of the surrogate and encodes the geometry of the training-point ensemble in parameter space.
For the regression problem to be well defined, \(V_{AB}\) must have full rank.
This is guaranteed if the vectors in \(\mathcal V\) are chosen such that all independent linear and symmetric quadratic monomials in \(\bc\) are sampled.
A convenient choice is to consider all vectors \(\bc\) with non-negative components satisfying
\[
c_a\geq 0\, , \qquad \sum_{a=1}^N c_a \leq 2 \, .
\]
This set contains exactly
\(N+\tfrac{1}{2}N(N+1)\)
non-zero points, corresponding to all independent linear terms \(c_a\) and symmetric quadratic terms \(c_a c_b\).
This is precisely the number of coefficients required to determine a quadratic polynomial in \(N\) dimensions whose constant term is fixed to unity at \(\bc=\bzero\).
This variant of the BIT algorithm has already been applied in a CMS measurement of Wilson coefficients up to quadratic order in the SMEFT expansion~\cite{CMS:2024ksn}.

The regression is performed on the full set of 16 reconstructed observables listed in Table~\ref{tab:features}.
We use a shallow-tree configuration in order to regularize the surrogate and to avoid learning statistical fluctuations of the nominal Monte Carlo sample.
Specifically, the maximum tree depth is set to \texttt{max\_depth}=4, which limits the complexity of the partition of feature space, while the minimum node size is set to \texttt{min\_size}=50, such that nodes with too few events are not further subdivided.
This choice stabilizes the training and ensures that the learned response is controlled by statistically well-populated regions of phase space.

To determine the split positions, the algorithm uses a binned representation of each input feature.
Each of the 16 observables is discretised into \texttt{n\_bins}=256 bins, and the candidate splits are scanned on this grid.
This binning is sufficiently fine, while keeping the optimisation of the tree structure computationally efficient.
Single-tree binning artifacts are smoothed out after a few boosting iterations.

As in standard boosting, the final predictor is constructed as a sum of weak learners.
At each boosting iteration \(b\), the tree fitted to the current residual is multiplied by a learning rate \(\eta^{(b)}\leq 1\), which controls the size of the update.
Smaller values of \(\eta^{(b)}\) make the fit more gradual and therefore act as an additional regularisation of the boosting sequence. We find that a constant learning rate of $0.2$ provides excellent convergence in all cases.
We monitor over-training by computing the loss function on held-out validation data amounting to 10\% of the total training sample. No sign of overtraining is observed, and we stop after 200 boosting iterations.
The resulting setup provides a robust compromise between flexibility and stability for learning the detector-level PDF dependence encoded in the ratios \(R(\bx,\bc)\) by means of
\begin{align}
\hat R(\bx,\bc)=1+ c_A\hat R_A(\bx)\, ,
\qquad
\hat R_A(\bx)
=
\sum_{b=1}^B \eta^{(b)}
\sum_{J\in\mathcal J^{(b)}} \bone_J(\bx)\,\hat R^{(b)}_{A,J}\, ,
\end{align}
where \(\mathcal J^{(b)}\) denotes the phase-space partition found at boosting iteration \(b\), and \(\hat R^{(b)}_{A,J}\) are the corresponding polynomial coefficients in terminal node \(J\). The index function satisfies $\bone_J(\bx)=1$ if $\bx\in J$ and is zero otherwise.

\begin{figure}[htbp]
    \centering
\includegraphics[width=0.49\linewidth]{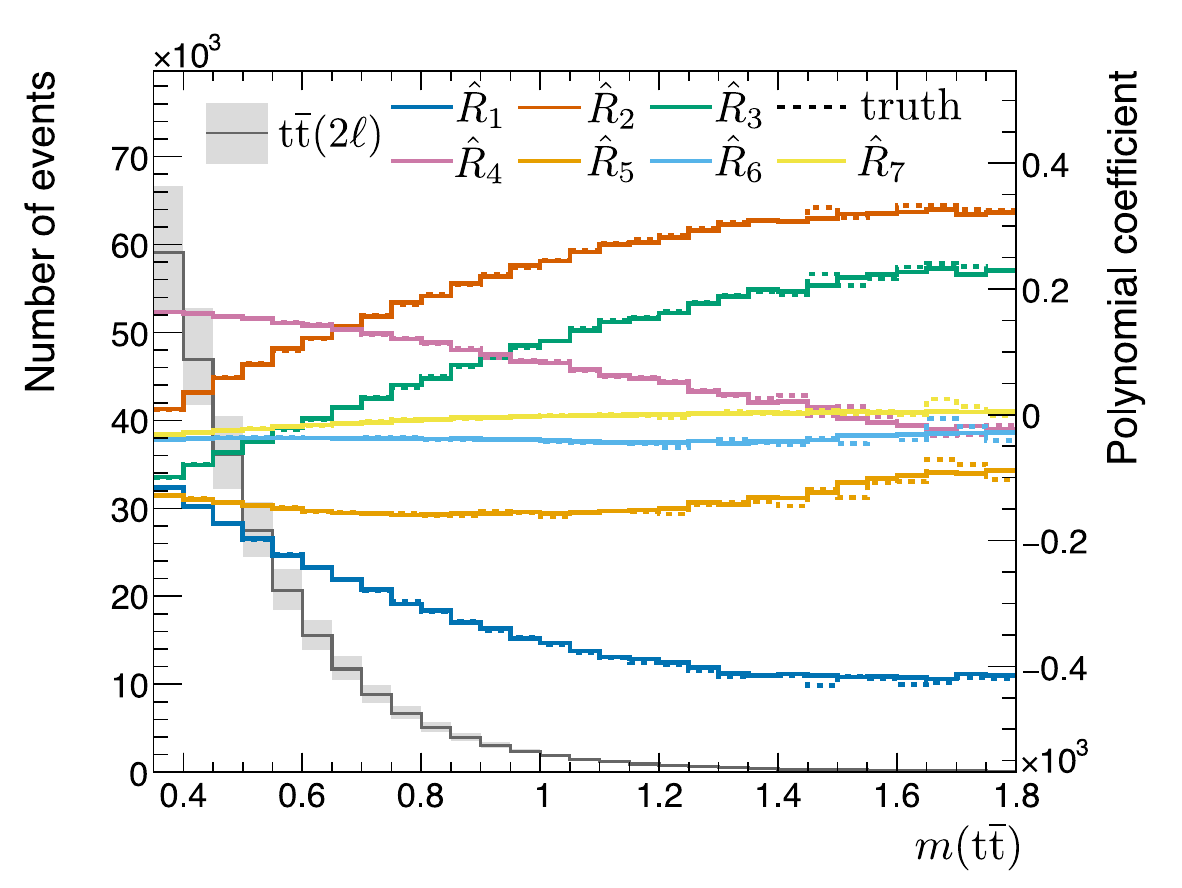}
\includegraphics[width=0.49\linewidth]{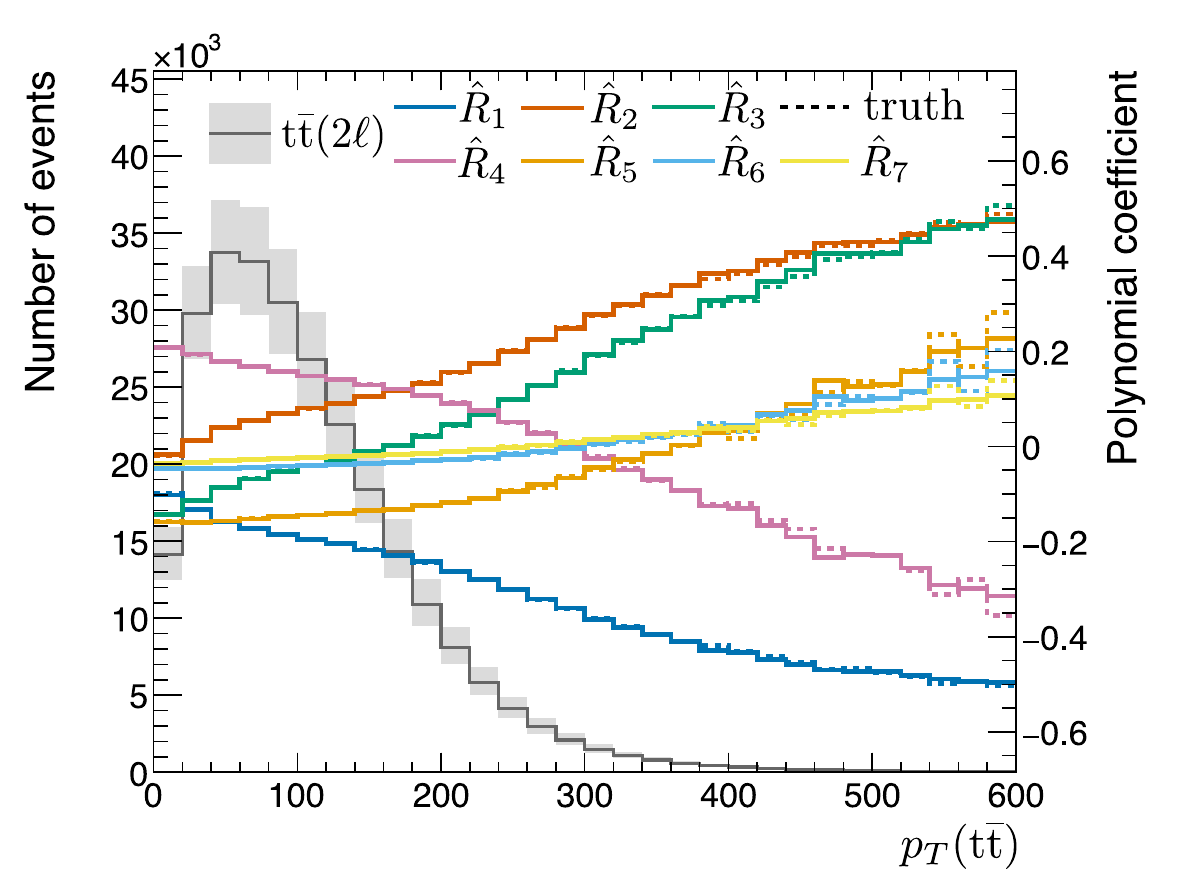}\\
\includegraphics[width=0.49\linewidth]{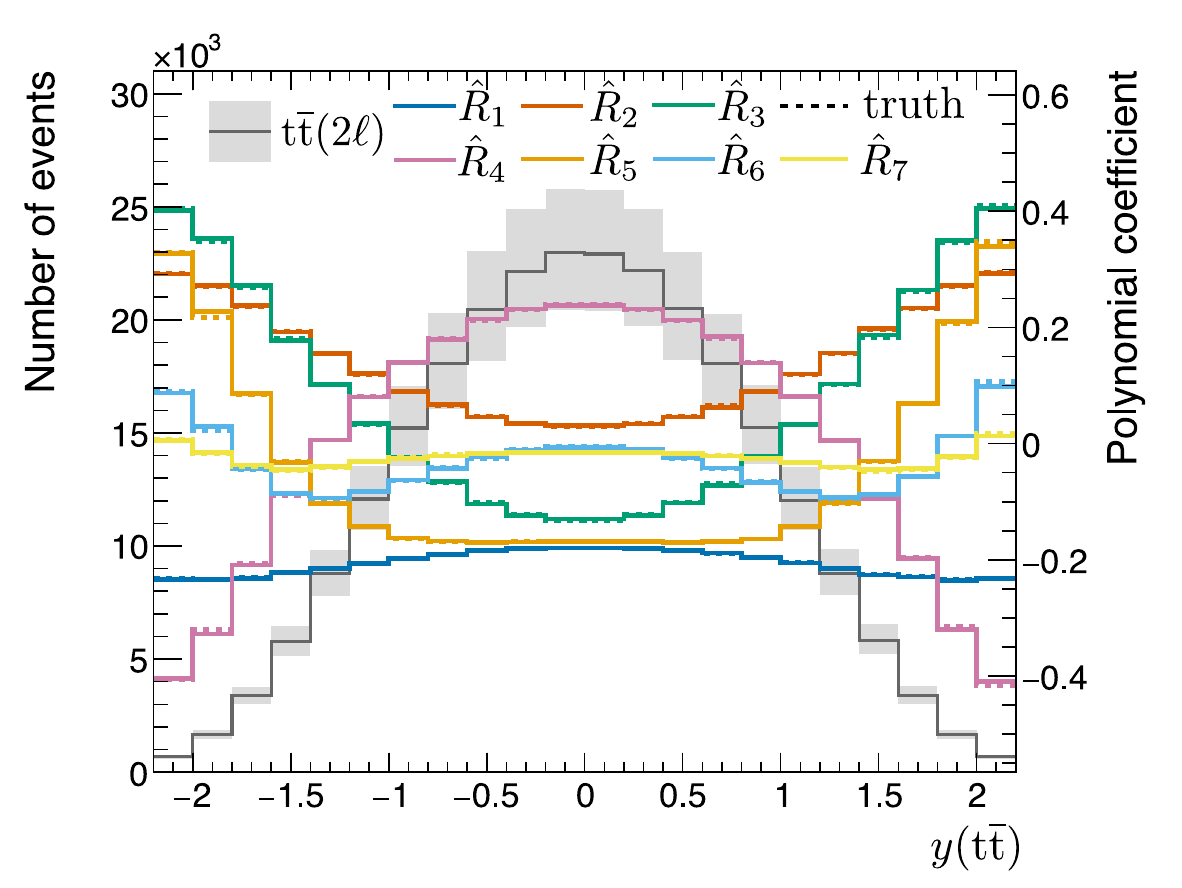}
\includegraphics[width=0.49\linewidth]{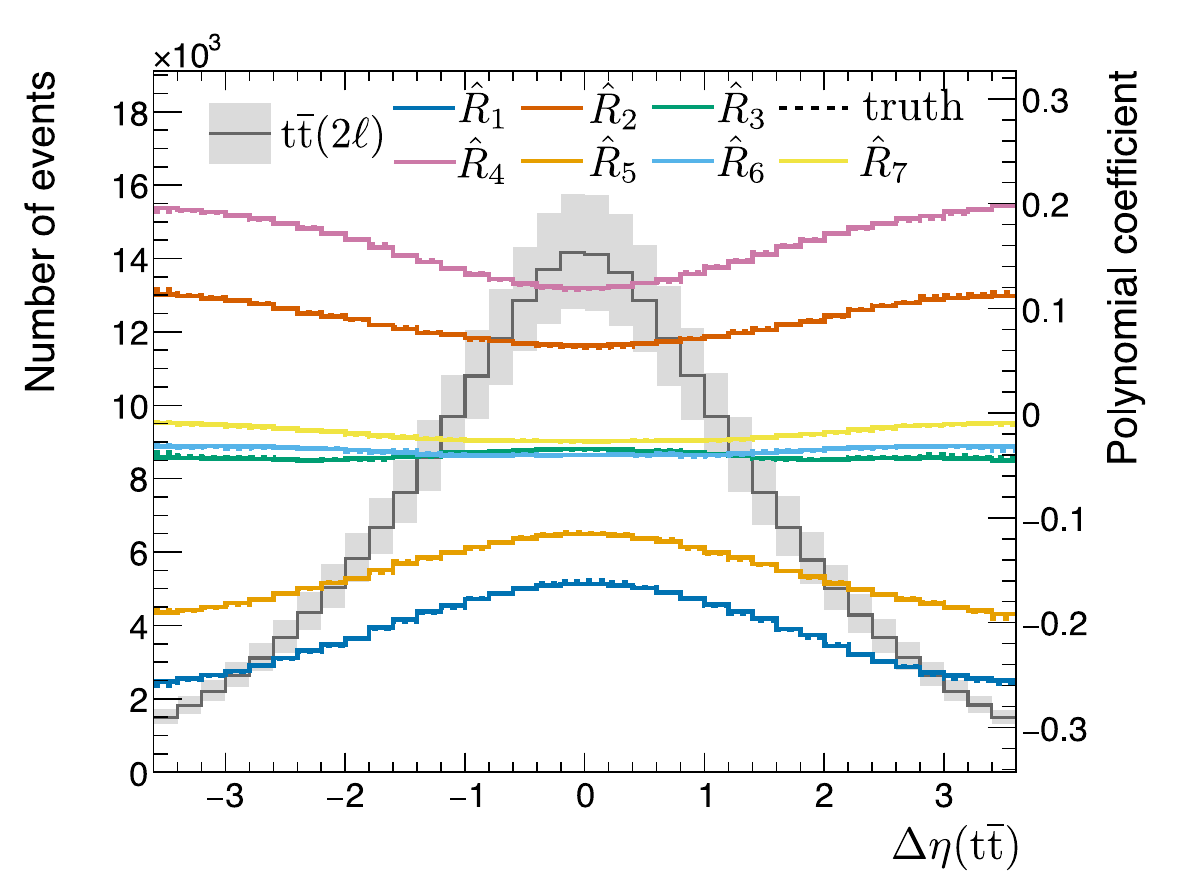}\\
\includegraphics[width=0.49\linewidth]{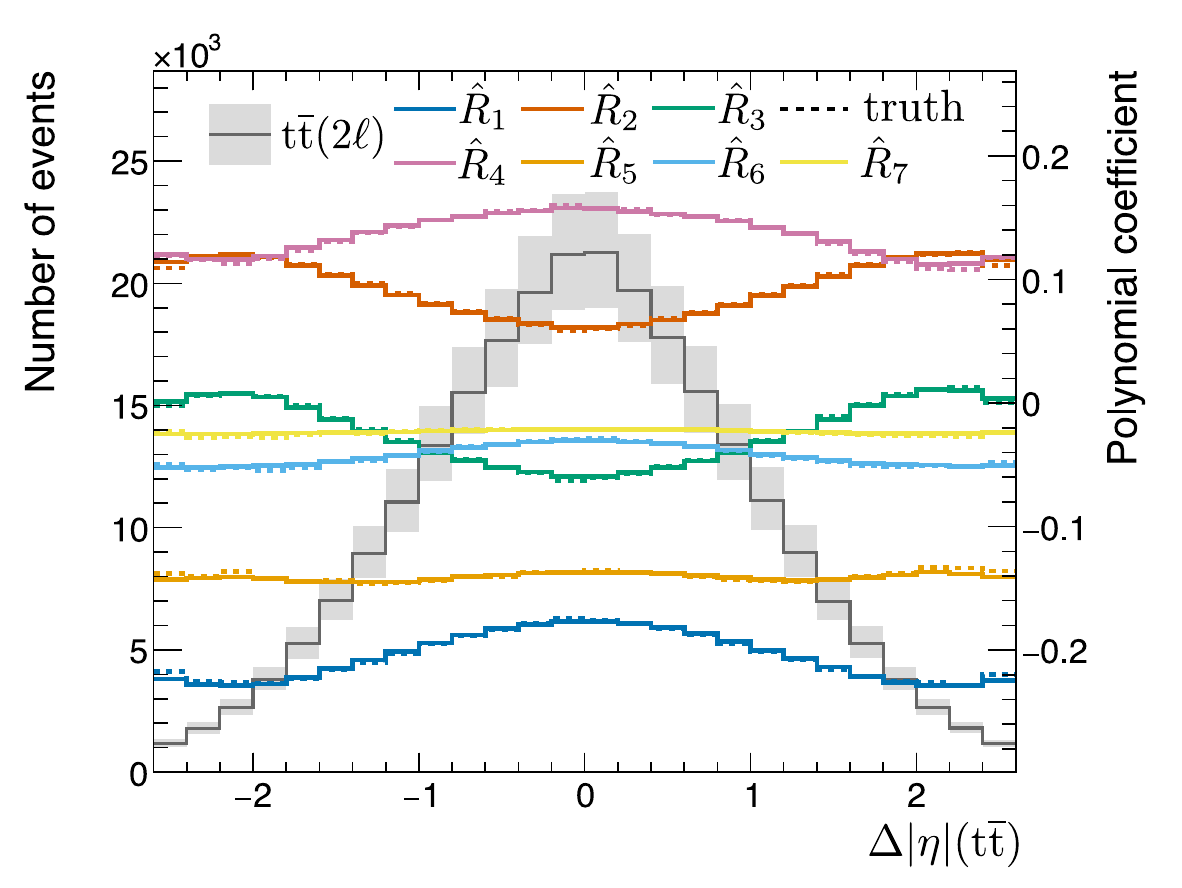}
\includegraphics[width=0.49\linewidth]{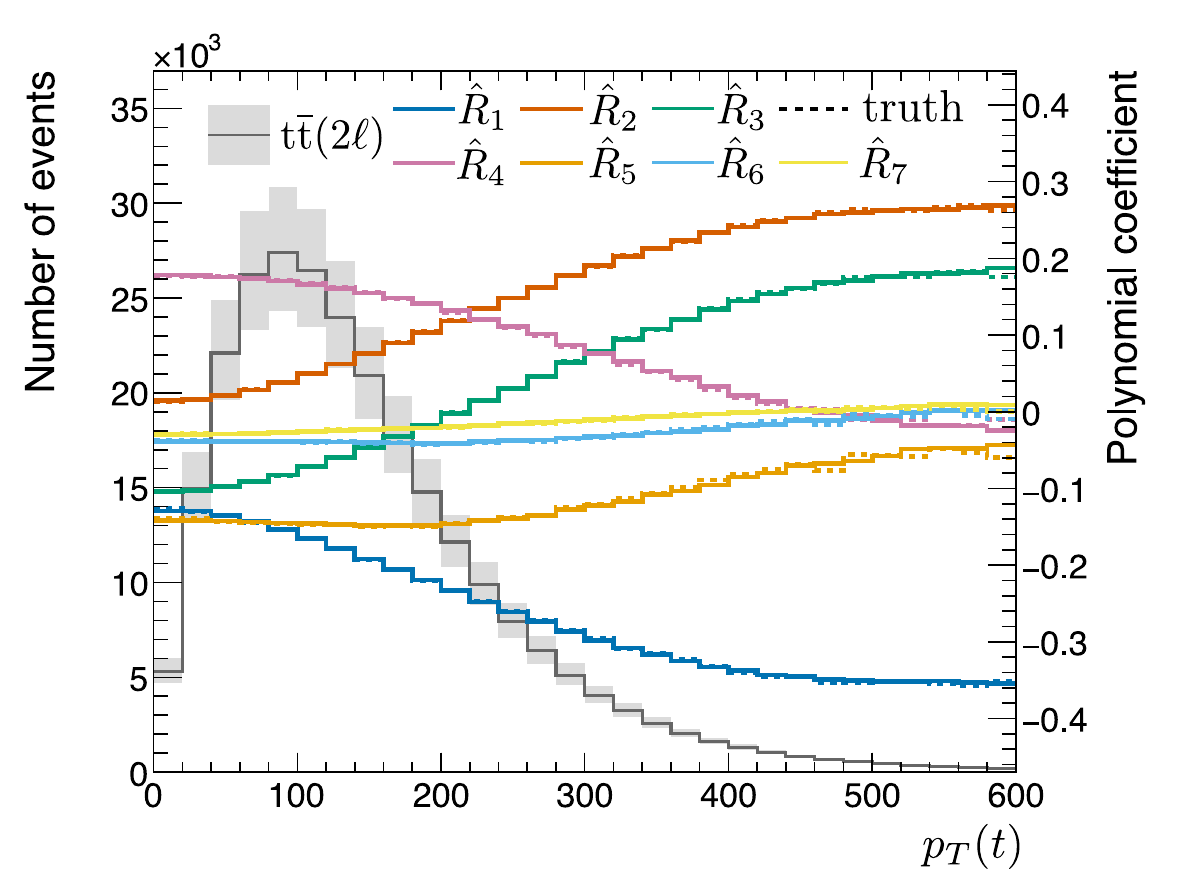}
    \caption{Distributions of detector-level training features in our event sample (black histograms) and the total systematic uncertainty~(shaded bands) associated to them: invariant mass and $\pt$ of the top quark pair (top); rapidity of the top quark pair (center left); difference of $\eta$ of the $\ttbar$ system (center right); difference of $|\eta|$ of the $\ttbar$ system (bottom left); $\pt$ of the top quark (bottom right).
The colored curves indicate the relative variations of these distributions with respect to the first \(N=7\) elements of the linear PDF model.
For each of the linear model variations, the true result obtained from reweighting the Monte Carlo sample (dashed) is compared with the associated BIT predictions $\hat R_1,\ldots,\hat R_7$ (solid).
The BIT predictions accurately reproduce both the shape distortions and the normalisation shifts induced by the linear PDF model.
    }
    \label{fit-impact-pod-1}
\end{figure}

\begin{figure}[htbp]
    \centering
\includegraphics[width=0.49\linewidth]{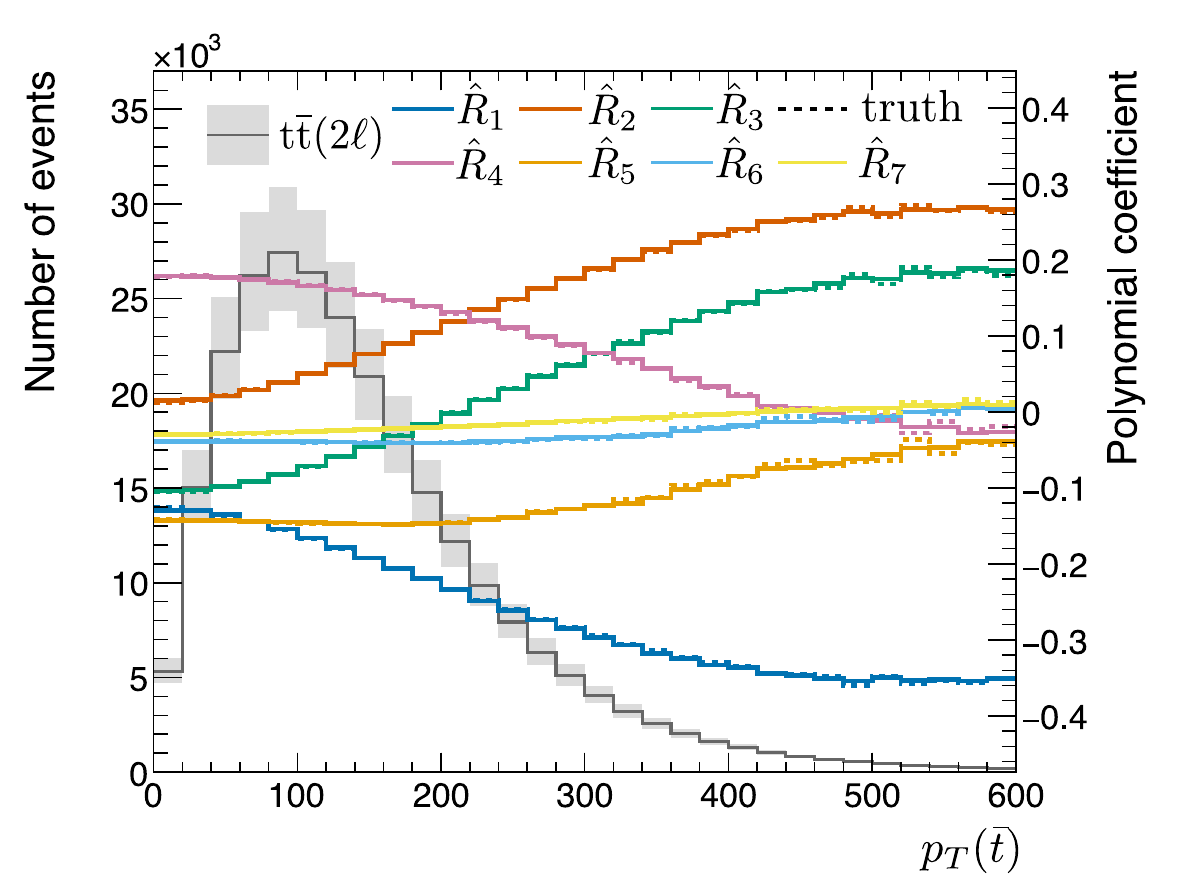}
\includegraphics[width=0.49\linewidth]{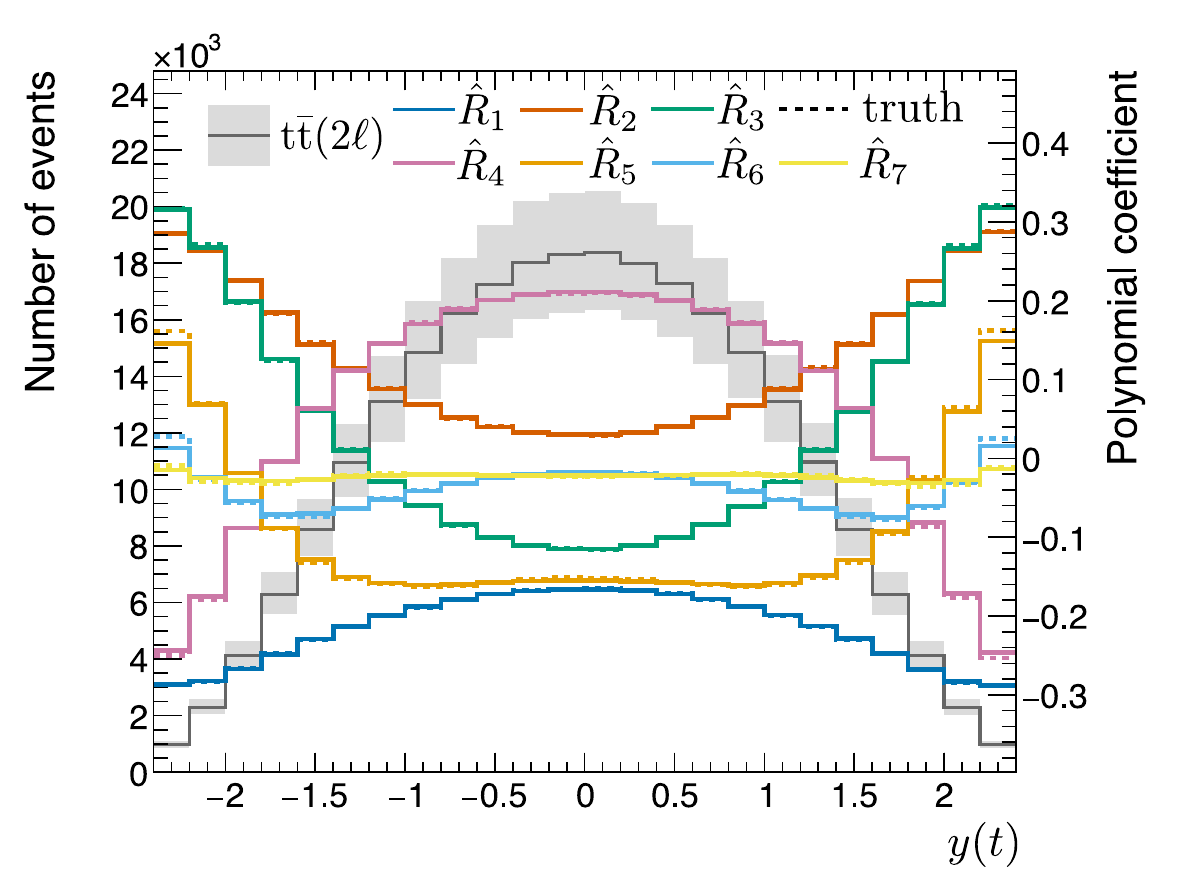}\\
\includegraphics[width=0.49\linewidth]{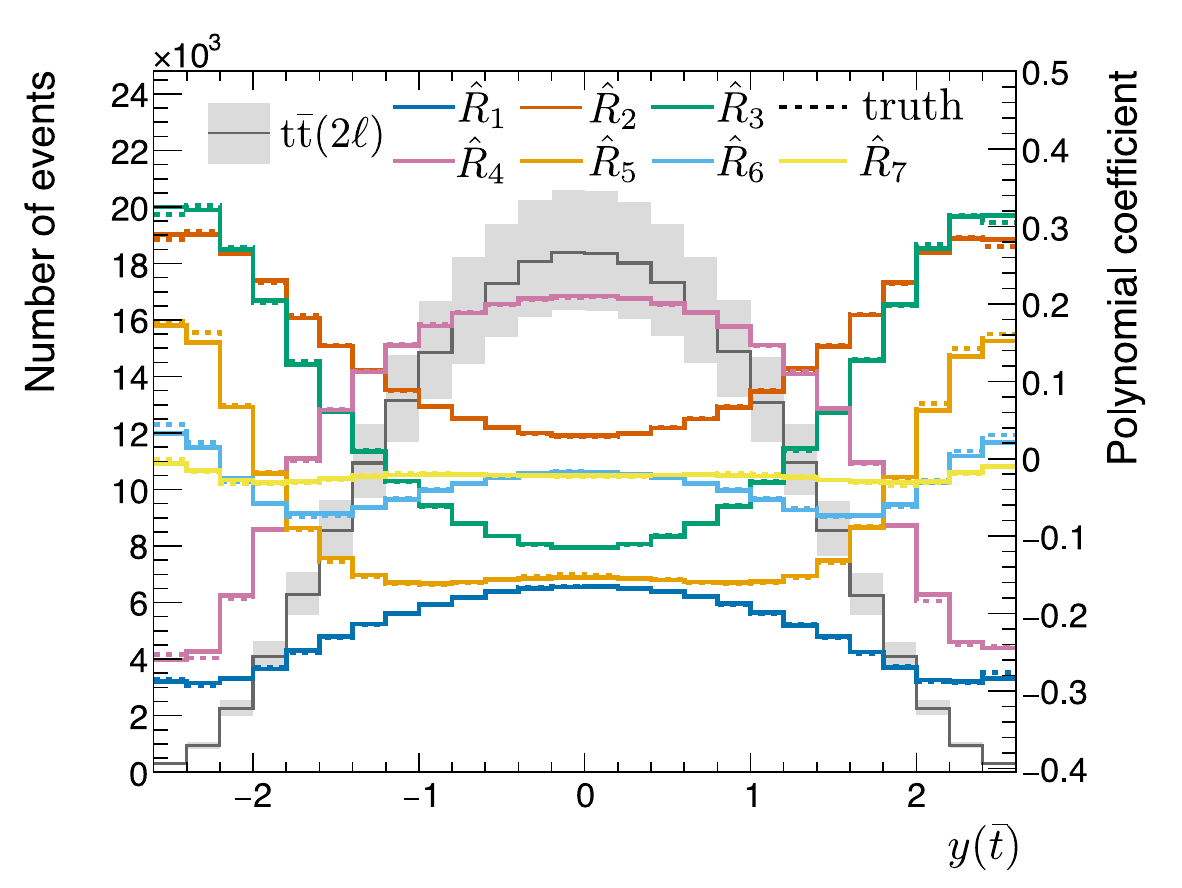}
\includegraphics[width=0.49\linewidth]{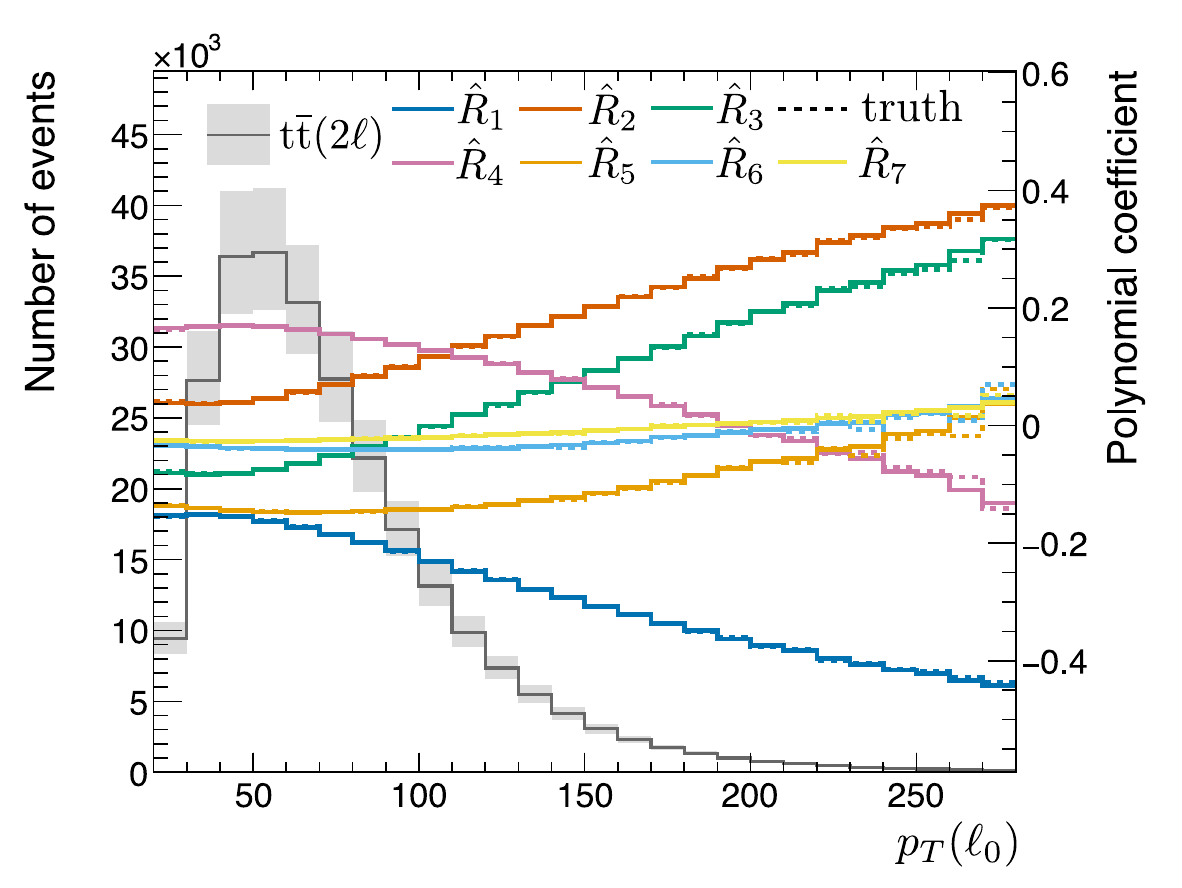}\\
\includegraphics[width=0.49\linewidth]{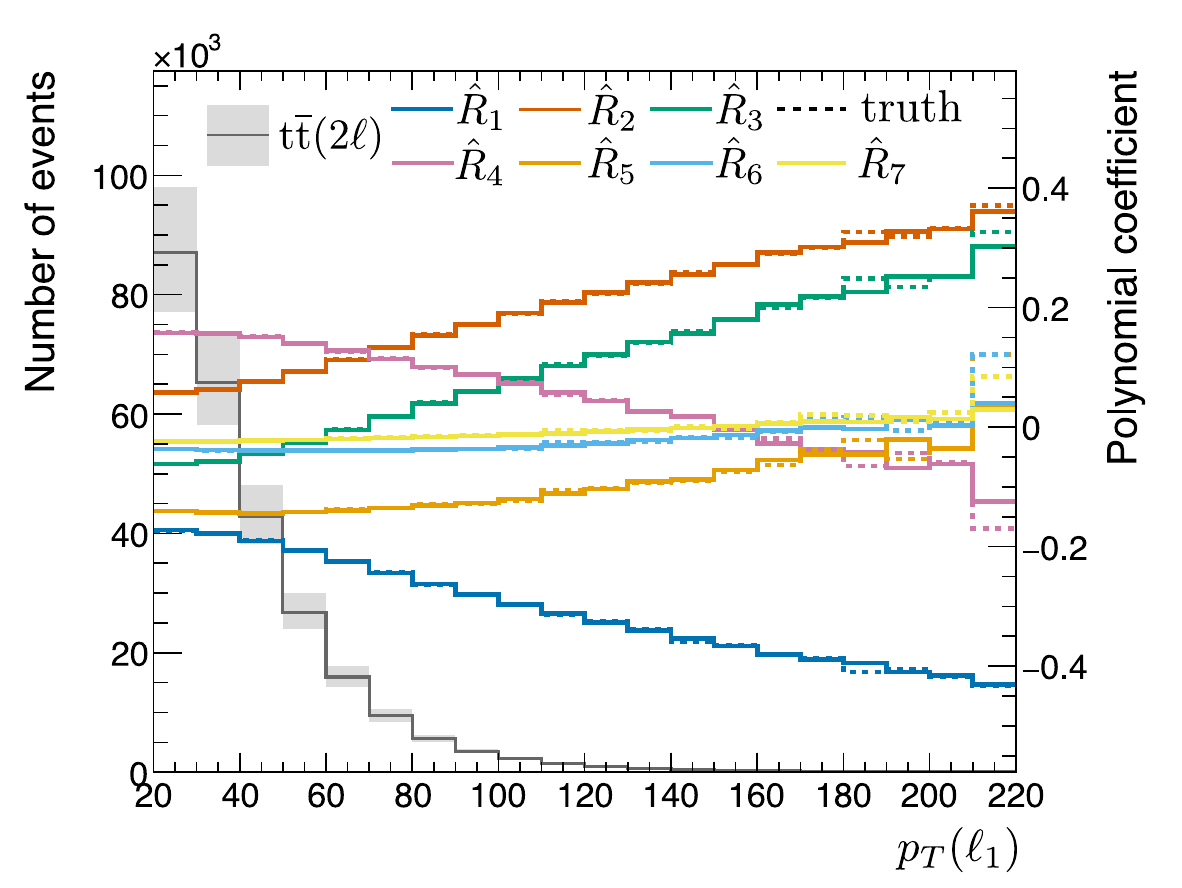}
\includegraphics[width=0.49\linewidth]{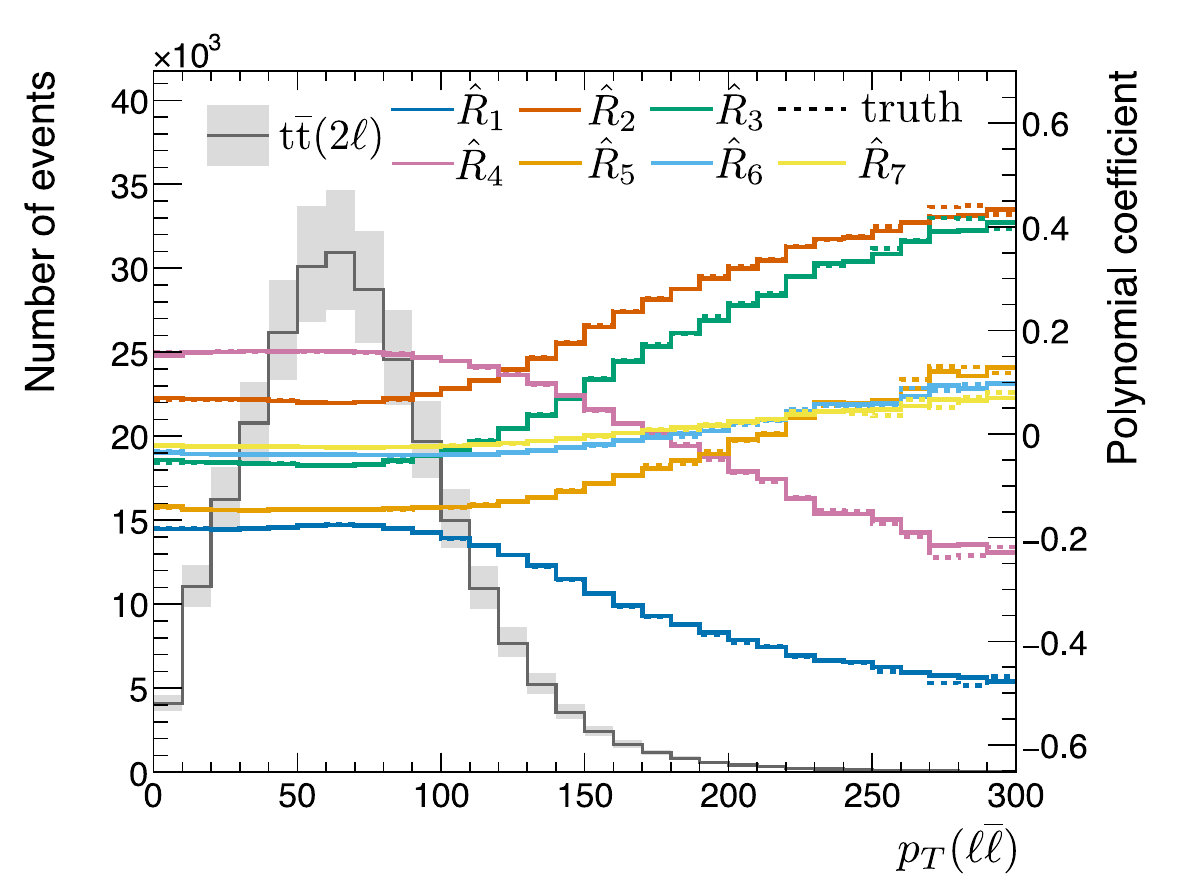}
    \caption{Distributions of detector-level training features in our event sample (black histograms) and the total systematic uncertainty~(shaded bands) associated to them: $\pt$ of the anti-top quark (top left); rapidity of the top quark (top right) and of the anti-top quark (center left); $\pt$ of the leading lepton (center right) and of subleading lepton (bottom left); $\pt$ of the $\ell\overline{\ell}$ system (bottom right).
    The colored curves indicate the relative variations of these distributions with respect to the first \(N=7\) elements of the linear PDF model.
    For each of the linear model variations, the true result obtained from reweighting the Monte Carlo sample (dashed) is compared with the associated BIT predictions $\hat R_1,\ldots,\hat R_7$ (solid).
    }
    \label{fit-impact-pod-2}
\end{figure}

\begin{figure}[t]
    \centering
\includegraphics[width=0.49\linewidth]{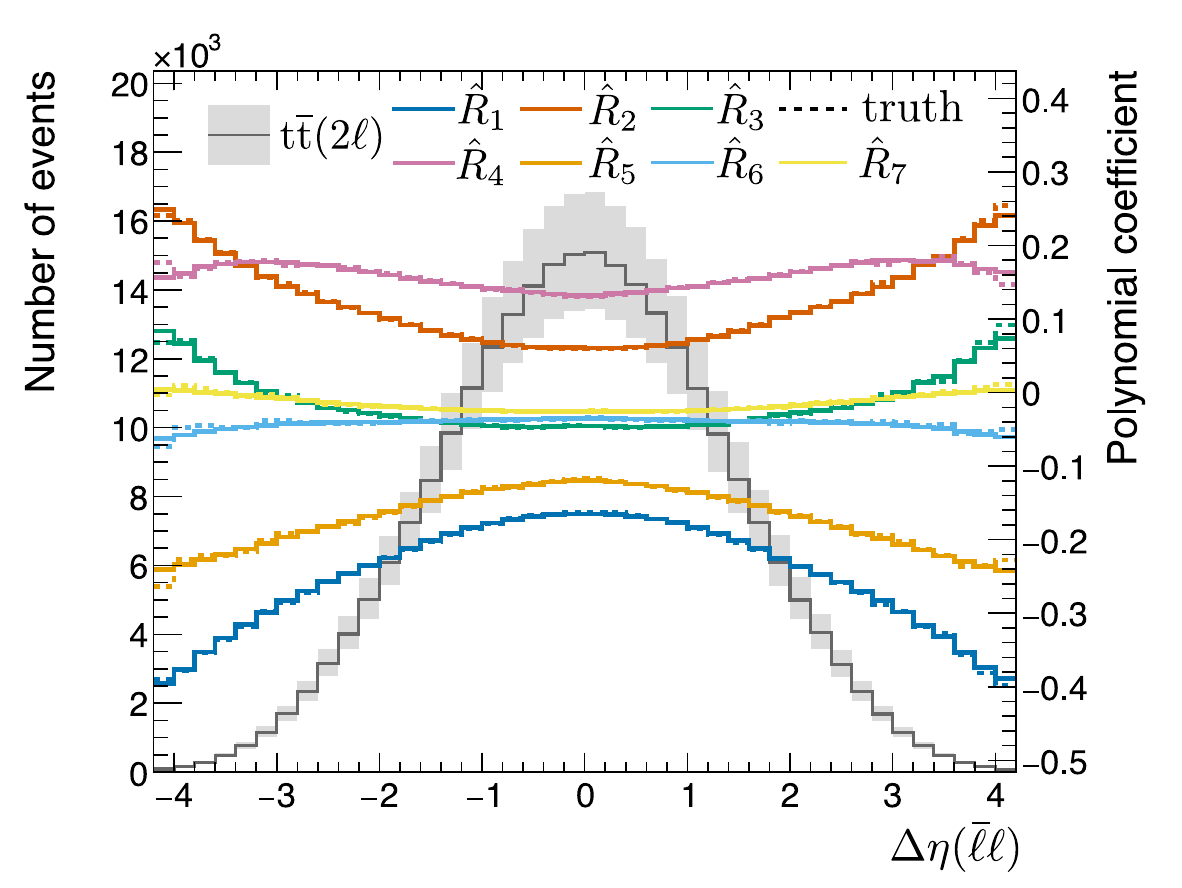}
\includegraphics[width=0.49\linewidth]{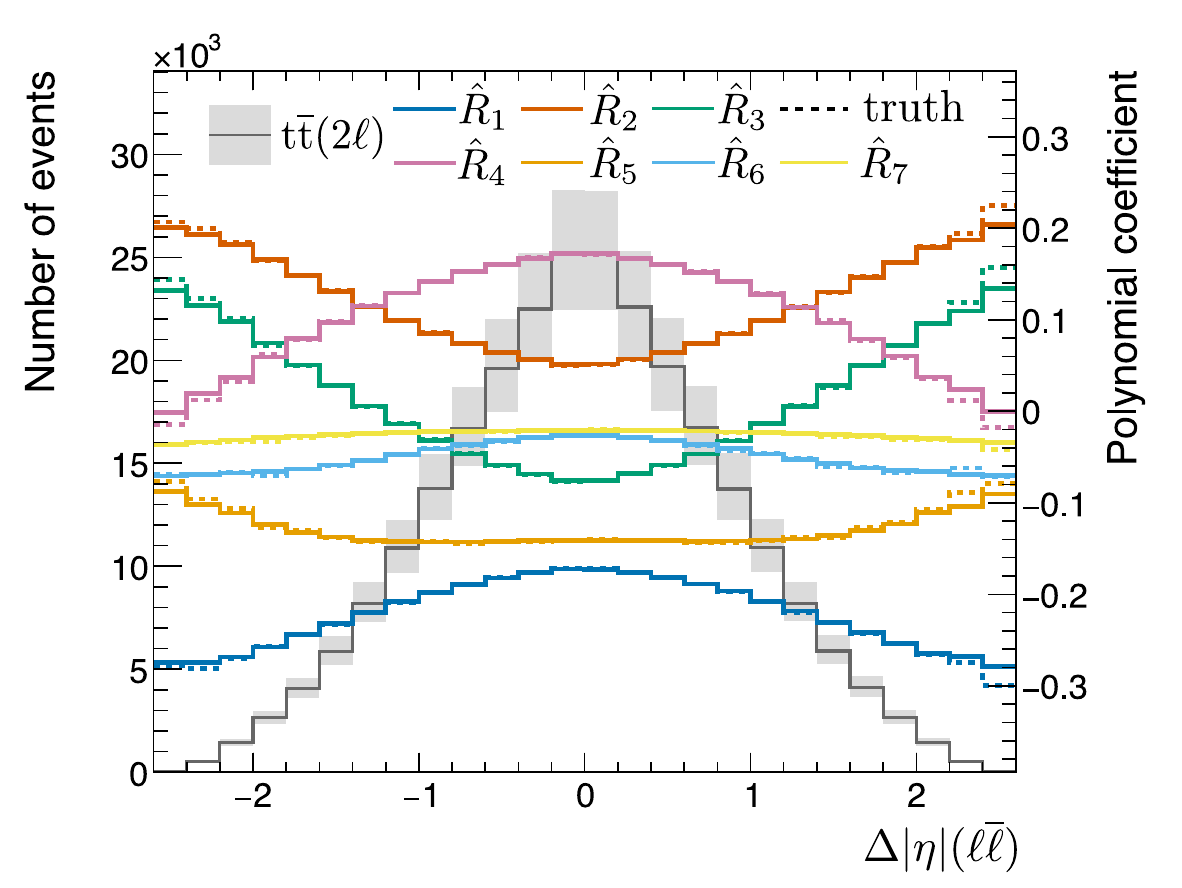}
\includegraphics[width=0.49\linewidth]{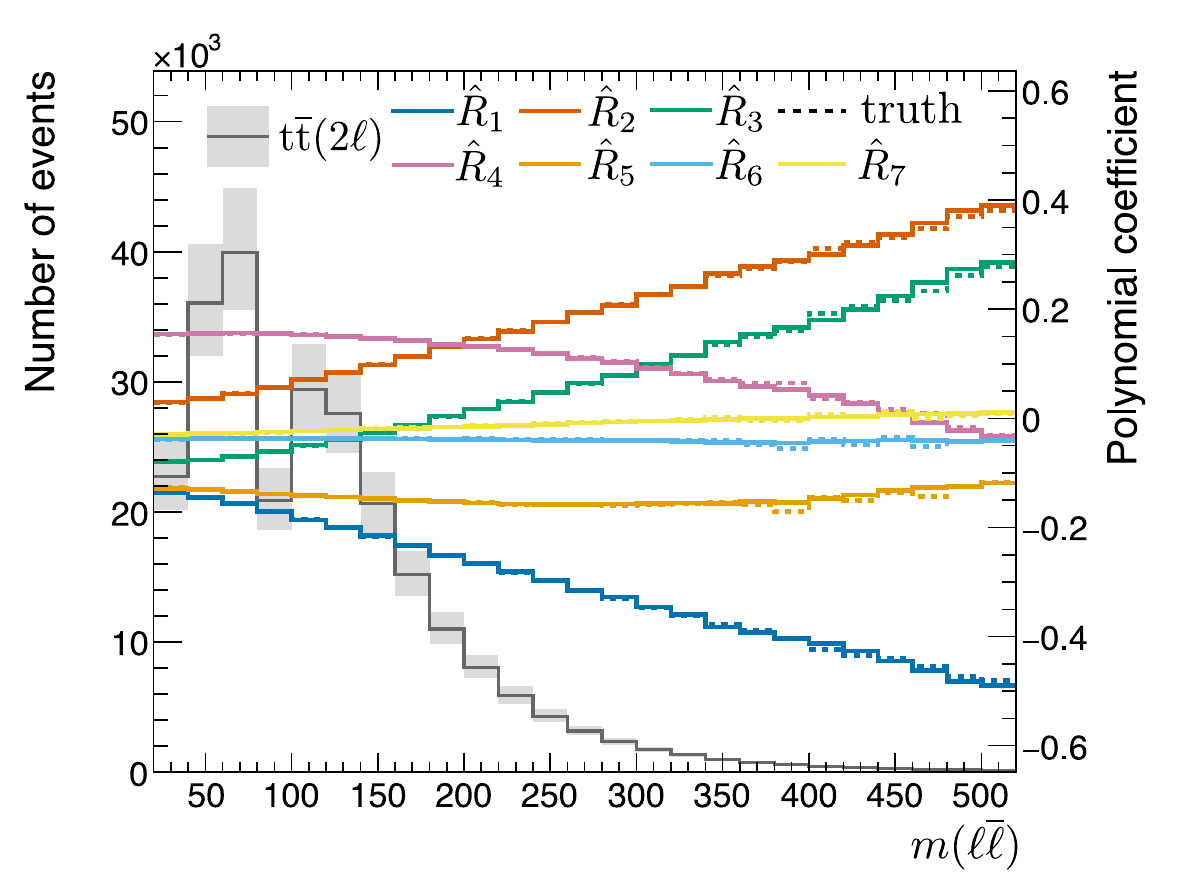}
\includegraphics[width=0.49\linewidth]{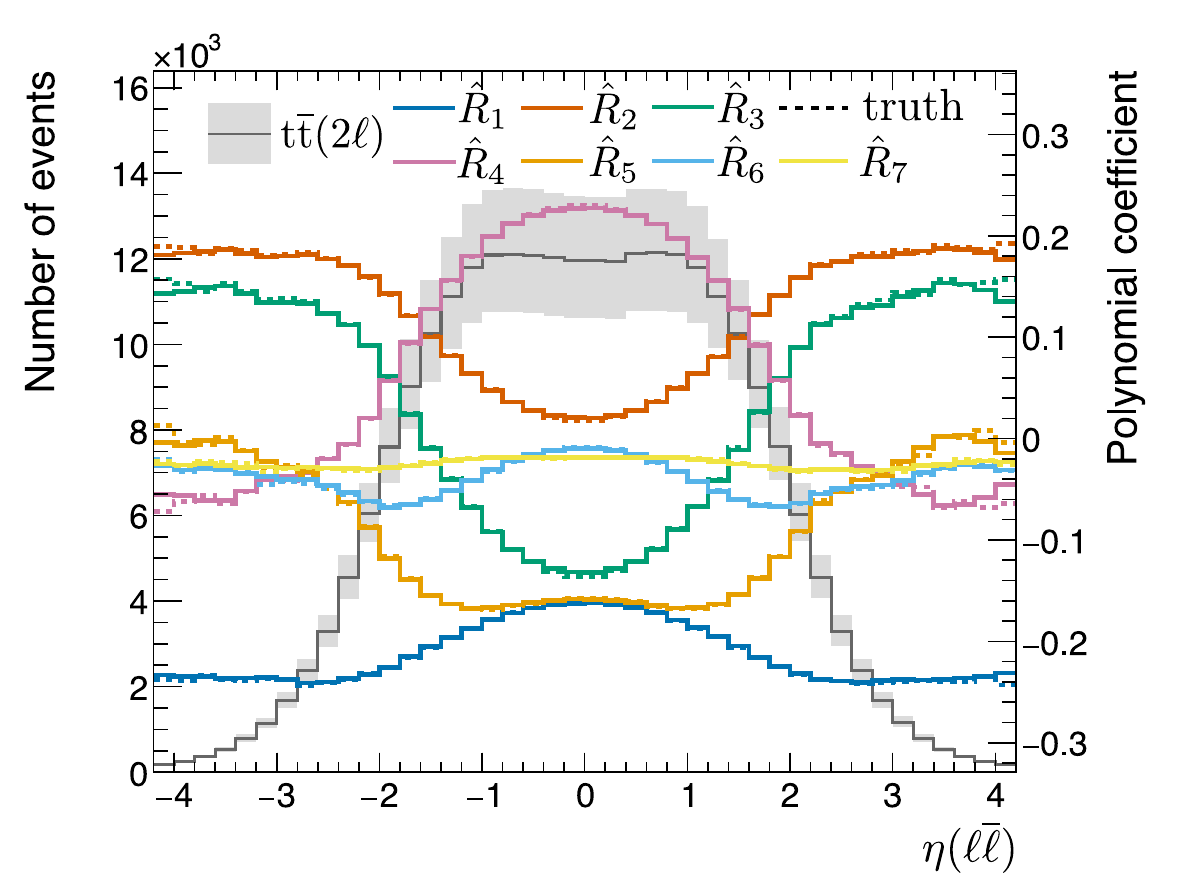}
\caption{Distributions of detector-level training features in our event sample (black histograms) and the total systematic uncertainty~(shaded bands) associated to them: difference of $\eta$ of the $\ell\overline{\ell}$ system (top left); difference of $|\eta|$ of the $\ell\overline{\ell}$ system (top right); subleading lepton (bottom left); invariant mass and pseudorapidity of the $\ell\overline{\ell}$ system (bottom).
    The colored curves indicate the relative variations of these distributions with respect to the first \(N=7\) elements of the linear PDF model.
    For each of the linear model variations, the true result obtained from reweighting the Monte Carlo sample (dashed) is compared with the associated BIT predictions $\hat R_1,\ldots,\hat R_7$ (solid).
    }
    \label{fit-impact-pod-3}
\end{figure}

We illustrate now how the linear model for the gluon PDF constructed in Sec.~\ref{sec:linear_model} and parametrized by the \(\{ \varphi^{(g)}_k(x)\}\) basis functions modifies the differential distributions used to construct the unbinned observables.
While the generic $\bc$-dependence is quadratic, the linear polynomial coefficients at the level of the differential cross section are dominant in all cases.
Figures \ref{fit-impact-pod-1}--\ref{fit-impact-pod-3} therefore only show the first six linear coefficients in the polynomial expansion of the differential cross section. The dashed lines show the true distributions $R_a(\bx)$, and the solid lines the BIT regression $\hat R_a(\bx)$. The agreement is excellent in the whole phase space.
The distribution of each feature from the nominal sample is also shown for reference.
This analysis demonstrates that the different basis elements induce different normalisation- and shape-dependent effects, indicating that an unbinned extraction of $\bc$ should be highly sensitive to the linear model parameters.

\paragraph{Validation of the learned $\hat R(\bx,\bc)$.}
We validate the performance of the BIT regression by comparing the polynomial coefficients of the true differential cross section ratio, \(R_A(\bx,\bc)\), with the learned quantities \(\hat R_A(\bx,\bc)\) defined by the ansatz in Eq.~(\ref{eq:rhat-ansatz}), using held-out validation data.
We first choose a value of \(a\) and then bin the distribution of \(\hat R_a(\bx,\bc)\), focusing on the linear terms. For each bin, we compute the expectation value \(\mathbb E\!\left[\omega(\bz,\bc)\mid \bx \in \text{bin}\right]\), effectively integrating over the latent space \(\bz\). If the prediction \(\hat R_a(\bx)\) is precise, the corresponding linear coefficient in \(\bc\) of the \textit{true} differential cross section ratio will agree with this binned quantity.

\begin{figure}[t]
    \centering
\includegraphics[width=0.55\linewidth]{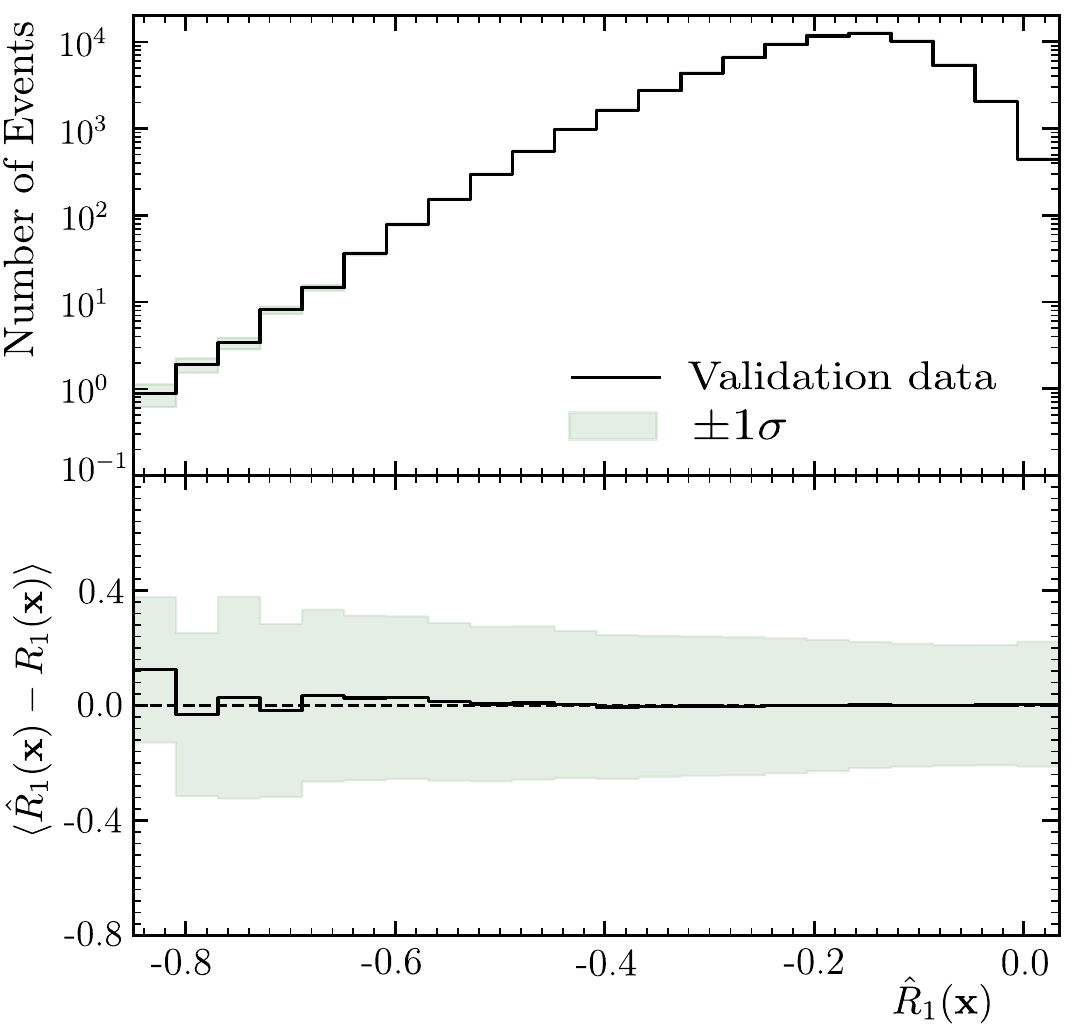}
\caption{Validation of the learned $\hat R(\bx,\bc)$.
    We show, as a representative example, the distribution of $\hat R_1(\bx)$ (top panel) and the associated residual (bottom panel).
   The uncertainty band corresponds to the $1\sigma$ spread of the residual and reflects the latent uncertainty. }
    \label{fig:R_validation}
\end{figure}

As a representative example, we choose \(a=1\) and show the distribution of \(R_1(\bx)\) in Fig.~\ref{fig:R_validation}.
The residual in the lower panel refers to the mean difference, in each bin, between the expectation value of the true quantity \(R_1(\bx)\) and the prediction \(\hat R_1(\bx)\). It is consistent with zero, especially in regions with many events, indicating that \(R_1(\bx)\) is learned well.
We also display the spread of the residual~(colored area), a measure of the latent-dependent variability.
We have verified that the rest of the coefficients of the linear model show similar behavior.

\subsection{Principal component analysis}
\label{sec:PCA}

The POD method described in Sec.~\ref{sec:linear_model} provides basis elements of the PDF space $\widehat{\mathcal{H}}$ whose coefficients in the linear model need to be determined by the data.
Therefore, the construction of this linear model for the gluon PDF is not influenced by the sensitivity of the $\ttbar(2\ell)$ data set to each of its coefficients.
From Figs.~\ref{fit-impact-pod-1}--\ref{fit-impact-pod-3} it is evident that the PDF variations modify both the overall rate and the shape of the detector-level predictions. It is conceivable that parameter combinations exist which alter the prediction in such a way that the data can not constrain it.
The directional derivative of the likelihood in this direction then approximately vanishes.
Such flat (or quasi-flat) directions are hence a source of numerical instability in minimisation methods based on gradient descent.
To bypass this potential limitation, we can use the trained surrogate for the POIs, $\hat R(\bx,\bc)$, to identify flat directions beforehand and remove them from the fit.

To this purpose we employ a principal component analysis~(PCA) starting from the detector-level Fisher information matrix.
This matrix is evaluated at the nominal value for the nuisance parameters, $\bc=0$, which we, therefore, drop from the notation for the sake of brevity.
Following~\cite{Benato:2025rgo}, we write the Fisher information as an integral over the space of variable-length data sets and insert the likelihood function in Eq.~(\ref{eq:ext-likelihood}) for the probability distribution on this space. The resulting Fisher information matrix $I_{ab}$ for the POIs is
\begin{align}
I_{ab}(\bc)
&=
\int \dd\mathcal D\, p(\mathcal D|\bc)\,
\frac{\partial}{\partial c_a}\log p(\mathcal D|\bc)\,
\frac{\partial}{\partial c_b} \log p(\mathcal D|\bc)
\nonumber\\
&=
\mathcal{L}_0
\int \dd \sigma(\bx|\bc)\,
\left(\frac{\partial}{\partial c_a}\log R(\bx,\bc)\right)
\left(\frac{\partial}{\partial c_b}\log R(\bx,\bc)\right) \, ,
\label{eq:Fisher-general}
\end{align}
where $a,b$ run over the $N$ coefficients of the linear model.
We evaluate this expression using the ML-trained surrogate \(R(\bx,\bc)\simeq \hat R(\bx,\bc)\) and perform the PCA at the reference PDF, \(\bc=\bzero\). The derivatives in Eq.~(\ref{eq:Fisher-general}) act on the polynomial expansion in Eq.~(\ref{eq:rhat-ansatz}), and the quadratic coefficient functions \(\hat R_{ab}(\bx)\) do not contribute in this limit. The empirical estimator of the Fisher matrix therefore becomes
\be
I_{ab}(\bzero)
\simeq
\hat I_{ab}
=
\sum_{\{\bx_i,w_{i,0}\}\in\mathcal D_{\bzero}^{\rm sim}}
w_{i,0}\,\hat R_a(\bx_i)\hat R_b(\bx_i)\, .
\label{eq:empirical_Fisher_matrix}
\ee
To identify the combinations of PDF coefficients (the POIs) that are best constrained by the $\ttbar(2\ell)$ unbinned measurement, we diagonalize the Fisher matrix by solving
\begin{align}
\label{eq:PCA_decomposition}
\sum_{b=1}^{N}\hat I_{ab}\,v_b^{(k)}
=
\lambda_k\,v_a^{(k)}\, ,
\qquad
a,k=1,\ldots,N\, ,
\end{align}
with \(N\) orthonormal eigenvectors \(\bv^{(k)}\) and associated non-negative eigenvalues \(\lambda_k\). The eigenvectors define principal components in the parameter space through
\begin{align}
\label{eq:principal_components}
d_k=\sum_{a=1}^{N} v_a^{(k)}\,c_a\, ,
\qquad
\sum_{a,b=1}^{N} c_a\,\hat I_{ab}\,c_b
=
\sum_{k=1}^{N}\lambda_k\,d_k^2\, .
\end{align}
The eigenvalues \(\lambda_k\) therefore quantify the sensitivity of the unbinned measurement along the corresponding principal directions.
Large eigenvalues \(\lambda_k\) identify combinations of gluon PDF coefficients that are well constrained by the \(\ttbar(2\ell)\) data set, while small \(\lambda_k\) correspond to weakly constrained directions.

\begin{figure}[t]
    \centering
\includegraphics[width=0.47\linewidth]{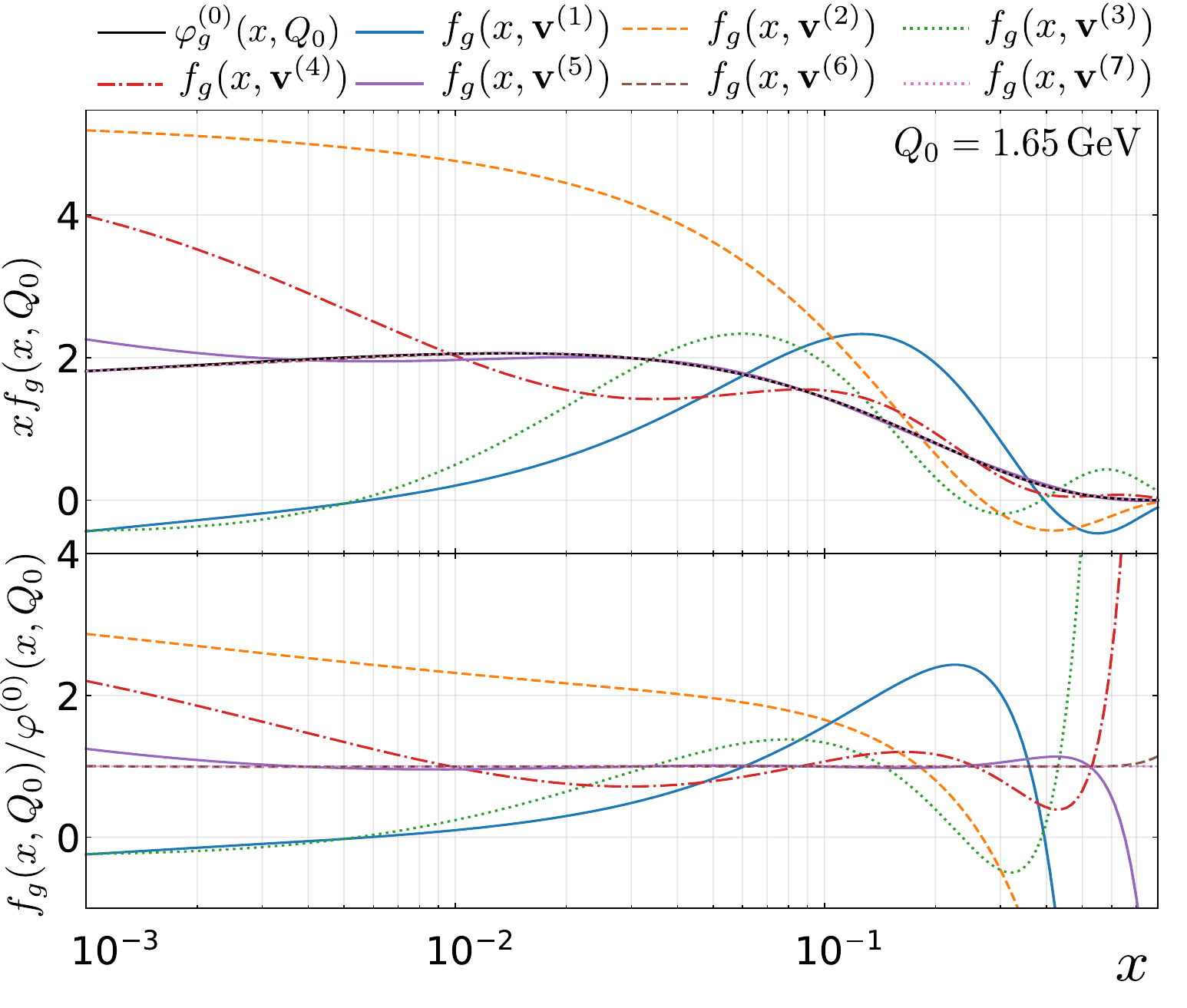}
\includegraphics[width=0.52\linewidth]{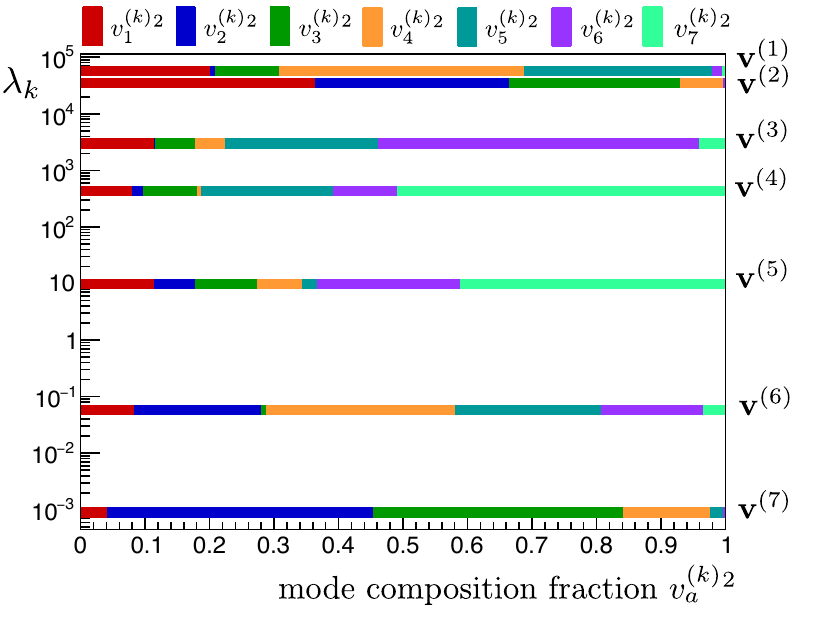}
\caption{The gluon PDF $xf_g(x,Q_0)$ in the linear model for the first 7 eigenvectors from the PCA of the expected unbinned Fisher information matrix Eq.~(\ref{eq:empirical_Fisher_matrix}) evaluated for $\bc=\bv^{(k)}$~(left).
The bottom panel displays the ratio to the central gluon of the linear model.
The mode decomposition fraction $v_a^{(k)2}$ for the same eigenvectors~(right). The $y$-axis indicates the corresponding eigenvalue.
All eigenvectors receive sizeable contributions from multiple basis elements of the linear model.
}
\label{fig:unbinned-basis}
\end{figure}

In the left panel of Fig.~\ref{fig:unbinned-basis} we show the gluon PDF $f_g(x,Q_0)$ in the linear model for the first 7 eigenvectors  obtained from the PCA (evaluated by setting $\bc=\bv^{(k)}$) of the expected unbinned Fisher information matrix Eq.~(\ref{eq:empirical_Fisher_matrix}) applied to the  \(\ttbar(2\ell)\) data set.
To visualize the decomposition of each principal direction, in the right panel of Fig.~\ref{fig:unbinned-basis} we display the corresponding squared eigenvector components $v_a^{(k)2}$, which quantify the fractional contribution of the original PDF basis coefficient $c_a$ to the \(k\)-th mode and add up to unity for every eigenvector.
The height on the $y$-axis indicates the corresponding eigenvalue.
All eigenvectors receive sizeable contributions from multiple basis elements of the linear
model, which implies that the unbinned $\ttbar(2\ell)$ data sets constrains non-trivial combinations of these basis elements.
Therefore, the POD eigenvector native hierarchy (Fig.~\ref{fig:pod_eigenvalues}) is not maintained when the coefficients $\bc$ are fitted to the data.

\begin{figure}[t]
    \centering
\includegraphics[width=0.80\linewidth]{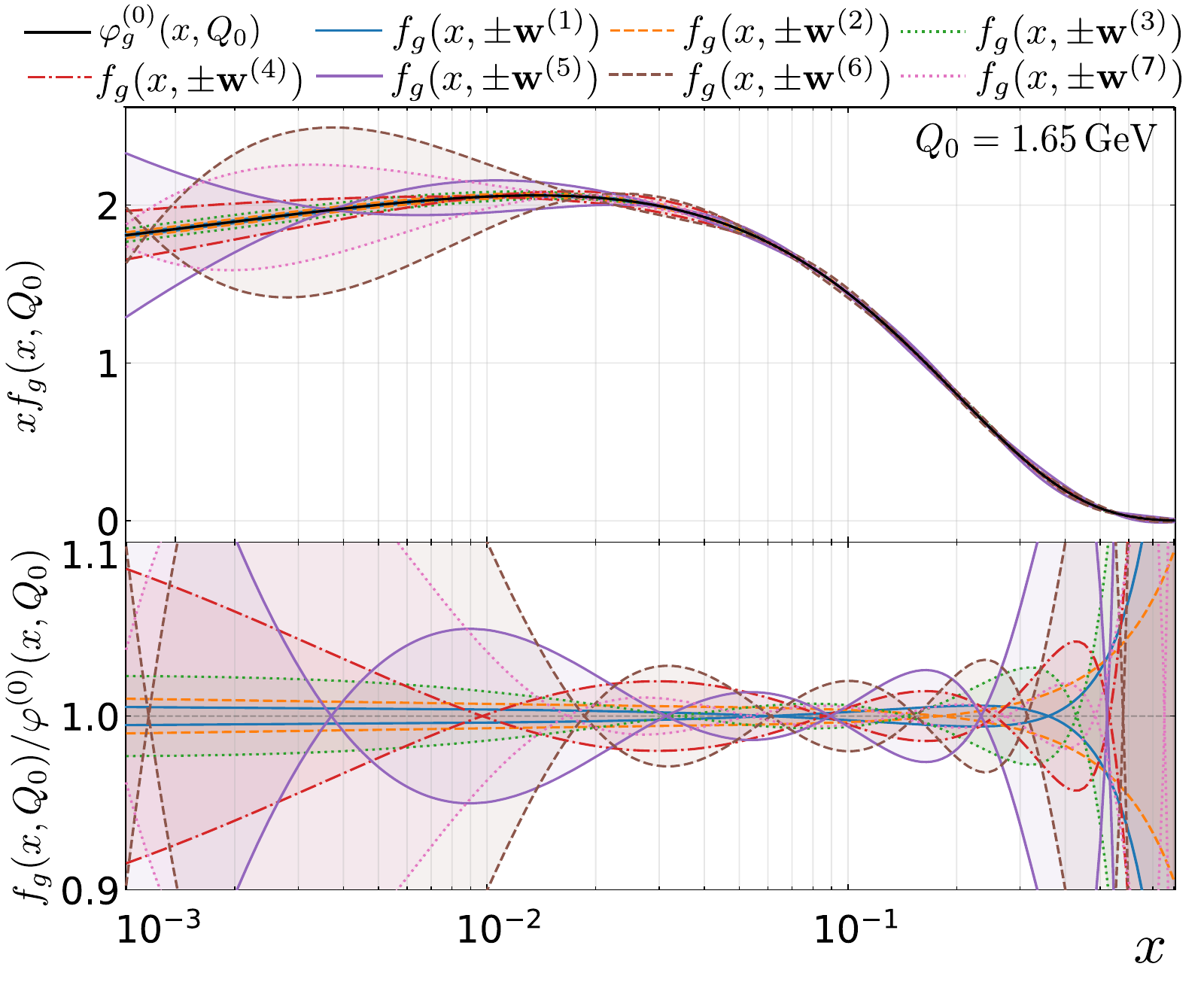}
    \caption{The gluon PDF $xf_g(x,Q_0)$ in the linear model for the first 7 eigenvectors from the PCA of the expected unbinned Fisher information matrix Eq.~(\ref{eq:empirical_Fisher_matrix}) evaluated for the normalised eigen-directions, $\bc=\pm\mathbf{w}^{(k)}$.
    The lower panel shows the ratio to the reference PDF.
    When evaluated along these normalised eigen-directions of the Fisher information matrix, the variations of the linear model are clearly smaller in the kinematic region ($2\times 10^{-2}\lsim x \lsim 0.3$) to which the $\ttbar{(2\ell)}$ unbinned data set has more sensitivity than in the small-$x$ and large-$x$ regions.
    }
\label{fig:modes-sigma}
\end{figure}

Figure~\ref{fig:modes-sigma} shows the same comparison as in the left panel of Fig.~\ref{fig:unbinned-basis}, now with the parameters of the linear model set to the normalised eigen-directions, that is
\be
\bc=\pm\mathbf{w}^{(k)}=\lambda_k^{-1/2}\,\bv^{(k)}\, ,
\ee
such that the displacement in parameters space approximately corresponds to unit Fisher distance.
 When evaluated along these normalised eigen-directions of the Fisher information matrix, the variations of the linear model are clearly smaller in the kinematic region ($2\times 10^{-2}\lsim x \lsim 0.3$, see the left panel of Fig.~\ref{fig:x_muF_density_id_21}) to which the $\ttbar{(2\ell)}$ unbinned data set has sensitivity, than in the small-$x$ and large-$x$ extrapolation regions.
Therefore, Fig.~\ref{fig:modes-sigma} highlights how after PCA rotation the $\ttbar{(2\ell)}$ data set will only constrain those directions in the linear model parameter space to which one has direct sensitivity.
Quasi-flat directions are identified by the large condition number of the associated eigenvalue and can be weeded out to improve the numerical stability of the fit (see also Sec.~\ref{subsec:stability}).

\subsection{Machine-learning systematic uncertainties}
\label{sec:learning-systematis}

For the calibration of reconstructed detector-level objects such as jets, missing transverse momentum, and leptons, the ATLAS and CMS open-data projects~\cite{CMS-Open-Data,ATLAS-Open-Data} provide sufficient information on systematic uncertainties to allow a realistic estimate of the experimental errors that would arise in an actual data analysis.
We focus on the main sources of systematic uncertainty that would enter a realistic analysis, but do not aim to reproduce the complete experimental systematic models adopted by ATLAS or CMS, which would be beyond the scope of this work.
Instead, we follow closely the procedures developed in Ref.~\cite{Schofbeck:2024zjo} for a simplified treatment.
For comparison, extensive information on the systematic uncertainties in binned measurements of the differential cross section of the $\ttbar(2\ell)$ process is provided by ATLAS~\cite{ATLAS:2023gsl} and CMS~\cite{CMS:2024ybg}.

For the integrated luminosity, which is associated with the overall normalisation, the central value $\mathcal{L}_0=137\,$fb$^{-1}$ corresponds to the full Run~II data set of the CMS experiment, and we assign a log-normal uncertainty
\begin{align}
\mathcal{L}(\bn)=\mathcal{L}_0\,\alpha_{\textrm{lumi}}^{\nu_{\textrm{lumi}}},
\end{align}
with $\alpha_{\rm lumi}=1.0073$ corresponding to 0.73\%~\cite{CMS:2021xjt,CMS:2025wux}.

\paragraph{Machine-learning systematic effects.}
Except for the uncertainty in the luminosity, all other systematic effects can depend on the features $\bx$ and therefore must be predicted with machine-learned surrogates.
We parametrize systematic effects according to Eq.~(\ref{eq:S-training-task}) where the log-polynomial coefficient functions $\hat\Delta_A(\bx)$ are implemented with the parametric neural network developed in Ref.~\cite{Benato:2025rgo}. Here, we provide a summary of the procedure.

For each source of systematic uncertainty, we first construct the systematically varied simulated event samples defined in Eq.~\ref{eq:def-variied-sim-data}, which we use as training data.
These samples correspond to a discrete set of nuisance parameter values $\bn\in\mathcal V$.
The nominal sample at $\bn=\bzero$ is kept separate and used as the common reference. The goal is to learn, directly at the level of the reconstructed event features $\bx$, the detector-level ratio $S(\bx,\bn)$ from Eq.~(\ref{eq:true-S-R}) which generalizes to the unbinned setting the bin-by-bin response model used in conventional treatments of systematic uncertainties.

To this end, we formulate the problem as a parametric binary-classification task between the nominal sample $\mathcal D_{\bzero}$ and the systematically varied samples $\mathcal D_{\bn}$ for nuisance-parameter points $\bn\in\mathcal V$. Summing the cross entropy over all nuisance points, we minimize
\begin{align}
L_{\rm CE}
=
-\sum_{\bn\in\mathcal V}
\left[
\sum_{\{\bx_i,w_i\}\in\mathcal D_{\bzero}}
w_i\log \hat f_{\bn}(\bx_i)
+
\sum_{\{\bx_i,w_i\}\in\mathcal D_{\bn}}
w_i\log \!\left(1-\hat f_{\bn}(\bx_i)\right)
\right]\label{eq:CE-loss-summed}
\end{align}
where it is important to note that $\mathcal{D}_0$ is the same for all $\bn\in\mathcal{V}$ in the first term. We also dropped the dependence on $\bc$ according to our factorisation assumption in Eq.~(\ref{eq:systematic_PoI_factorisation}) from the notation and train systematic effects for $\bc=\bzero$.
The minimum of Eq.~(\ref{eq:CE-loss-summed}) for an arbitrarily expressive function is formally attained at
\begin{align}
    f^\ast_{\bn}(\bx)=\frac{1}{1+S(\bx,\bn)}
\end{align}
which motivates ansatz for $\hat S$ in terms of the neural network outputs $\hat\Delta_A$,
\begin{align}
\hat f_{\bn}(\bx)
=
\frac{1}{1+\hat S(\bx,\bn)}
=
\frac{1}{1+\exp\left(\nu_A\hat\Delta_A(\bx)\right)}.
\end{align}
As before, the coefficients $\nu_A$ denote the linear, quadratic, and mixed monomials constructed from the nuisance parameters. The functions $\hat\Delta_A(\bx)$ are the primary quantities learned from simulation. This parametrisation ensures that the nuisance dependence remains continuous in $\bn$, while the non-trivial kinematic dependence is encoded in the detector-level coefficient functions $\hat\Delta_A(\bx)$.

Substituting this ansatz yields
\begin{align}
L[\hat\Delta_A]
=
\sum_{\bn\in\mathcal V}
\Bigg[
\sum_{\{\bx_i,w_i\}\in\mathcal D_{\bzero}}
w_i\,\mathrm{Soft}^+\!\left(\nu_A\hat\Delta_A(\bx_i)\right)
+
\sum_{\{\bx_i,w_i\}\in\mathcal D_{\bn}}
w_i\,\mathrm{Soft}^+\!\left(-\nu_A\hat\Delta_A(\bx_i)\right)
\Bigg] ,
\end{align}
with $\mathrm{Soft}^+(x)=\log(1+\exp(x))$. At its functional minimum, this objective returns
\begin{align}
\hat S(\bx,\bn)
=
\exp\!\left(\nu_A\hat\Delta_A(\bx)\right)
\simeq
\frac{\dd\sigma(\bx|\bn)}{\dd\sigma(\bx|\bzero)}\, ,
\end{align}
such that the network provides a smooth surrogate for the effect of each nuisance parameter on the fully differential event distribution.

In practice, we train separate surrogates for the different classes of systematic effects discussed below. Depending on the structure of the nuisance variation, we retain only the polynomial terms required by the available simulated variations and by the observed size of non-linear effects. For the dominant uncertainties, this procedure captures both normalisation shifts and shape distortions across the full feature space $\bx$, while providing a compact parametric representation that can be evaluated event by event in the likelihood.

The coefficient functions $\hat\Delta_A(\bx)$ are implemented with fully connected neural networks that take as input the 16 reconstructed event features. We train separate networks for the different classes of systematic effects, using the same architecture in all cases; only the number of output nodes changes according to the number of polynomial coefficients $\hat\Delta_A(\bx)$ required by the corresponding nuisance-parameter ansatz.

In the following, we discuss theoretical uncertainties associated with the modeling of the partonic scattering, and then consider experimental uncertainties associated with the jet-energy calibration, b-tagging efficiencies, and lepton efficiencies before we finally turn to the specifics of the training setup.

\paragraph{Theoretical uncertainties.}
Missing higher order uncertainties (MHOUs) associated to the NLO truncation of the QCD perturbative calculation of {\sc\small POWHEG} are estimated by varying the event-wise renormalisation $\mu_R$ and factorisation $\mu_F$ scales around the central choice Eq.~(\ref{eq:scale_choice}) within a given range.
Accounting for MHOUs in the perturbative calculation is required to achieve a faithful PDF uncertainty estimate~\cite{NNPDF:2024dpb} since for many of the measurements that are used as input to the PDF fit they are comparable or larger than experimental uncertainties.
The importance of accounting for MHOUs is also highlighted by their role in joint extractions of the PDFs with SM parameters, such as the strong coupling constant~\cite{Ball:2025xgq}.

In our analysis we parametrize MHOUs with two nuisance parameters $\nu_R$ and $\nu_F$, associated to the variations of the central scales which we denote by $\mu_{R,0}$ and $\mu_{F,0}$.
Here we consider $\nu_R=\pm1$ and $\nu_F=\pm1$, which correspond to the usual factor-two variations around the central scales:
\begin{align}
\mu_R(\nu_R)=2^{\nu_R}\mu_{R,0}\qquad
\mu_F(\nu_F)=2^{\nu_F}\mu_{F,0} \, .
\end{align}
{\sc\small POWHEG} internal reweighting provides event weights corresponding to the scale variations necessary to evaluate the MHOUs using the resulting 9-point prescription:
\begin{align}
\label{eq:MHOU_variations}
(\nu_R,\nu_F)\in\mathcal{V}=
\{(-1,-1),\,(-1,0),\,(-1,1),\,(0,-1),\,(0,1),\,(1,-1),\,(1,0),\,(1,1)\},
\end{align}
with the nominal prediction reproduced with $\nu_R=\nu_F=0$.
This prescription is somewhat different from the popular 7-point prescription, which avoids variations that fall outside the condition imposed by
\be
\frac{1}{2} \le \lp \frac{\mu_R(\nu_R)}{\mu_F(\nu_F)}\rp  \le 2 \, ,
\ee
see~\cite{NNPDF:2019ubu} for an overview of scale variation prescriptions.
Beyond scale variations, other approaches have been considered to estimate MHOUs, e.g.~\cite{Tackmann:2024kci,Bonvini:2020xeo,Duhr:2021mfd,Lim:2024nsk,Gambhir:2025lka}, but these are less developed and so far they do not have a universal formulation that can be applied to all processes entering the PDF fit.

We model the scale dependence of the NLO partonic cross sections up to quadratic accuracy in the nuisance parameters $\mu_R$ and $\mu_F$ by using the following expression
\begin{align}
\hat S_{\textrm{mhou}}(\bx,\nu_R,\nu_F)=
\exp\left(
\nu_R\hat\Delta_{R}(\bx)
+\nu_F\hat\Delta_{F}(\bx)
+\nu_R^2\hat\Delta_{RR}(\bx)
+\nu_F^2\hat\Delta_{FF}(\bx)
+\nu_R\nu_F\hat\Delta_{RF}(\bx)
\right) \, ,
\end{align}
where the five functions $\hat{\Delta}_A(\bx)$ labeled by $A=\{R,F,RR,FF,RF\}$ correspond to the linear, quadratic, and mixed contributions in the nuisance parameters associated to the renormalisation and factorisation scales.
The 8 simulated variations of Eq.~(\ref{eq:MHOU_variations}) therefore over-constrain the parameterisation (as would also be the case with the 7-point prescription).
The functions $\hat{\Delta}_A(\bx)$ are used to train a parametric neural network~(PNN) as described in Ref.~\cite{Benato:2025rgo} to the {\sc\small POWHEG} simulation.
Correlated scale variations with $\nu_R=\nu_F=\pm1$ lead to effects at the level of about $10\%$ in several observables, which could be reduced by using an event sample generated at NNLO accuracy~\cite{Mazzitelli:2020jio}.
The quadratic model reproduces these trends well, with only small residual deviations in the extreme kinematic regions.

In addition to MHOUs, we also consider theoretical uncertainties associated with the $\alpha_s$ value provided by the nominal PDF set. The variations of the predictions are implemented as an additional variation of $\alpha_s$ using the corresponding PDF replicas as nuisance parameter, keeping the underlying correlations. We parametrize the effect using a single nuisance parameter $\nu_{\alpha_s}$. The up- and down-varied training data correspond to $\alpha_s(m_\PZ)=0.119$ associated with $\nu_{\alpha_s}=+1$ and to $\alpha_s(m_\PZ)=0.117$ associated with $\nu_{\alpha_s}=-1$. The central in $\mathcal{D}_\bzero$ is $\alpha_s(m_\PZ)=0.118$ and associated with $\nu_{\alpha_s}=0$.
The resulting surrogate is given by
\begin{align}
\hat S_{\alpha_s}(\bx,\nu_{\alpha_s})=
\exp\left(
\nu_{\alpha_s}\hat\Delta_{\alpha_s}(\bx)
\right).
\end{align}

\paragraph{Jet energy calibration.}
To evaluate the impact of uncertainties in reconstructed jet transverse momenta, we vary the reconstructed values according to the ``total'' jet energy scale uncertainty of the CMS experiment provided in Ref.~\cite{CMS-Open-Data}. These variations affect the event selection, missing energy, top quark kinematic reconstruction, and other event features in~\bx. Since the per-jet variations depend on the nominal $p_{\textrm{T}}$ and pseudo-rapidity, which are latent (not included in \bx), the resulting function $J_{\nu_{\textrm{JES}}}(\bx,\bz)$ also depends on the latent event configuration. Because the total JES uncertainty comprises several systematic effects~\cite{CMS:2016lmd} whose effects should, in principle, be fit separately, we devise an ad-hoc procedure to decorrelate the effects of jet energy scale variations according to six equally spaced bins of $-2.4\leq\eta_{\text{jet}}\leq 2.4$. In each bin of $\eta_{\text{jet}}$, we vary the jet-related inputs to the reconstruction and separately fit separate log-linear surrogates,
\begin{align}
\hat S_{\textrm{JES}}(\bx,\bn_{\textrm{JES}})=
\exp\left(\sum_k\nu_{\textrm{JES},k}\hat\Delta_{\textrm{JES},k}(\bx)\right),\qquad k=1,\ldots,6.
\end{align}
Most observables show only weak shape dependence on the jet energy calibration, while $p_{\textrm{T}}(\ttbar)$ exhibits variations of a few percent.
In the tails of the $m(\ttbar)$ distribution, small asymmetries in the simulated variations are partially symmetrized by the log-linear model. Extending the surrogate to higher order could capture such effects but is not required for the current precision.
The indirect effects of JES variations from jets outside the acceptance are negligible. Examples of the induced variations are provided in Ref.~\cite{Schofbeck:2024zjo}.

\paragraph{Jet energy resolution.}
Jet-energy resolution~(JER) uncertainties are applied to simulated events after the jet-energy calibration and affect the width, rather than the mean, of the jet momentum response. In practice, if a reconstructed jet can be matched to a generator-level jet, its momentum is corrected using the corresponding JER scale factor; otherwise, the jet momentum is stochastically smeared according to the nominal resolution and its uncertainty. Since these variations modify the reconstructed jet momenta, they propagate to the event selection, missing transverse momentum, top quark reconstruction, and therefore to the full set of observables~$\bx$.

We follow~\cite{CMS-Open-Data} and introduce six JER nuisance parameters, corresponding to the regions $|\eta_{\rm jet}|<1.93$, $1.93<|\eta_{\rm jet}|<2.5$, $2.5<|\eta_{\rm jet}|<3$ with $p_{\mathrm T}<50~\GeV$ and $p_{\mathrm T}>50~\GeV$, and $3<|\eta_{\rm jet}|<5$ with $p_{\mathrm T}<50~\GeV$ and $p_{\mathrm T}>50~\GeV$. For each nuisance parameter we construct $\pm1\sigma$ variations and train the corresponding detector-level surrogate, analogously to the JES case, as
\begin{align}
\hat S_{\rm JER}(\bx,\bn_{\rm JER})
=
\exp\!\left(\sum_k\nu_{\text{JER},k} \hat\Delta_{{\rm JER},k}(\bx)\right),
\end{align}
where $k=1,\ldots,6$ labels the JER components. In this way, the effect of the JER uncertainty is included as a smooth event-wise deformation of the differential cross section.
It is found that JER effects are negligible for this analysis, with typical relative variations below $3\times10^{-3}$.

\paragraph{Jet tagging efficiencies.}
Uncertainties in the b-tagging efficiency for jets, along with their application, are provided in Ref.~\cite{CMS-Open-Data}. This approach relies on $p_{\textrm{T}}$, pseudo-rapidity, and a nominal binary b-tag label from \Delphes. To apply variations we also require the generator-level jet flavor $f \in \{\textrm{udsg}, \cPqc,\cPqb\}$. Using the nominal simulation, we parametrize the $p_{\textrm{T}}$ and $\eta$-dependent tagging efficiencies $\varepsilon_f(p_{\textrm{T}},\eta)$ for each flavor.
Two systematic uncertainties are considered with scale factors $\textrm{SF}_f(p_{\textrm{T}},\eta)$ and variations $\Delta\textrm{SF}_f(p_{\textrm{T}},\eta)$. The heavy-flavor (HF) tagging uncertainty modifies the b- and c-jet tagging efficiencies in a correlated way through the nuisance parameter $\nu_{\textrm{HF}}$. The light-flavor (LF) mistagging uncertainty affects the tagging rates for light-quark and gluon jets and is associated with $\nu_{\textrm{LF}}$.

The reweighting function for synthetic data is
\begin{align}
r(\bx_i,\bz_i|\nu_k,0)
=
\frac{F(\nu_k,\textrm{jets in event }i)}
{F(0,\textrm{jets in event }i)},
\end{align}
where
\begin{align}
F(\nu_k,\textrm{jets})&=
\prod_{\textrm{tagged jets}}
\varepsilon_{f}(p_{\textrm{T}},\eta)
\left(\textrm{SF}_f+\nu_k\Delta\textrm{SF}_{f,k}\right)
\nonumber\\
&\times
\prod_{\textrm{untagged jets}}
\left(
1-\varepsilon_{f}(p_{\textrm{T}},\eta)
(\textrm{SF}_f+\nu_k\Delta\textrm{SF}_{f,k})
\right).
\end{align}
Using $\nu_k=\pm1$, we construct synthetic data sets and fit linear surrogates
\begin{align}
\hat S_{\textrm{HF}}(\bx,\nu_{\textrm{HF}})
=
\exp\left(\nu_{\textrm{HF}}\hat\Delta_{\textrm{HF}}(\bx)\right),
\qquad
\hat S_{\textrm{LF}}(\bx,\nu_{\textrm{LF}})
=
\exp\left(\nu_{\textrm{LF}}\hat\Delta_{\textrm{LF}}(\bx)\right).
\end{align}
The HF  tagging uncertainty variations typically modify distributions at the few-percent level, while the LF variations can reach slightly larger values due to the larger multiplicity of light jets in the selected events.

\paragraph{Lepton efficiencies.}
Uncertainties in lepton efficiencies are detailed in Refs.~\cite{open-data-ws-1,open-data-ws-2,open-data-ws-3} and are incorporated using an event-weighting function. Since the efficiency scale factors depend on the lepton pseudo-rapidity, which is not included in $\bx$, we retain the dependence on the latent configuration \bz via
\begin{align}
r(\bx_i,\bz_i|\nu_\ell)
=
\prod_{\ell=1}^{2}
\left(
1+\frac{\Delta_\ell\textrm{SF}(\ell)}
{\textrm{SF}(\ell)}
\right)^{\nu_\ell}.
\end{align}
Using the variations $\mathcal{V}=\{-1,1\}$, we train the surrogate separately for $\ell=\mu$ and $\ell=e$, so in summary we have
\begin{align}
\hat S_\ell(\bx,\nu_\mu,\nu_e)
=
\exp\left(\nu_\mu\hat\Delta_\mu(\bx)+\nu_e\hat\Delta_e(\bx)\right).
\end{align}
These variations are typically below the percent level and show little dependence on the event kinematics.

\paragraph{Systematic uncertainties summary.}
In summary, our model for the systematic uncertainties approximates $\ttbar(2\ell)$, modeled as a single region, at detector level through the differential cross section ratio
\begin{align}
\frac{\mathcal L(\bn)}{\mathcal L(\bzero)}
\frac{\dd\sigma(\bx|\bc,\bn)}{\dd\sigma(\bx|\bzero,\bzero)}
\;\simeq\;&
\alpha_{\rm lumi}^{\nu_{\rm lumi}}\,
\hat R(\bx,\bc)\,
\hat S_{\rm mhou}(\bx,\nu_R,\nu_F)\,
\hat S_{\alpha_s}(\bx,\nu_{\alpha_s})\,
\hat S_{\rm JES}(\bx,\bn_{\rm JES})\,
\hat S_{\rm JER}(\bx,\bn_{\rm JER})\,\nonumber\\
&\times
\hat S_{\rm HF}(\bx,\nu_{\rm HF})\,
\hat S_{\rm LF}(\bx,\nu_{\rm LF})\,
\hat S_{\ell}(\bx,\nu_\mu,\nu_e)\, \equiv 1+\hat T(\bx;\bc,\bn) \, ,
\label{eq:model-summary-factorized}
\end{align}
with all factors defined above.
This suffices to compute the unbinned test statistic defined in Sec.~\ref{sec:likelihood-eval}.

We display the total systematic uncertainty band associated to our model as shaded areas in Figs.~\ref{fit-impact-pod-1}--\ref{fit-impact-pod-2}.
This band is constructed from 1000 nuisance-parameter vectors sampled from the prefit distribution \(\mathcal N(\bn|\bzero,\bone)\). Each draw defines one point in nuisance-parameter space at which we evaluate Eq.~(\ref{eq:model-summary-factorized}) and reweight the simulated events accordingly on an event-by-event basis. For each draw, this yields a varied histogram.
The shaded band is then obtained from the bin-by-bin 16\% and 84\% quantiles of the resulting histogram ensemble.

\paragraph{Training of surrogates for systematic uncertainties.}
The surrogate for the nuisance-parameter dependence is implemented with the parametric neural-network approach introduced in~\cite{Benato:2025rgo}.
For each systematic effect, we train a separate network on the 16 reconstructed features \(\bx\), whose outputs are $\hat\Delta_A(\bx)$, while the nuisance parameters enter only through $\nu_A$ in Eq.~(\ref{eq:S-training-task}). Each output node corresponds to a coefficient in the polynomial expansion of the logarithm of the differential cross section.
The neural network architecture consists of two hidden layers with 128 nodes each, using ReLU activation functions.
The output layer is initialized at zero, such that the network starts from the nominal prediction and learns only the systematic deviations from it.

To ensure a reproducible separation of training and validation, events are assigned deterministically to non-overlapping subsets using unique identifiers together with a fixed random seed.
The network parameters are optimized with the \texttt{AdamW} optimizer~\cite{AdamW-optimizer} using an initial learning rate of \(1.5 \times 10^{-4}\).
Training is performed for at most 500 epochs, with a gradual phase-out of the learning rate from the beginning of training.
To control over-training, we employ early stopping based on the validation loss, evaluated on held-out data, with a patience of 20 epochs and no minimum improvement threshold beyond strict monotonic improvement.
The training is therefore terminated once the validation loss ceases to improve and the best-performing epoch is retained.

\paragraph{Validation of surrogate training.}
The trained surrogates for the systematic uncertainties are validated with classifier two-sample tests~(C2ST)~\cite{Friedman:2003id,lopezpaz2018revisiting,Das:2023ktd}, following the prescription of Ref.~\cite{Schofbeck:2024zjo}. Such tests compare high-dimensional distributions directly, without relying on low-dimensional projections. The underlying idea is that a well-trained surrogate $\hat S_i(\bx,\nu_i)$ correctly describes the effect of a nuisance variation, so that reweighting the varied sample with $\hat S_i^{-1}$ should render it statistically indistinguishable from the nominal one. To test this, we train a binary neural classifier to separate two samples and use the area under the ROC curve~(AUC) as metric. An AUC of $0.5$ corresponds to indistinguishable samples, while an AUC of $1$ corresponds to perfect separation.

\begin{figure}[t]
    \centering
    \hfill
\includegraphics[width=0.49\linewidth]{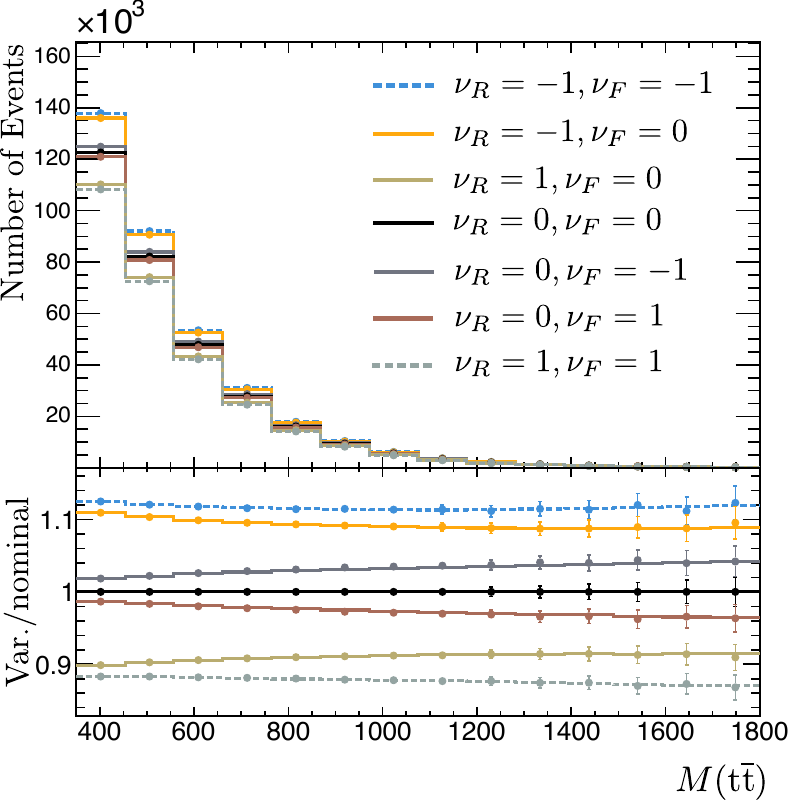}\hfill
\includegraphics[width=0.49\linewidth]{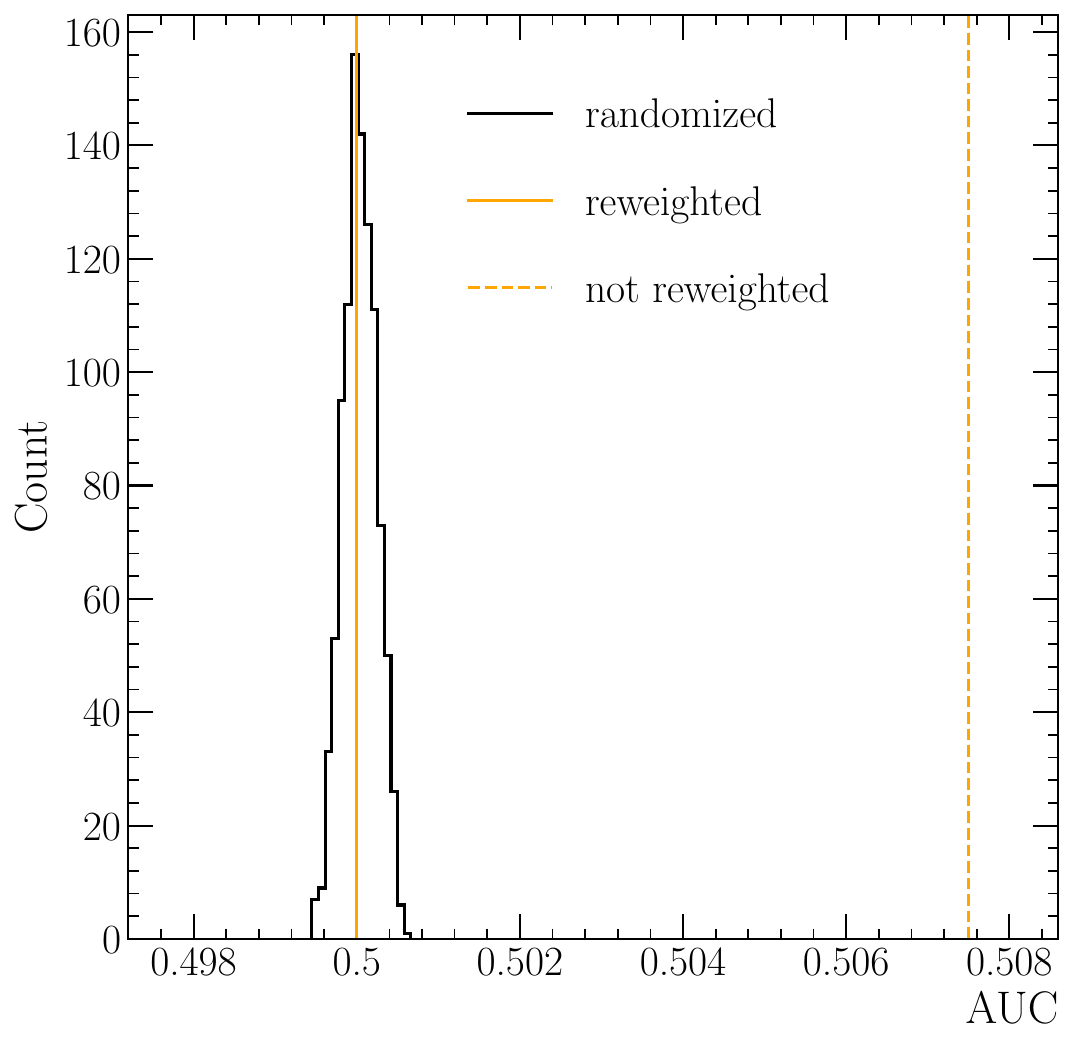}
\hfill
    \caption{Validation of the learned MHOU surrogate $\hat S_{\rm mhou}(\bx,\nu_R,\nu_F)$. The variations of the reconstructed $m(\ttbar)$ distribution~(markers) for the shifts in  $(\nu_{R},\nu_{F})$ listed in Eq.~(\ref{eq:MHOU_variations}), compared with the corresponding surrogate predictions shown as lines~(left).
    The C2ST~(right), performed for $(\nu_{R},\nu_{F})=(1,1)$, comparing the nominal sample with the shifted sample before reweighting (orange dashed) and after reweighting with the learned surrogate to the nominal point (orange solid). The black histogram shows the distribution of AUC values obtained from 1000 repetitions with shuffled labels, corresponding to the expected spread around $0.5$ for indistinguishable samples.}
    \label{fig:S_validation}
\end{figure}

To demonstrate the quality of the surrogate training, in Fig.~\ref{fig:S_validation} we consider as a representative example the MHOU surrogate $\hat S_{\rm mhou}(\bx,\nu_R,\nu_F)$.
The left panel shows the variation of the reconstructed $m(\ttbar)$ distribution around the nominal sample at $(\nu_{R},\nu_{F})=(0,0)$ for the true and learned predictions for the scale-varied shifts listed in Eq.~(\ref{eq:MHOU_variations}).
For the C2ST test, we choose the variation $(\nu_{R},\nu_{F})=(1,1)$. Note that the test is sensitive only to shape differences; the visible normalisation offsets in the one-dimensional projection do not provide discriminating power to the classifier. Conversely, the good agreement in this projection alone does not guarantee agreement in the full 16-dimensional feature space.

The right panel of Fig.~\ref{fig:S_validation} summarizes the C2ST result.
Using the original event weights, the classifier separates the nominal and shifted samples with an AUC of $0.5078$ (orange dashed), confirming that the classifier architecture is sufficiently expressive. The fact that this value is close to $0.5$ reflects the modest effect of the shape variation from the shifted scales.
Next we reweight the shifted sample according to
\begin{align}
w_i \;\to\; w_i'=\hat S_i^{-1}(\bx_i,\nu_i)\,w_i\, ,
\end{align}
where $w_i$ and $\bx_i$ denote the event weights and features of the shifted sample. If the surrogate is accurate, the reweighted sample should follow the nominal distribution. Indeed, we observe that the AUC moves to $0.5$ (orange solid). Finally, we repeat the same test after shuffling the class labels between the weighted nominal and shifted samples. By construction, this yields AUC values centered at $0.5$; repeating the procedure 1000 times gives the black histogram, which quantifies the expected spread from the classifier architecture and training procedure.

The classifier used for these surrogate validation tests has four hidden layers with $(512,512,256,128)$ units and is trained with the {\tt Adam} optimizer, while all reported AUCs are evaluated on independent held-out data. We therefore observe that the classifier detects the unreweighted difference, but fails to distinguish the nominal and reweighted samples, demonstrating that the learned surrogate removes the nuisance-induced distortion. We have performed this test on several other trainings with identical (successful) results.

\subsection{Reference binned analysis}
\label{sec:binned-reference}

As a reference for the unbinned analysis, we consider a binned measurement following the choices of measured variables and binning taken in~\cite{CMS:2019esx}, namely the CMS measurement of $\ttbar{(2\ell)}$ at $\sqrt{s}=13$ TeV based on an integrated luminosity of $\mathcal{L}=35.9$ fb$^{-1}$.
We define a two-dimensional binning in the observables $m(\ttbar)$ and $y(\ttbar)$ with bin boundaries given by
\begin{align}
m(\ttbar)\in [300,\,400,\,500,\,650,\,1500]~\GeV\quad\text{and}\quad
|y(\ttbar)|\in [0,\,0.4,\,0.8,\,1.2,\,2.5]\, .
\end{align}
The resulting binned measurement thus consists of a \(4\times 4\) grid, double differential in \(m(\ttbar)\) and \(|y(\ttbar)|\), for a total of 16 data points.

The statistical treatment adopted for this binned reference measurement follows the general methodology of Ref.~\cite{Benato:2025rgo}, which we now summarize.
For the dependence on the POIs we use the same quadratic ansatz as in the unbinned case,
\begin{align}
\hat R_I(\bc)=1+\sum_a c_a \,\hat R_{I,a}+ \sum_{a,\,b\leq a}c_a c_b \, \hat R_{I,ab}=1+ c_A\hat R_{I,A}
\end{align}
where \(I\) labels the analysis bins and the index \(A\) again runs over the linear and symmetric quadratic monomials in the POIs.
Hence, the number of POIs and the polynomial structure in \(\bc\) are the same as in the unbinned model. The coefficients are obtained from varied detector-level yields according to the linear PDF model. Because PDF variations produce an exact quadratic polynomial, this polynomial detector-level interpolation is also exact.

For the nuisance parameter dependence, we also use the same parameterisation as in the unbinned analysis, with the same choice of parameters, prescriptions, and polynomial structures for the various sources of systematic uncertainty.
The only difference is that the surrogates \(\hat\Delta\) do not depend on the event features \(\bx\).
Instead of learning functions \(\hat\Delta_A(\bx)\), in the binned case we determine bin-wise coefficients \(\hat\Delta_{I,A}\) such that
\begin{align}
\hat S_I(\bn)=\exp\!\left(\nu_A \hat\Delta_{I,A}\right)\, .\label{eq:binned-predictions}
\end{align}
Likewise as in the unbinned case, we determine the constants $\hat\Delta_{I,A}$ from systematically varied data. The binned ``loss function'' is a multi-Gaussian $\chi^2$ computed from the differences of the predicted values in Eq.~(\ref{eq:binned-predictions}) and the binned training data variations at the same parameter points $\bn\in\mathcal{V}$ that we also use for training the unbinned surrogates.
A more detailed description is provided in Ref.~\cite{Benato:2025rgo}.

Therefore, the binned reference model differs from the unbinned analysis only in the replacement of the fully differential event-wise surrogates by a coarse bin-by-bin representation in the two observables \(m(\ttbar)\) and \(|y(\ttbar)|\).
Otherwise, the assumptions on factorisation and the modeling of POI effects and systematic uncertainties are identical. Additionally, in App. \ref{app:benchmarking_binned}, we benchmark the precision of our binned PDF determination against variants of the NNPDF global fit based on comparable gluon-sensitive datasets. This ensures a meaningful comparison between the gluon PDF determination from binned and unbinned analyses presented next.

\section{Results}
\label{sec:results}

Here we present the main results of this work, namely the determination of the gluon PDF from an NSBI analysis of unbinned simulated observables in top-quark pair production at the LHC.
First, in Sec.~\ref{sec:gluonPDF_results} we assess the expected precision of the NSBI approach as compared to its binned counterpart.
Subsequently, Sec.~\ref{subsec:stability} studies the stability of our results with respect to
a number of methodological variations.
Finally, we present in Sec.~\ref{sec:pheno} an initial exploration of the implications of our analysis for Higgs production in gluon fusion at the LHC.
Appendix~\ref{app:benchmarking_binned} then provides a benchmarking of the binned gluon PDF determination baseline with global PDF fits, as well as with dedicated variants of NNPDF4.0 based on synthetic data.

For the results presented in this section we follow the pipeline presented in Sec.~\ref{sec:unbinned-obs}.
First, for a given set of $N$ basis elements, we construct the eigenbasis as described in Sec.~\ref{sec:PCA}.
Then a complete set of surrogates, obtained with the methods described in Sec.~\ref{sec:learn-logratio} for $\hat R(\bx,\bc)$ and in Sec.~\ref{sec:learning-systematis} for $S(\bx,\bn)$, determine the differential cross section ratio in Eq.~(\ref{eq:model-summary-factorized}) from which the extended likelihood ratio in Eq.~(\ref{eq:ext-like-mse-2}) can be computed.
Finally, the minimisation, needed for the profiling of the nuisance parameters, is performed with \textsc{Minuit}~\cite{James:1975dr}.

\subsection{The gluon PDF from unbinned NSBI measurements}
\label{sec:gluonPDF_results}

In the following we present results based on the linear model for the gluon PDF with $N=7$ basis elements as the default and with the experimental and theoretical systematic uncertainties described in Sec.~\ref{sec:unbinned-obs}.
Asimov data are generated from the nominal {\sc\small POWHEG} sample based on NNPDF3.1 and reweighted to the central element of the linear PDF model, such that the best fit values of the coefficients satisfy $c_a=0$ with $a=1,\ldots,7$ by construction.
Stability upon variation of these settings is studied in Sec.~\ref{subsec:stability}.
We present results both at $Q_0=1.65$ GeV, the reference scale for the linear model of the gluon PDF, and for $Q=175$ GeV as representative scale for $\ttbar$ production at the LHC.
Uncertainties are reported as 68\% CL intervals.

\begin{figure}[t]
    \centering
\includegraphics[width=0.99\linewidth]{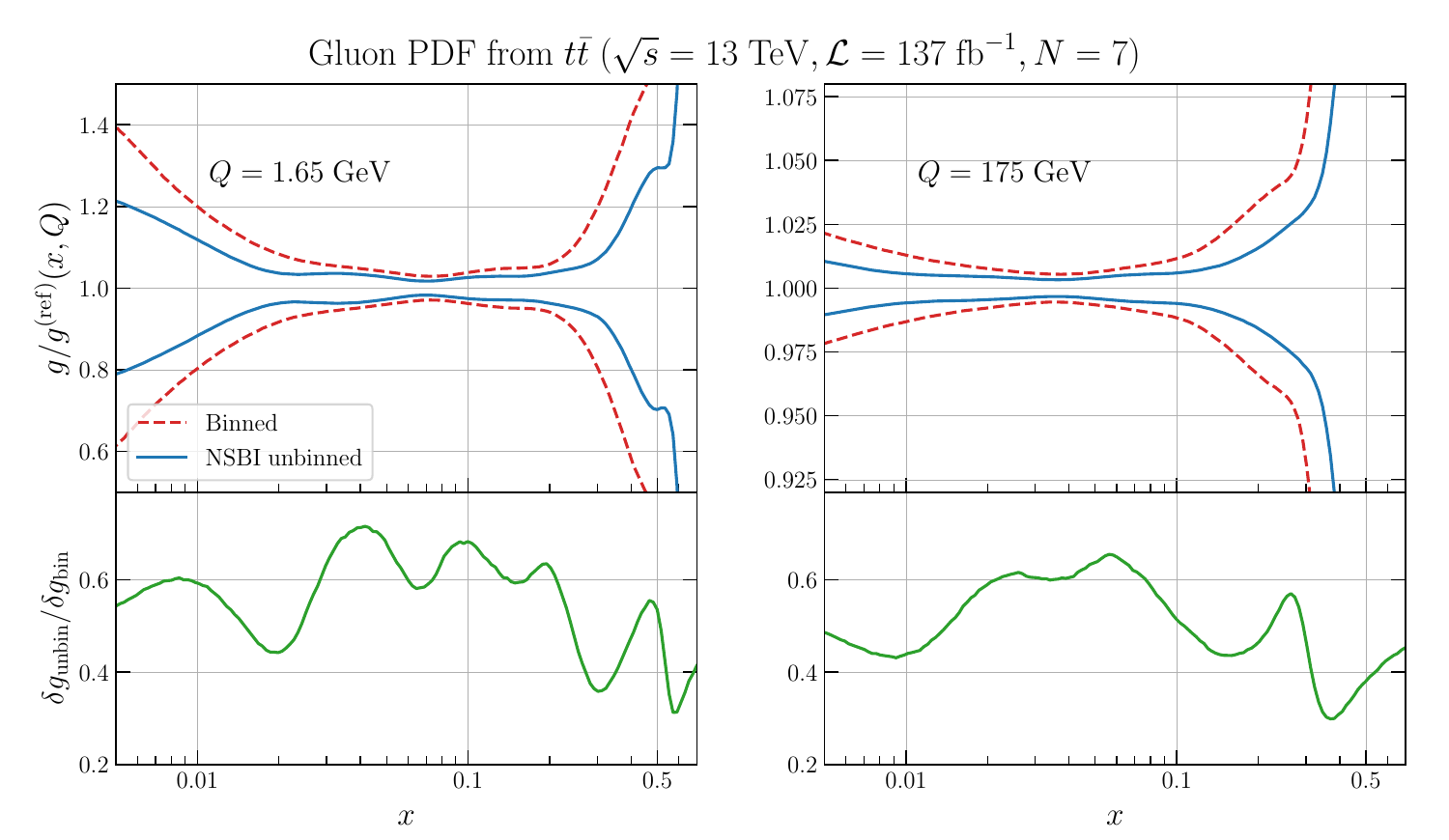}
\vspace{-0.1cm}
\caption{The expected precision (68\% CL intervals) of the gluon PDF from top quark pair production at the LHC, comparing the outcome of the binned and NSBI unbinned analyses at both $Q_0=1.65$ GeV (left) and $Q=175$ GeV (right panels).
We use the baseline linear model with $N=7$ elements.
The bottom panels display the ratio of PDF uncertainties between the binned and unbinned determinations.
  }
\label{fig:Sec6-Fig1-NSBIg-bin-vs-unbin}
\end{figure}

Figure~\ref{fig:Sec6-Fig1-NSBIg-bin-vs-unbin} displays our determination of the gluon PDF from Asimov top-quark pair production at the LHC, comparing the outcome of the binned and NSBI unbinned analyses at both $Q_0=1.65$ GeV and $Q=175$ GeV.
Results are shown normalized to the central value of the reference gluon PDF in the region of $x$ matching the kinematic coverage of the measurement, and the bottom panels display the ratio of
PDF uncertainties between the unbinned and binned determinations, $\delta g_{\rm unbin}/\delta g_{\rm bin}$.
The improvement in sensitivity in going from a binned to an unbinned analysis is visible both at low and at high scales.
Indeed, $\delta g_{\rm unbin}/\delta g_{\rm bin}$ ranges between 0.3 and 0.7 depending on the value of $x$; on average, the precision of the NSBI unbinned analysis improves by around a factor 2 as compared to the binned reference.
At $Q=175$ GeV, the effects of DGLAP evolution result in an overall reduction of PDF uncertainties mostly at small-$x$.
As discussed below, the stability of the linear PDF model upon variations of $N$ is improved as the scale $Q$ at which we evaluate the gluon increases.

Having established the feasibility and impact of a NSBI unbinned determination of the gluon PDF from top-quark pair production data at the LHC, we  demonstrate its stability with respect to a number of methodological choices.

\subsection{Stability analysis and methodological variations}
\label{subsec:stability}

We now demonstrate the stability and robustness of our procedure with respect to a number of methodological choices.
First, we assess the stability of our results with respect to the dimensionality of the linear model function basis.
Then, we quantify the interplay between statistical and systematic uncertainties in the determination of the gluon PDF.
Finally, we show how the method can accurately reconstruct different PDFs used to generate the synthetic data.

\begin{figure}[t]
    \centering
\includegraphics[width=0.99\linewidth]{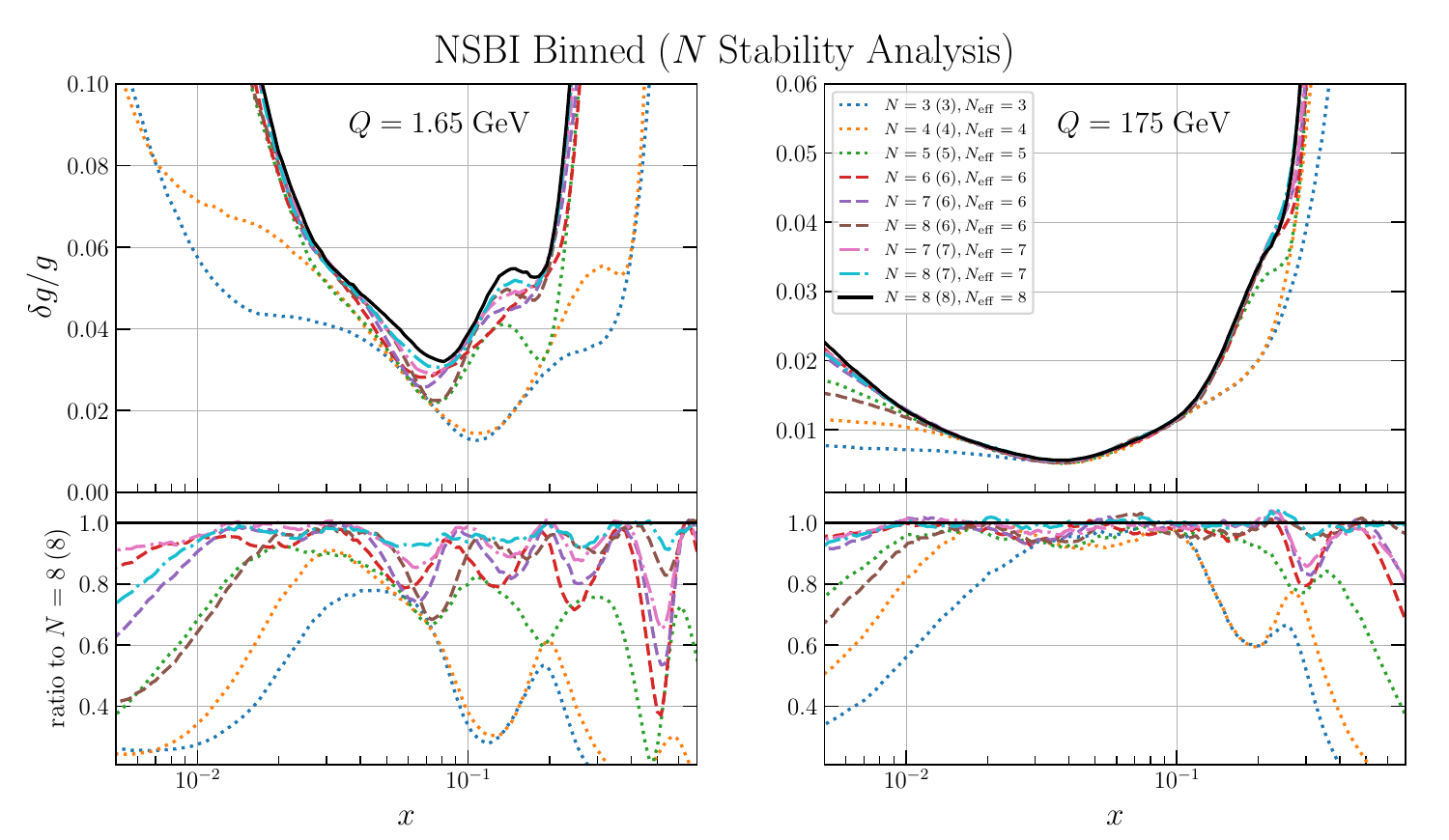}
\vspace{-0.01cm}
    \caption{Comparison of the relative gluon PDF uncertainties $\delta g/g$ for $Q=1.65$ GeV (left) and $Q=175$ GeV (right) in the binned determination for linear models of different dimensionality $N$.
    The numbers in parenthesis indicate the number of PoIs after PCA rotation and the removal of quasi-flat directions, which result into the effective number of PoIs $N_{\rm eff}$.
    The bottom panels display the ratio of the PDF uncertainties with respect to $N=8~(8)$.
    }
\label{fig:Sec6-Fig4-Nstability}
\end{figure}

\begin{figure}[t]
    \centering
\includegraphics[width=0.99\linewidth]{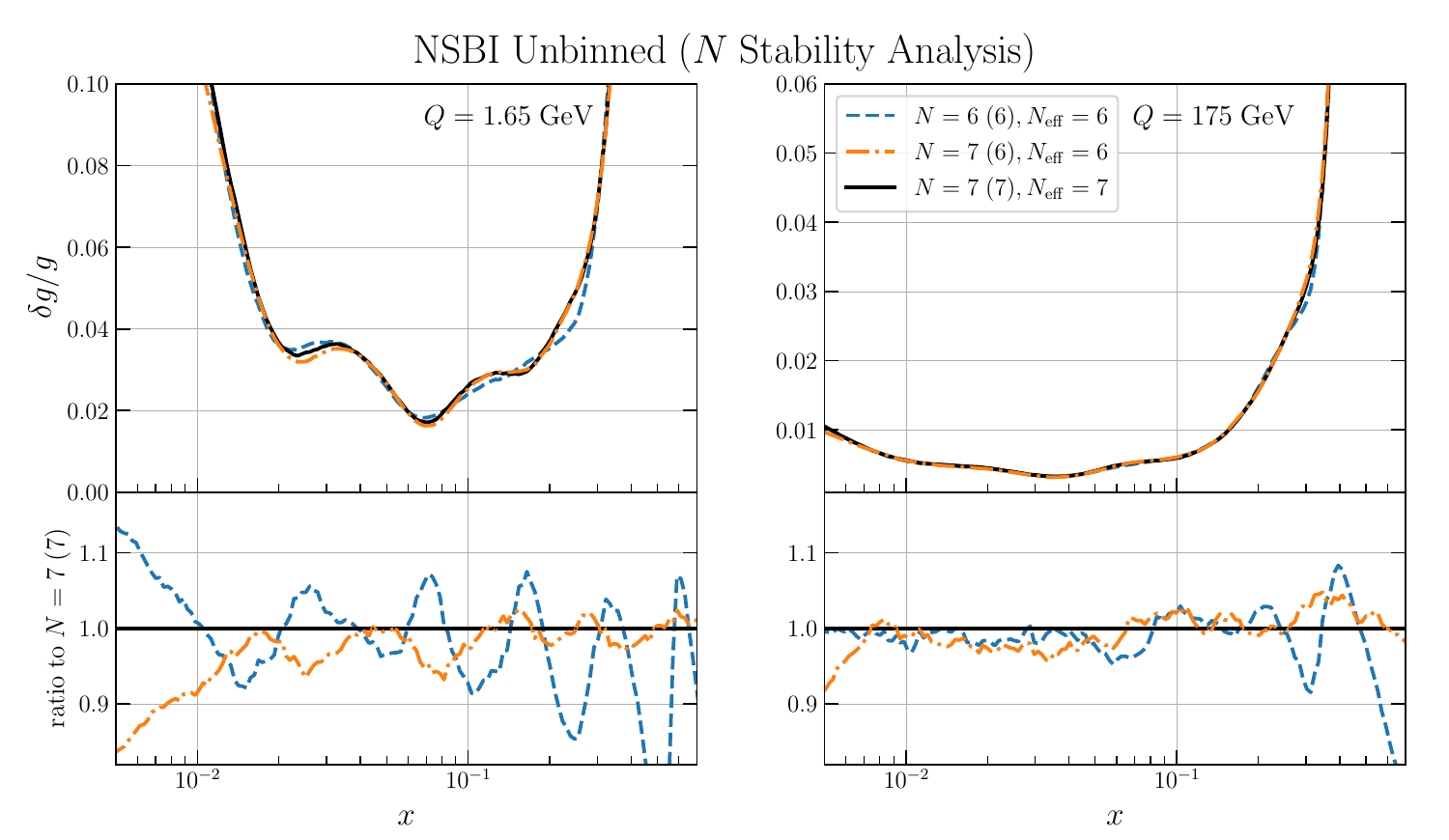}
\vspace{-0.01cm}
    \caption{Same as Fig.~\ref{fig:Sec6-Fig4-Nstability} for the gluon PDFs determined from the NSBI binned analyses.
    }
\label{fig:Sec6-Fig4-Nstability-Unbinned}
\end{figure}

\paragraph{Linear basis dimensionality.}
Figures~\ref{fig:Sec6-Fig4-Nstability} and~\ref{fig:Sec6-Fig4-Nstability-Unbinned} compare the relative uncertainties, $\delta g/g$, for the gluon PDF in the binned and unbinned analyses respectively for linear models based on different values of $N$, including our baseline choice of $N=7$, both at $Q=1.65$ GeV and at $Q=175$ GeV.
The bottom panels display the ratio of PDF uncertainties with respect to the $N=8~(8)$ (binned) and $N=7~(7)$ (unbinned) results.
The numbers in brackets indicate the number of fitted degrees of freedom entering the fit after PCA rotation and removal of quasi-flat directions, which define $N_{\rm eff}$, the effective number of PoIs.
Results for different variants of the linear model are ordered with respect to their value of $N_{\rm eff}$.

From the results of Figs.~\ref{fig:Sec6-Fig4-Nstability} and~\ref{fig:Sec6-Fig4-Nstability-Unbinned} one observes that in the region where the bulk of the data set has sensitivity on the gluon PDF, $0.01\le x \le 0.4$, the results of both the binned and unbinned fits are stable both at low and at high scales once $N_{\rm eff}$ is large enough.
While low values of $N_{\rm eff}$ are as expected insufficient to achieve a robust PDF uncertainty estimate, there  is a clear convergence in the gluon PDF uncertainty as $N_{\rm eff}$ is increased as measured by the differences with respect to $N=8~(8)$ (binned) and $N=7~(7)$ (unbinned analysis), specially at the scales where the synthetic data is available.
Large differences are observed in the small-$x$ and large-$x$ extrapolation regions (highlighting large model dependence) at $Q=1.65$ GeV, which are partly reduced after DGLAP evolution to $Q=175$ GeV.

We conclude that our baseline linear model with $N=7$ is sufficiently stable, both in the binned and unbinned analysis, upon the addition of more elements to the basis functions, provided that quasi-flat directions with low eigenvalues after PCA rotation are weeded out.
Without this removal, increasing the dimensionality of the linear model eventually leads to numerical instabilities.

\begin{figure}[t]
    \centering
\includegraphics[width=0.99\linewidth]{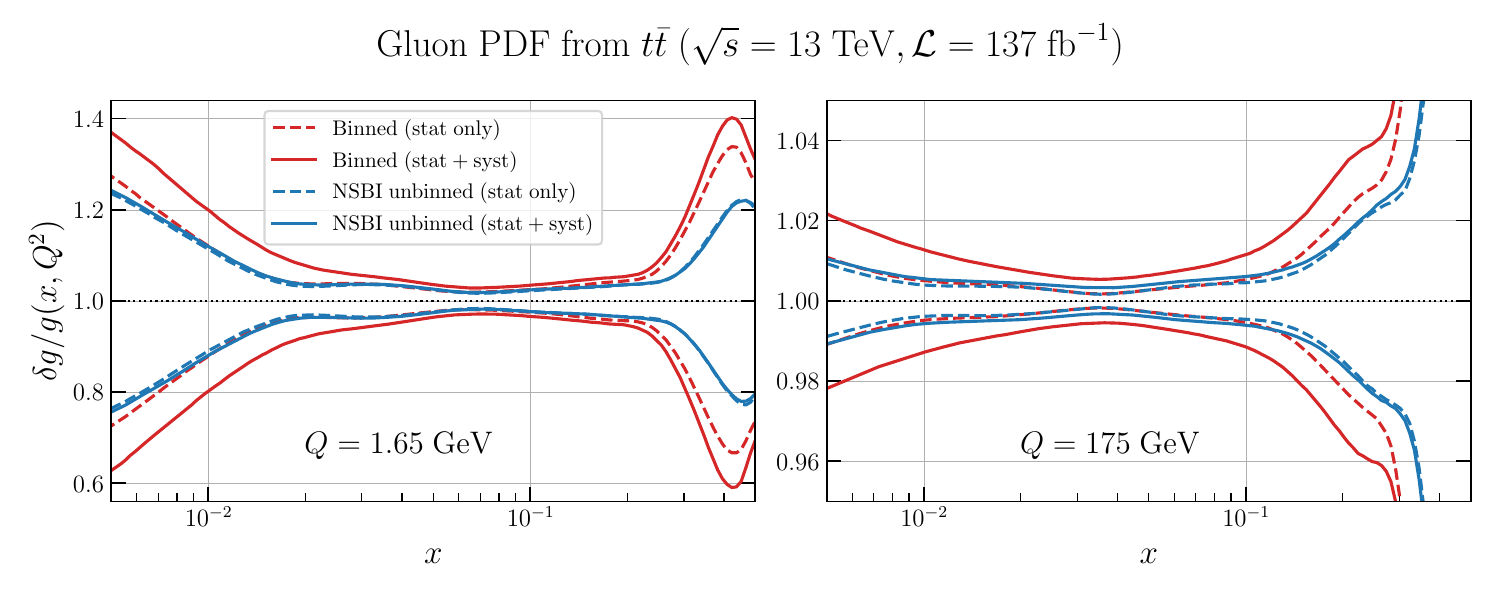}
\vspace{-0.5cm}
    \caption{Comparison of the binned and NSBI unbinned uncertainties for $N=6$ without systematic uncertainties~(dashed) to the nominal result~(solid), shown for $Q=1.65~$GeV (left) and $Q=175~$GeV (right).
    }
\label{fig:Sec6-Fig5-stat-vs-syst}
\end{figure}

\paragraph{Uncertainty breakdown.}
The baseline results presented in Sec.~\ref{sec:gluonPDF_results} are obtained with an uncertainty model for the Asimov data which includes both statistical errors and the experimental and theoretical systematic uncertainties encoded by the corresponding nuisance parameters.
In order to quantify the interplay between statistical and systematic errors in this determination of the gluon PDF, Fig.~\ref{fig:Sec6-Fig5-stat-vs-syst} shows a similar comparison as that of Fig.~\ref{fig:Sec6-Fig4-Nstability} now for the results obtained where only statistical errors are taken into account, compared to the baseline result where also experimental and theoretical systematic uncertainties enter in the definition of the test statistic.
We show results for $N=6$, though we have verified that conclusions remain unaltered for the baseline $N=7$ model.

From Fig.~\ref{fig:Sec6-Fig5-stat-vs-syst}, one sees that in the binned fit the worsening in precision as compared to the statistical-only result is sizeable for $x\lsim 0.3$.
For instance, around $x\sim 0.1$ the gluon PDF uncertainty at $Q=175$ GeV increases by a factor two due to the effects of the systematic nuisances.
These differences are markedly reduced in the case of the unbinned analysis, showing that unbinned measurements are more efficient to constrain the systematic uncertainties from the data.
One also observes from this comparison that the precision achieved with the binned analysis neglecting all systematic uncertainties is comparable with the unbinned result which accounts for the full uncertainty model of the data.

The overall message of Fig.~\ref{fig:Sec6-Fig5-stat-vs-syst} is that the NSBI unbinned determination of the gluon PDF benefits from improved constraining potential for the systematic uncertainties entering the fit as nuisance parameters.
Although, as discussed in Sec.~\ref{sec:unbinned-obs}, here we adopt a heuristic model for the dominant sources of theoretical and experimental uncertainties, it is likely that this feature remains if the analysis is carried out at the level of experimental data with a fully realistic prescription of the systematic effects.

\paragraph{Reconstructing alternative PDFs.}
The results shown in Sec.~\ref{sec:gluonPDF_results} are obtained from Asimov data generated using the same reference gluon PDF as in the linear model.
We verify that in this case the best-fit values of the linear model parameters are consistent with zero, as expected for a successful closure test.
Here we demonstrate that also when the Asimov data are generated with a different hypothesis for the gluon PDF, the linear model can accurately reconstruct it through non-zero best-fit values of its parameters.

\begin{figure}[t]
    \centering
\includegraphics[width=0.45\linewidth]{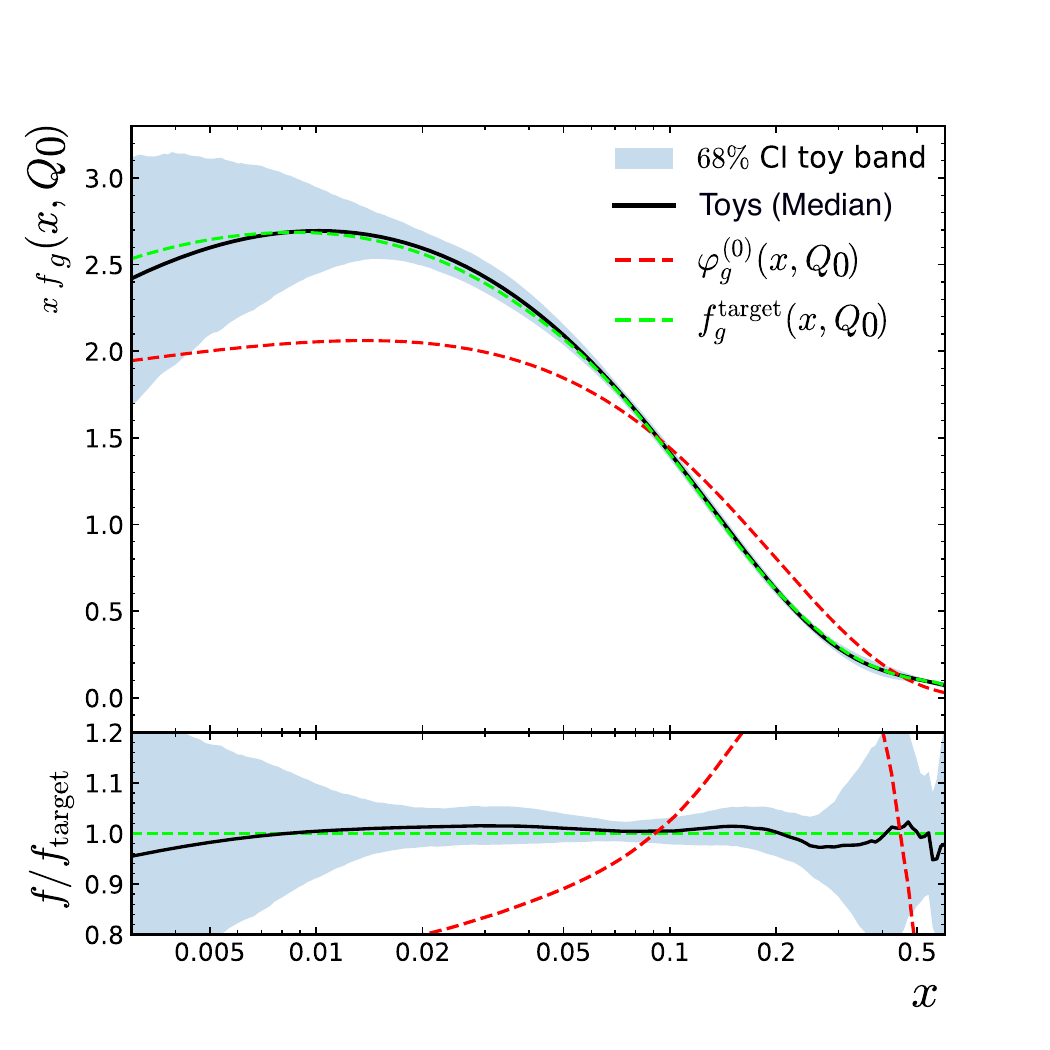}
\includegraphics[width=0.49\linewidth]{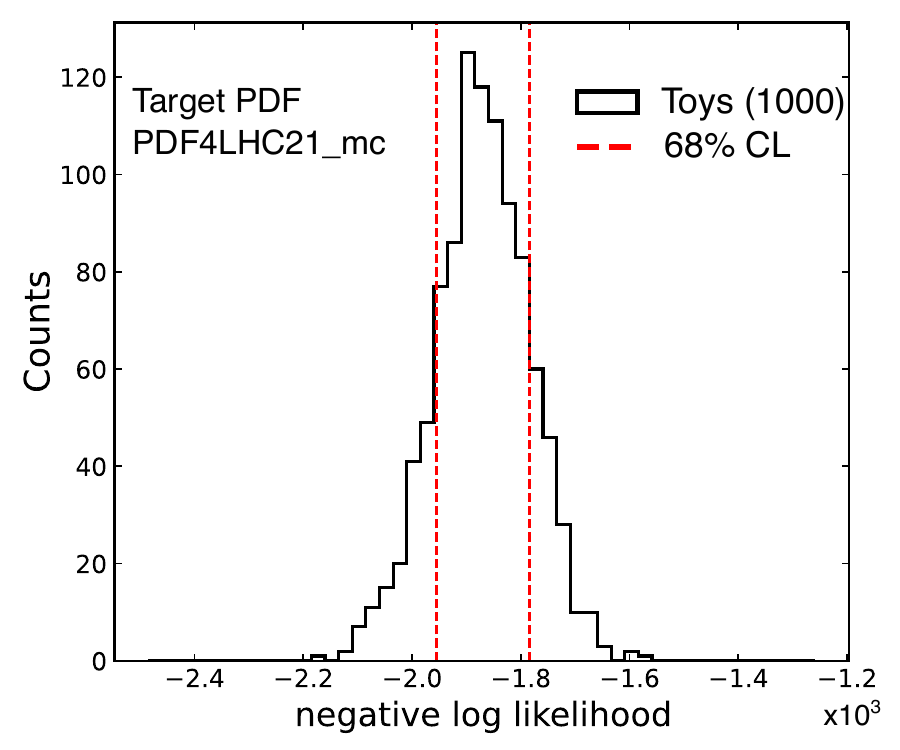}
\includegraphics[width=0.49\linewidth]{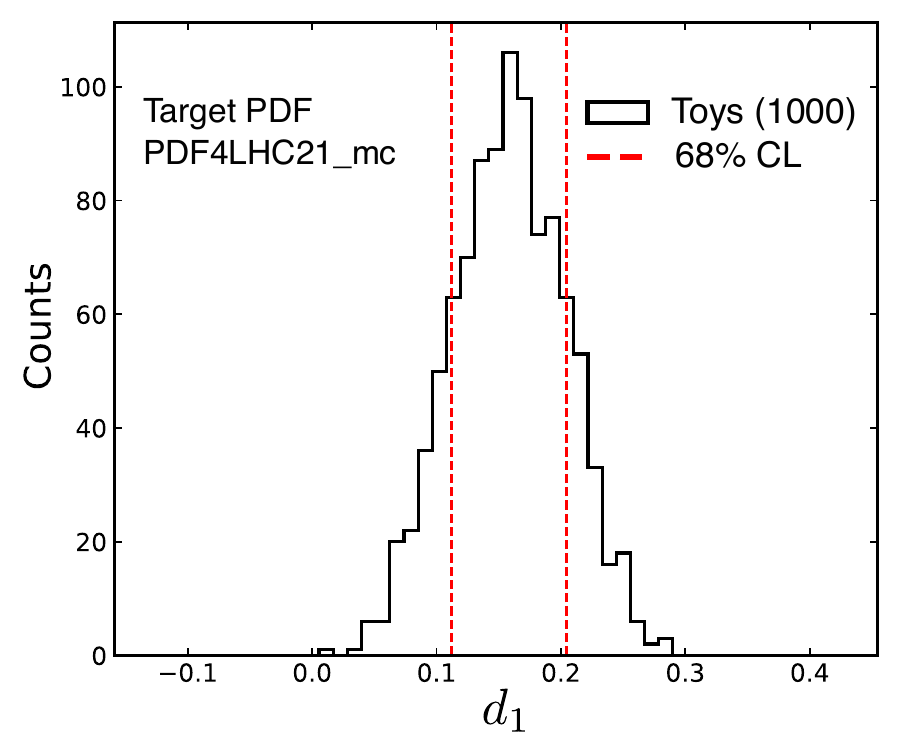}
\includegraphics[width=0.49\linewidth]{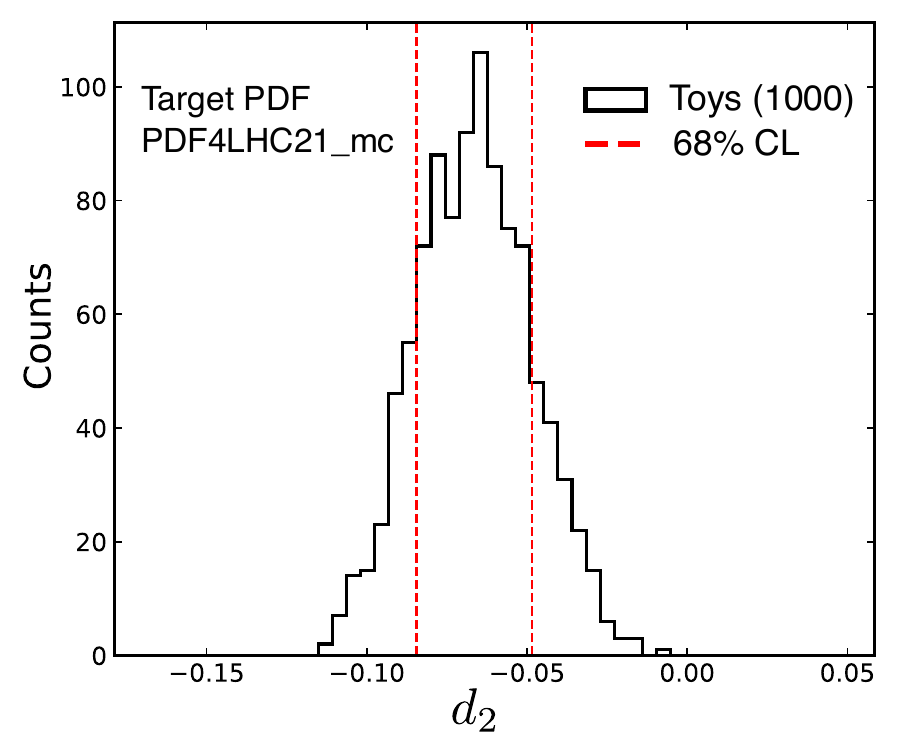}
\vspace{-0.1cm}
    \caption{Fit of the linear gluon PDF model to toy data sampled from a simulated data set~(top left) according to the central element of PDF4LHC21~($f_{\rm target}^{(g)}$).
    The mean of the toy distribution~(solid black) is in good agreement with the target gluon PDF~(dashed green).
    The distribution of the negative log-likelihood~(top right) and of the first two principal components~(bottom) in Eq.~(\ref{eq:principal_components}) are also shown.
    }
\label{fig:gluon-stability-inputPDF-run2}
\end{figure}

To this end, Fig.~\ref{fig:gluon-stability-inputPDF-run2} shows the results for a determination of the gluon PDF using a linear model with $f_{\rm ref}^{(g)}=\varphi_0^{(g)}(x)$ (the average over the elements of $\mathcal{H}$), see Eq.~(\ref{eq:linear_model_definition}), fitted to an Asimov data set reweighted with $f_{\rm target}^{(g)}\ne f_{\rm ref}^{(g)}$ and corresponding to the central element of PDF4LHC21.
The reweighting provides per-event probabilities and an inclusive cross section for the target PDF. We use these probabilities to obtain unweighted toy random samples, each representing a hypothetical measurement. The mean over the $N_{\rm toys}=10^3$ toys, equivalent to repeating the measurement $N$ times, is in good agreement with the target gluon PDF, demonstrating reconstruction accuracy also for an alternative hypothesis.
The top right and bottom panels show histograms for the model parameters over the $N_{\rm toys}=10^3$ generated using the PDF4LHC21 alternative hypothesis for the gluon PDF.
The distribution of fitted model parameters is clearly non-zero, as expected since now the best-fit values are different from those corresponding to the reference PDF set in the linear model.
We conclude that our method can be reliably used to reconstruct PDFs different from the central ones of the original linear PDF model.

\subsection{Implications for Higgs production in gluon-fusion}
\label{sec:pheno}

As a first phenomenological exploration of the potential impact of the results of this work for LHC processes, we consider Higgs production in gluon fusion, double differentially in the rapidity ($y_h$) and transverse momentum ($p_T^h$) of the produced Higgs boson.
As well known, the inclusive cross section for Higgs production in gluon fusion is sensitive to the gluon PDF around $x\approx 10^{-2}$, which corresponds to the lower edge of the kinematic coverage of top-quark pair production at the LHC.
However, double differential measurements in $(y_h,p_T^h)$ can cover the bulk of the kinematic region in $(x,Q^2)$ constrained by $\ttbar$ data, in particular by going to either high-$p_T^h$ or forward rapidities (or both).

\begin{figure}[t]
    \centering
\includegraphics[width=0.99\linewidth]{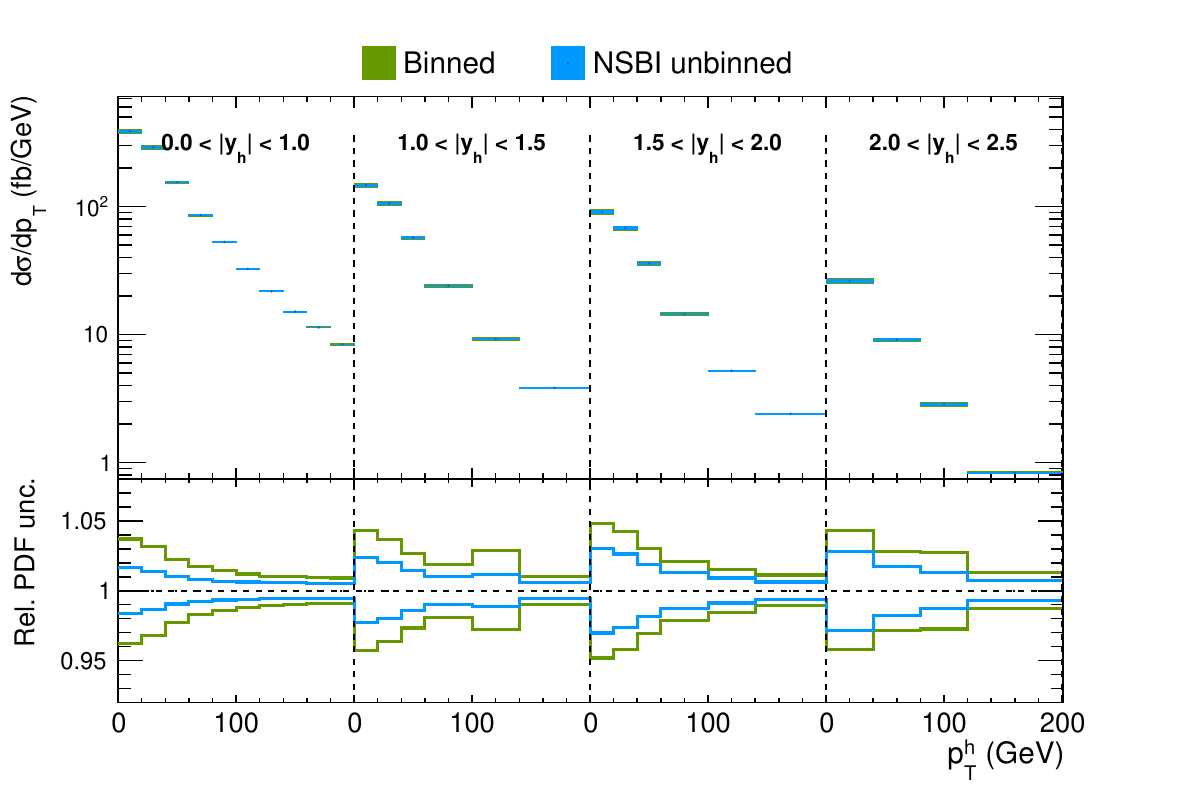}
    \caption{Double-differential cross section of Higgs production in gluon fusion, evaluated at NLO in QCD with {\sc\small Madgraph5\_aMC@NLO} as a function of absolute rapidity $|y_h|$ and transverse momentum $p_T^h$ of the Higgs boson.
    PDF uncertainties are evaluated for for the binned~(green) and NSBI unbinned~(blue) determinations of the gluon PDF presented in this work (with $N=7)$.
     }
    \label{fig:Higgs-pheno-distribution}
\end{figure}

With this motivation, Fig.~\ref{fig:Higgs-pheno-distribution} presents predictions for double-differential distributions in Higgs production in gluon fusion, evaluated at NLO in QCD with {\sc\small Madgraph\_aMC@NLO}~\cite{Alwall:2014hca} in the $m_{t}\to \infty$ limit.
For each of the four bins in the absolute Higgs rapidity $|y_h|$, we show the associated transverse momentum $p_T^h$ distribution.
Results are presented for the binned and NSBI unbinned determinations of the gluon PDF from $\ttbar$ data presented in this work.
Note that, by construction, the two predictions share the same central value.
The bottom panels display the relative PDF uncertainties for each case.

The analysis of Fig.~\ref{fig:Higgs-pheno-distribution} indicates that there is a significant improvement in precision in the NSBI unbinned analysis as compared to the baseline binned predictions, consistent with the PDF uncertainty reduction observed in Fig.~\ref{fig:Sec6-Fig1-NSBIg-bin-vs-unbin}.
This finding suggests that, at least for Higgs production in gluon fusion, PDF inputs for theoretical cross sections could be calibrated internally within ATLAS or CMS using their own measurements (in this case, using $\ttbar$ production) as the control sample, hence reducing the dependence on external data sets or theoretical assumptions.
This result is particularly advantageous to fully exploit that the patterns of experimental and theoretical systematic correlated errors can be kept under precise control if the whole pipeline is based on data within a single experiment.

The extrapolation of this specific application to other LHC processes is, however, far from straightforward since there one has to take into account also the contribution from the quark PDFs.
Nevertheless, the general principle is likely to uphold also for these more general cases: unbinned NSBI offers the potential of a purely inter-experiment calibration of proton structure inputs.

\section{Summary and outlook}
\label{sec:summary}

While particle physics moves into the HL-LHC era with unprecedented statistics, the availability of new methods to fully exploit the information in unbinned data becomes instrumental to push forward the frontier of our field and to understand nature at the smallest possible distances.
In this work, we have proposed a novel strategy using NSBI to exploit unbinned high-dimensional observables and demonstrated how it can be deployed to constrain proton structure.
Our approach bypasses some of the limitations hindering  traditional methods to PDF determinations based on binned observables unfolded to theory-friendly low-dimensional parton-level distributions with a multi-Gaussian statistical model.
By combining a linear model for the gluon PDF with unbinned  $\ttbar$ simulations at the LHC in the fully leptonic final state, we demonstrate an increase in sensitivity as compared to traditional binned measurements.
This highlights the potential of this method to inform a new generation of proton structure analyses.
Furthermore, our initial phenomenological application to Higgs production in gluon fusion suggests that NSBI can be deployed to obtain purely intra-experiment calibration of the PDF inputs required for theoretical predictions.

Crucial to our analysis is the careful assessment of the stability and robustness of the NSBI unbinned methodology, including the validity of the training of the ML surrogates.
We have also determined the dimensionality $N$ of the linear model which ensures sufficient expressivity in the region relevant to describe top-quark pair production; studied the stability of the fit with respect to $N$; and demonstrated that the method also closes if we assume different underlying gluon PDFs as ground truth.

The next step in this programme is to carry out this measurement of the gluon PDF on real LHC data.
In this endeavour, NNLO Monte Carlo generators such as {\sc\small MiNNLO}$_{\rm PS}$~\cite{Mazzitelli:2020jio,Monni:2020nks}, {\sc\small UNNLOPS}~\cite{Hoche:2014uhw}, or {\sc\small Geneva}~\cite{Alioli:2013hqa} for PDF-sensitive processes will be essential to achieve state-of-the-art accuracy in the theoretical predictions.
Beyond the proof of concept presented here, the same methods can be applied to the determination of the quark PDFs. Provided that sufficient unbinned measurements become available, a complete PDF fit from unbinned data should be feasible.
A natural next step is to combine the gluon sensitivity of \ttbar production with the complementary information provided by single-top and low- and high-mass Drell--Yan measurements.
In particular, b-initiated single-top production is sensitive to the large-\(x\) light quark PDFs as well as the large-\(x\) gluon PDF via perturbative gluon splitting~\cite{Nocera:2019wyk}.
Likewise, low- and high-mass neutral- and charged-current Drell--Yan data provide a powerful handle on quark and antiquark PDFs, with fully differential information in invariant mass, rapidity, and angular observables helping to lift degeneracies with BSM phenomena and reduce the impact of PDF uncertainties beyond what is possible in single-differential analyses~\cite{Torre:2020aiz,Panico:2021vav}.
Inclusive-jet measurements should ultimately also enter this programme given their constraining power on  large-\(x\) PDFs, although in this case a fully NSBI-based implementation may have to wait for a NNLO-accurate event generator describing jet production.

More ambitiously, combined determinations of PDFs/SM and BSM parameters based on unbinned measurements offer potential for sensitivity improvements with respect to comparable analyses based on binned observables.
Examples of such combined determinations include the simultaneous determination of the strong coupling constant~\cite{Ball:2025xgq,Ablat:2025gbp}, the mass of the top quark~\cite{Ball:2026qno,Alekhin:2024bhs}, and SMEFT Wilson coefficients~\cite{Kassabov:2023hbm,Costantini:2024xae}.
This is especially relevant in top-quark and Drell--Yan production, where BSM effects can partially align with, or even be absorbed into, PDF deformations in conventional analyses~\cite{Torre:2020aiz,Panico:2021vav,Hammou:2023heg,Cole:2026eex}.
In this respect, the high-dimensional detector-level information retained by NSBI can help disentangle genuine modifications of the hard production process from distortions induced by anomalous interactions in the decay kinematics of the top quark, or in the leptonic angular structure of Drell--Yan events.
On a somewhat longer timescale, we envision a simultaneous in-situ determination of top-quark and Higgs observables in which the proton PDFs act as the mediator between the two sectors, allowing constraints from \ttbar production to propagate directly to Higgs measurements, tightening the interplay between the two sectors within a single experimental likelihood.

Finally, to achieve the integration of binned and unbinned measurements in global PDF fits, it is crucial that unbinned likelihoods are released by the experimental collaborations, using, e.g., a variant of ML-assisted parametrized likelihoods.
In this new paradigm for PDF fits combining binned and unbinned measurements, the NSBI method presented in this work should also be applicable to carry out the latter outside the LHC, for example for PDF studies at the EIC or at FASER.

\subsection*{Acknowledgments}
We are grateful to the NNPDF Collaboration for organizing its Morimondo 2025 meeting where this idea was first conceived.
We are grateful to Mark Costantini and James Moore for their work on the linear model and for contributing to the public code.
We thank Luca Rottoli for discussions on the treatment of PDF uncertainties in {\sc\small POWHEG}.
We thank Kamil Laurent for producing the NNPDF top-only fit shown in App.~\ref{app:benchmarking_binned}.
The computational results presented were obtained using the CLIP cluster (\url{https://clip.science}).
S.~S.~C's work was supported by the ``Ram\'on y Cajal'' program under Project No.
RYC2024-048719-I, funded by ICIU/AEI/10.13039/501100011033 and by the FSE+.
E.H. is supported by the supported by the Swiss National
Science Foundation.
J.t.H is supported by
the STFC grant award ST/X000494/1.
L.M. acknowledges support from the European Union under the MSCA fellowship (Grant agreement N. 101149078) {\it Advancing global SMEFT fits in the LHC precision era (EFT4ward)}.
 M. U. is supported by the European Research Council under the European Union’s Horizon
2020 research and innovation Programme (grant agreement n.950246) and partially supported by the STFC grant ST/T000694/1.

\appendix
\section{The gluon linear model: stability and interpretability}
\label{app:linear_models_PDF}

In this appendix we provide additional details on the construction of the linear PDF model presented in Sec.~\ref{sec:linear_model}.
Specifically, we study the stability of the linear model with respect to the choice of reference quark PDFs, and visualise the hierarchies present between the different basis elements of the linear model, which provide insights into its interpretability.

\subsection{Dependence on the reference quark PDFs}
\label{app:PDF_quark_assumption}

In this work we construct a linear representation for the gluon PDF at the scale $Q_0$, while the quark PDFs are set to some chosen fixed reference.
The momentum sum rule is then
enforced by appropriately rescaling all the generated gluon replicas before performing the POD.
In Fig.~\ref{fig:pod_validation} we compared the outcome of the POD validation for the gluon PDF constructed with two different samples, one with the quark PDFs set to PDF4LHC21 and the other with the quark PDFs set to NNPDF4.0.
The performance of the method was found to be identical in both cases.

\begin{figure}[htbp]
    \centering
\includegraphics[width=0.99\linewidth]{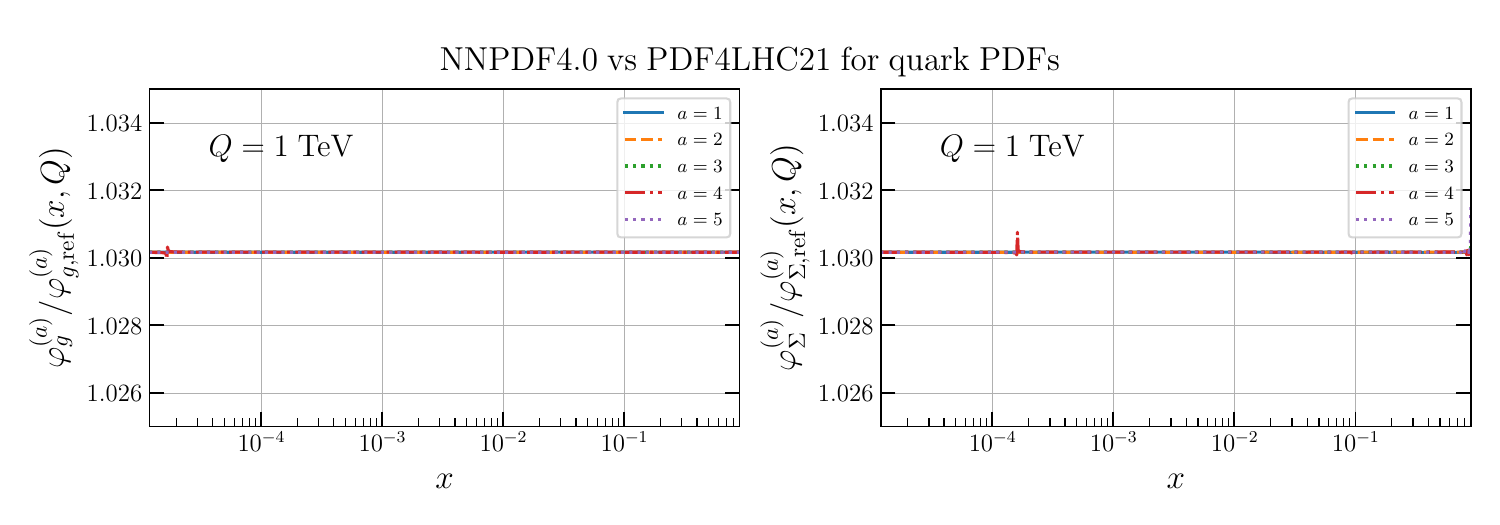}
    \caption{Same as Fig.~\ref{fig:gluon-basis-elements} at $Q=1$ TeV, now for the ratio of basis functions evaluated by means of the POD procedure applied to samples with two different assumptions of the quark PDFs (PDF4LHC21 and NNPDF4.0).
    The same behavior is observed for the rest of the eigenvectors.
    }
    \label{fig:gluon-basis-elements-checkQuarks}
\end{figure}

The reason for the stability of the POD procedure can be clearly seen in Fig.~\ref{fig:gluon-basis-elements-checkQuarks},
which displays the ratio of basis functions $\varphi_k(x)$ evaluated at $Q=1$ TeV with two different assumptions of the quark PDFs,
either PDF4LHC21 or NNPDF4.0.
The two sets of basis functions are identical up to a small overall rescaling of around 3\%, which can be fully reabsorbed when fitting the model coefficients.
This global offset is related to differences in the momentum integral carried by quarks in the two reference quark PDF sets.
We conclude that the linear model for the gluon PDF used here for the NSBI analysis is stable with respect to the choice of reference PDF set for the quarks.

\subsection{Model dimensionality and quasi-flat directions}
\label{app:basis_dimension}

We found in Fig.~\ref{fig:weight_bounds_PDF4LHC21} that the uncertainty of the parameters $c_k$ of the linear PDF model fitted to the PDF4LHC21 replicas drastically increases with the dimension $N$ of the linear model.
This result stems from the fact that increasing excessively the
number of model parameters opens quasi-flat directions in the theory parameter space and hence introduces numerical instabilities.

\begin{figure}[htbp]
    \centering
\includegraphics[width=0.8\linewidth]{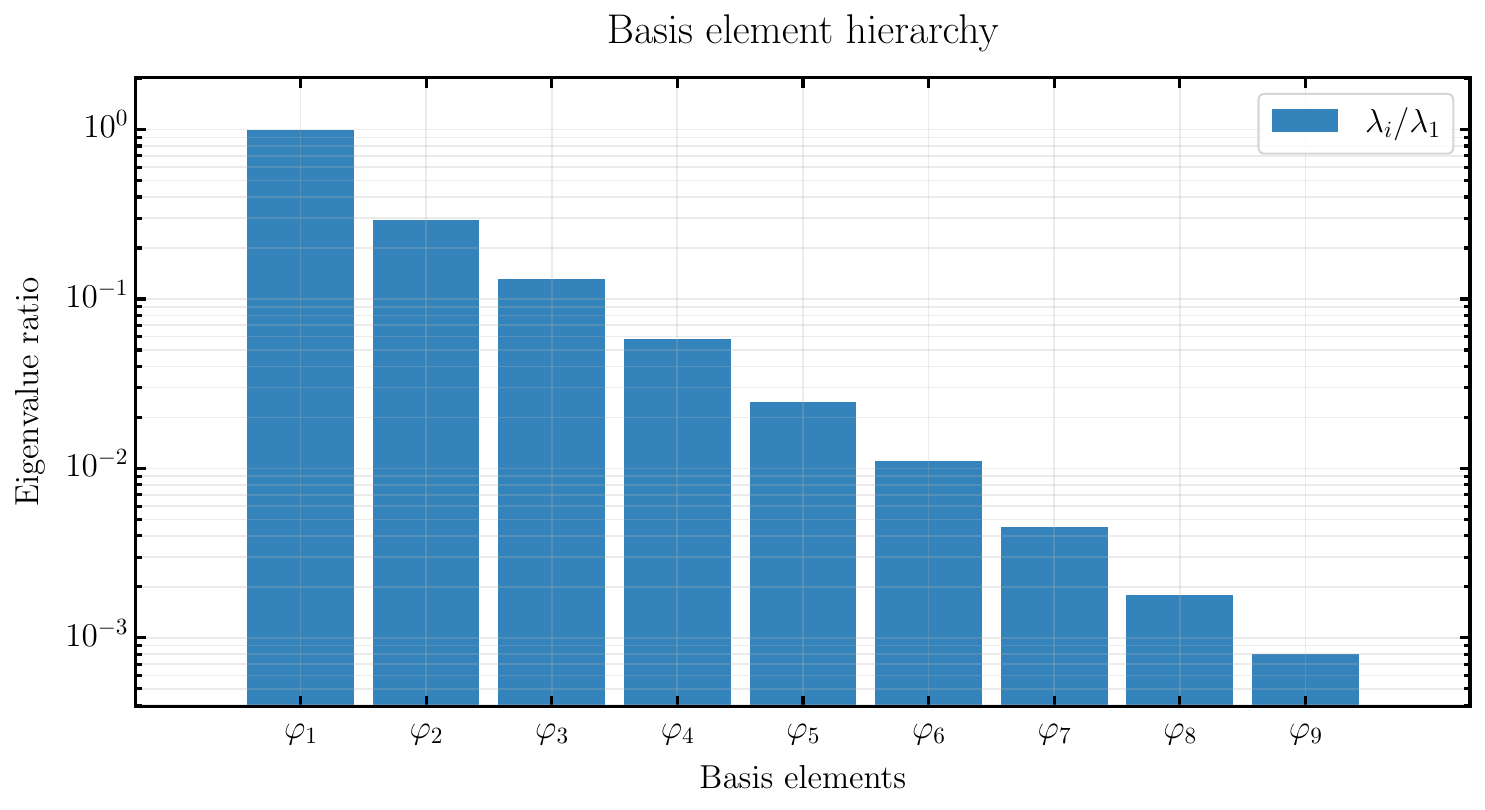}
    \caption{The eigenvalues of the linear model $\lambda_i$ (normalised to $\lambda_1$) for the first $N=9$ eigenvectors of the POD basis adopted in this study.
    A strong hierarchy among them is observed.
    }
    \label{fig:pod_eigenvalues}
\end{figure}

The appearance of quasi-flat directions becomes apparent upon inspection of the eigenvalues $\lambda_k$ associated to the eigenvectors $\varphi_k$ from the diagonalisation of the autocorrelation matrix $A$ presented
in Eq.~(\ref{eq:autocorrelation_matrix}).
These eigenvalues are plotted in Fig.~\ref{fig:pod_eigenvalues}, normalised to $k=1$ for the first $N=9$ members of the basis used throughout this study.
We observe a sharp and regular reduction of each $\lambda_k$ for each increment of the
index $k$, reaching $\lambda_9 / \lambda_1 < 10^{-3}$.
This hierarchy implies that considering only a small number of parameters of interest to describe the PDF
in the NSBI procedure presented in Sec.~\ref{sec:nsbi_for_pdfs} is well justified.
It also indicates that including too many basis elements in the analysis is bound to result in numerical instabilities.
As we saw in Fig.~\ref{fig:weight_bounds_PDF4LHC21}, the main consequence of increasing the dimension of the basis is the sizeable increase in the spread of the best-fit values of the theory parameters $c_k$.

\begin{figure}[t]
    \centering
\includegraphics[width=0.9\linewidth]{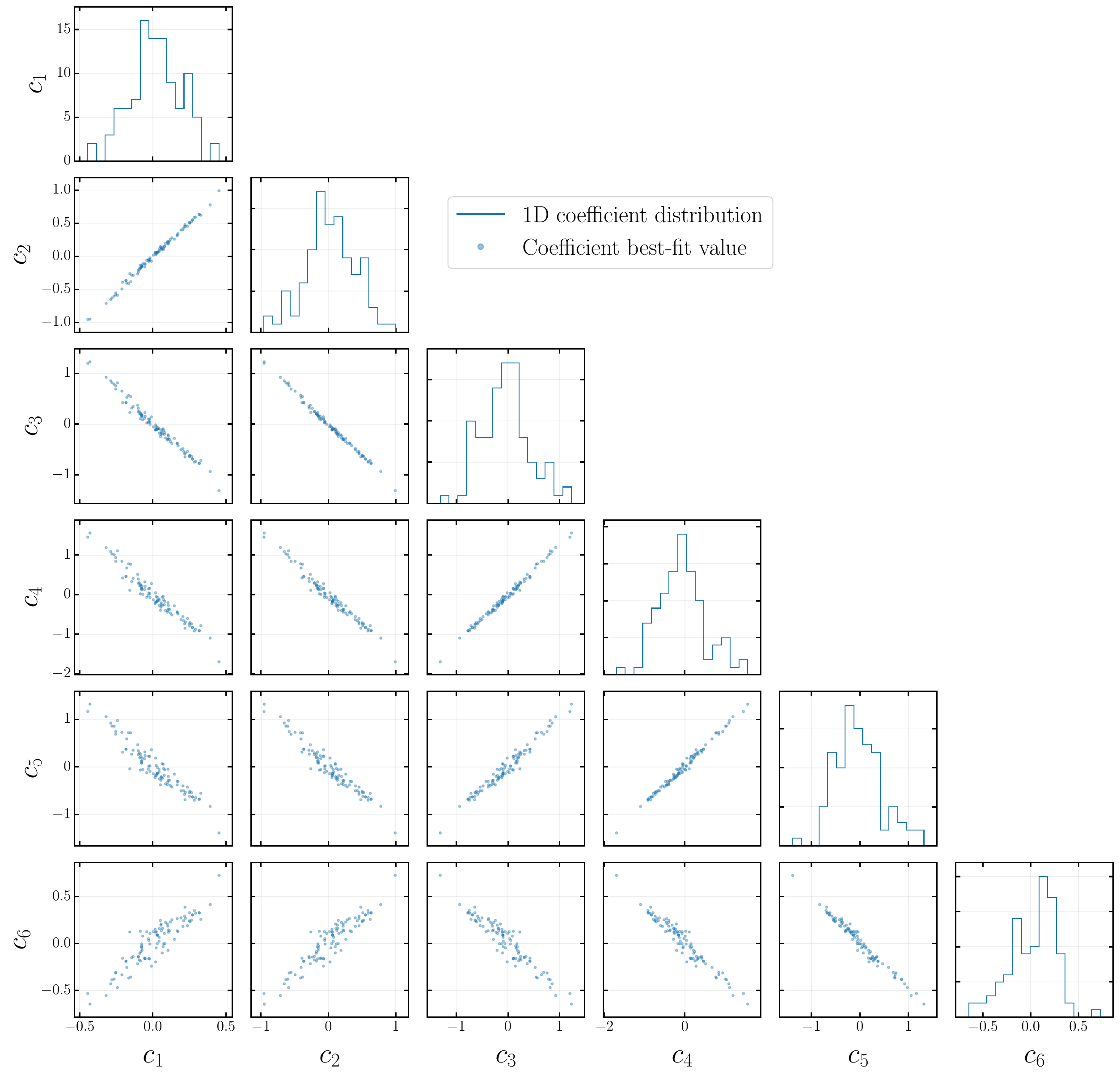}
    \caption{Scatterplot matrix associated to the results in
Fig.~\ref{fig:weight_bounds_PDF4LHC21} for the linear model consisting of $N=6$ elements.
    }
\label{fig:weight_corner_plot_PDF4LHC21}
\end{figure}

Furthermore, in Fig.~\ref{fig:weight_corner_plot_PDF4LHC21} we display the correlation between the different $c_k$ for $k=1$ to $k=6$ from the direct reconstruction of the target PDF4LHC21
presented at the end of Sec.~\ref{sec:linear_model}.
We observe, as expected, very strong correlations between the model parameters.
This justifies the careful consideration of which parameters of interest should be included in the NSBI analysis, as discussed in Sec.~\ref{sec:nsbi_for_pdfs}.

\section{Binned PDF determination benchmark}
\label{app:benchmarking_binned}

In this appendix we compare the binned determination of the gluon PDF presented in Sec.~\ref{sec:results} with the results from global PDF fits based on different datasets and methodologies.
The scope of this appendix is not to investigate in detail the quantitative role played by $\ttbar$ data in PDF sets, which has been studied at length in the literature, but rather to show that our binned PDF baseline provides a reasonable reference for the subsequent NSBI unbinned analyses.

\begin{figure}[t]
    \centering
\includegraphics[width=0.99\linewidth]{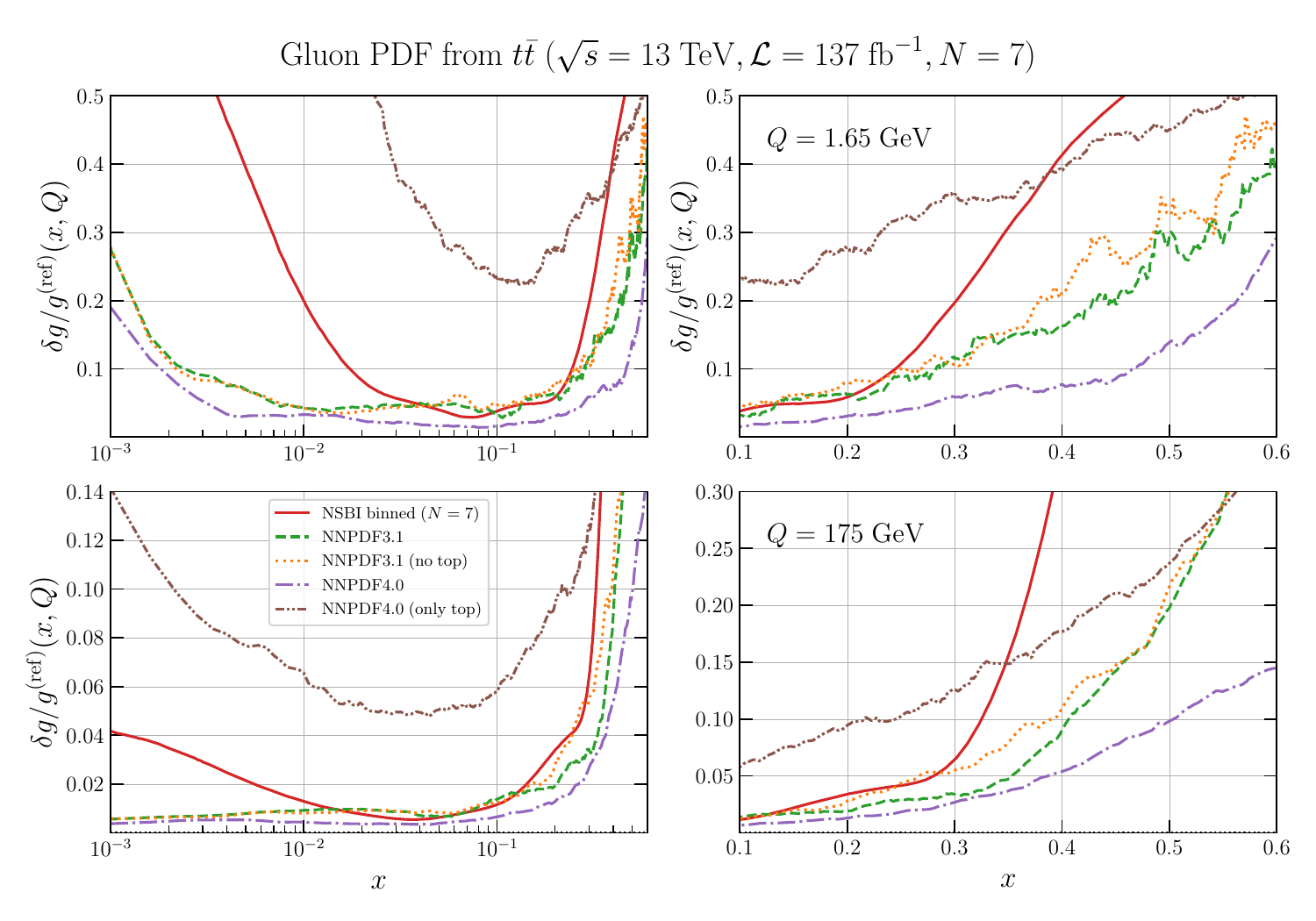}
    \caption{The relative 68\% CL uncertainties in different determinations of the gluon PDF at $Q=1.65$ GeV (top) and $Q=175$ GeV (bottom panels) in both logarithmic (left) and linear (right panels) scale.
    We compare the binned ($N=7$) result with the following global PDF fits: NNPDF3.1, a variant thereof where top-quark pair production data is removed, NNPDF4.0, and finally a variant thereof where the input data set is restricted to $\ttbar$ data exclusively.
    }
\label{fig:App-Fig-NSBIg-vs-NNPDF31nop-vs-NNPDF40onlytop}
\end{figure}

First, Fig.~\ref{fig:App-Fig-NSBIg-vs-NNPDF31nop-vs-NNPDF40onlytop} displays the relative 68\% CL uncertainties in different determinations of the gluon PDF at $Q=1.65$ GeV and $Q=175$ GeV in both logarithmic and linear scale.
We compare the binned baseline with the following global PDF fits: NNPDF3.1, a variant thereof where top-quark pair production data is removed, NNPDF4.0, and finally a variant thereof where the input data set is restricted to $\ttbar$ data.
From the latter, we see that fitting the gluon entirely from $\ttbar$ measurements leads as expected to large and not competitive PDF uncertainties.
The reason is that $\ttbar$ production on its own provides only mild constraints on the quark PDFs, and hence a simultaneous determination of all independent quark flavor combinations (7 in the case of NNPDF4.0) together with the gluon results in a poor precision for the gluon PDF.

For this reason, in our approach  we focus on the direction in the PDF space which is most constrained by $\ttbar$ data, namely the gluon, and fix the quark PDFs by means of external constraints.
Indeed, Fig.~\ref{fig:App-Fig-NSBIg-vs-NNPDF31nop-vs-NNPDF40onlytop} indicates how in the context of a global fit an extraction of the gluon PDF from $\ttbar$ data is only possible together with the input of other constraints, either other datasets or as in our approach assumptions on the quark PDFs.

\begin{figure}[t]
    \centering
\includegraphics[width=0.99\linewidth]{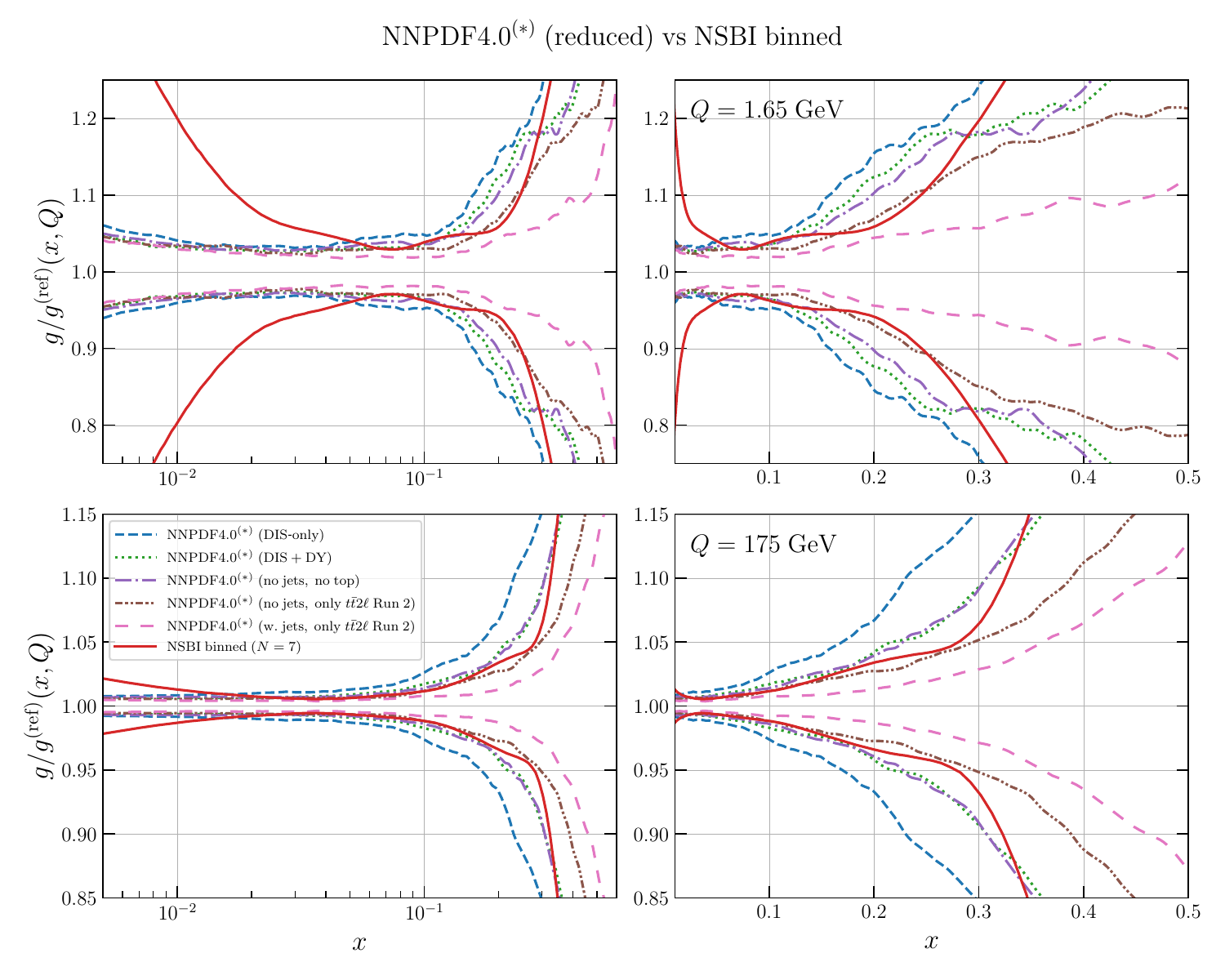}
    \caption{Comparison of PDF uncertainties in the binned determination of the gluon PDF presented in this work ($N=7$) with variants of
    NNPDF4.0  based on synthetic data (denoted by NNPDF4.0$^{(*)}$).
    These variants are based on increasingly large datasets, starting from DIS-only, then adding Drell-Yan data, then a global fit without any top or jet data, subsequently adding the CMS measurement of $\ttbar2\ell$ from Run 2, and finally the same fit adding also all jet and dijet data.
    We show results both at $Q=1.65$ GeV (top) and at $Q=175$ GeV (bottom panels).
    \label{fig:gluon-NNPDFsynth-NSBIbinned}
    }
\end{figure}

Figure~\ref{fig:gluon-NNPDFsynth-NSBIbinned} then compares the PDF uncertainties between our binned determination of the gluon  with variants of
NNPDF4.0 based on different (synthetic) datasets and denoted by NNPDF4.0$^{(*)}$.
These variants are based on an increasingly large datasets, starting from DIS-only, then adding Drell-Yan data, then a global fit without any top or jet data, subsequently adding the CMS measurement of $\ttbar2\ell$ from Run 2, and finally the same fit adding also all jet and dijet data.
This way, one can illustrate how in the context of a global PDF determination the various groups of processes contribute to the constraints on the gluon PDF.

In terms of input dataset, the NNPDF4.0$^{(*)}$ variant without jet production and where the only top dataset is $t\bar{t}2\ell$ from Run 2 should be compared to the gluon binned determination presented in this work.
Indeed, in both cases the large-$x$ gluon is mostly constrained by comparable observables, namely the binned $t\bar{t}2\ell$ analysis from Run 2, while quark PDFs are constrained separately.
While in the global fit other datasets also constrain the gluon PDF, {\it e.g.} via DGLAP evolution and higher-order corrections to DIS and Drell-Yan production, one may expect that this specific NNPDF4.0$^{(*)}$ variant should display gluon uncertainties which are similar, in the region of sensitivity, to those from our binned determination.

From Fig.~\ref{fig:gluon-NNPDFsynth-NSBIbinned} one observes that the addition of more data indeed leads to a monotonically decrease of the large-$x$ gluon PDF specially in the large-$x$ region, consistent with expectations.
In particular, we note that adding the CMS $\ttbar2\ell$ measurement provides non-trivial information on the large-$x$ gluon as compared to the corresponding fit without any top and jet data, and then the inclusion of jet and dijet production leads to a further improvement on the gluon PDF precision (consistently with previous studies {\it e.g.}~\cite{Czakon:2016olj}).
Furthermore, the binned baseline determination from this work displays PDF uncertainties which are comparable or worse than those of the corresponding NNPDF4.0$^{(*)}$ variant discussed above.
This qualitative agreement in the $\ttbar2\ell$ signal region indicates that our binned baseline provides a sensible reference to gauge the relative impact of the NSBI unbinned measurements in the gluon PDF determination.
We also note that this feature holds both at low and at high values of $Q$.
Outside the region covered by the $\ttbar 2\ell$ dataset, as expected, the PDF uncertainties in our binned determination are much larger in comparison with the corresponding NNPDF4.0$^{(*)}$ baseline .

\begin{figure}[htbp]
    \centering
\includegraphics[width=0.99\linewidth]{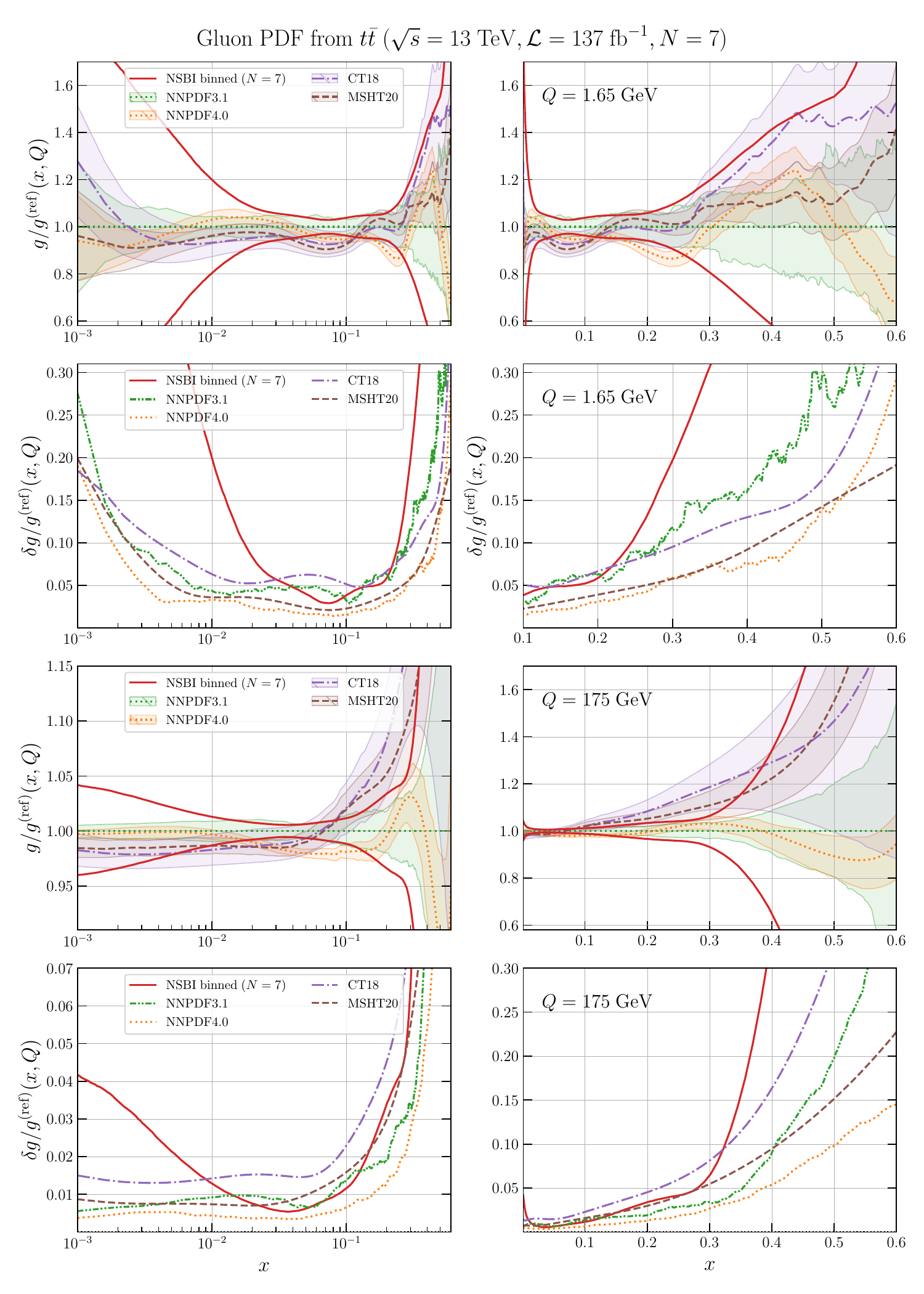}
\vspace{-0.5cm}
\caption{Same as Fig.~\ref{fig:Sec6-Fig1-NSBIg-bin-vs-unbin} now comparing our baseline binned determination of the gluon PDF with the results from four global PDF fits: NNPDF3.1, NNPDF4.0, CT18, and MSHT20, in all cases at NNLO and with $\alpha_s(m_Z)=0.118$.
We also show the relative 68\% CL PDF uncertainty associated with each gluon PDF (2${}^\text{nd}$ and 4${}^\text{th}$ rows).
The reference PDF set is NNPDF3.1.
    }
\label{fig:Sec6-Fig2-NSBIg-vs-globalfits}
\end{figure}

\begin{figure}[t]
    \centering
\includegraphics[width=0.99\linewidth]{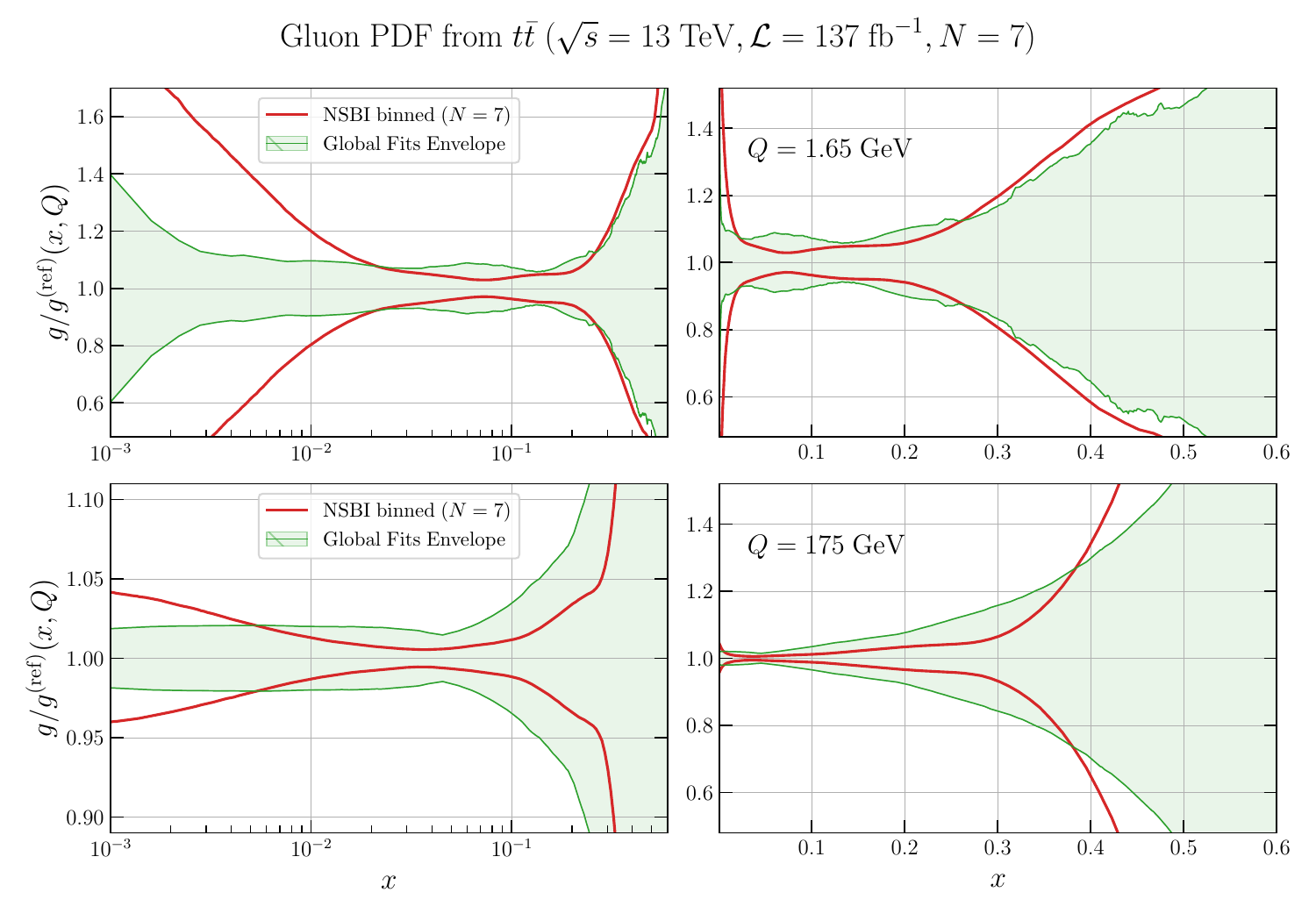}
\vspace{-0.5cm}
\caption{Comparison of the binned gluon PDF uncertainties to the symmetrized uncertainty envelope of NNPDF3.1, NNPDF4.0, CT18, and MSHT20 for $Q=1.65$~GeV~(top) and $Q=175$~GeV~(bottom), on a logarithmic~(left) and linear~(right) scale, see also Fig.~\ref{fig:Sec6-Fig2-NSBIg-vs-globalfits} for the comparison with the original PDF sets.
    }
\label{fig:Sec6-Fig3-NSBIg-vs-globalfits-envelope}
\end{figure}

Next, for completeness, we compare our binned baseline determination ($N=7)$ with global PDF fits.
Figure~\ref{fig:Sec6-Fig2-NSBIg-vs-globalfits} displays this binned determination alongside the results from NNPDF3.1, NNPDF4.0, CT18, and MSHT20, in all cases at NNLO and with $\alpha_s(m_Z)=0.118$.
We also display the relative 68\% CL PDF uncertainty associated with each of the gluon PDFs shown.
We do not display PDF4LHC21~\cite{PDF4LHCWorkingGroup:2022cjn}, since this PDF set is obtained from the unweighted statistical combination of NNPDF3.1, CT18, and MSHT20, and hence would not modify the qualitative discussion of this comparison.
Furthermore, we also show in Fig.~\ref{fig:Sec6-Fig3-NSBIg-vs-globalfits-envelope} the same comparison as Fig.~\ref{fig:Sec6-Fig2-NSBIg-vs-globalfits} now replacing the global fits with their (symmetrised) envelope to facilitate visualisation.
From this comparison, we see that in the region where the $\ttbar2\ell$ data has maximum sensitivity, $0.03 \lsim x \lsim 0.2$, at $Q=1.65$ GeV the precision of the binned gluon PDF baseline is bracketed by that of the global PDF fits (which differ not only in data and theory, but also in their methodologies for error estimate) while outside this region our binned baseline displays much larger uncertainties due to the lack of constraints on the gluon PDF there.

\begin{figure}[t]
    \centering
\includegraphics[width=0.90\linewidth]{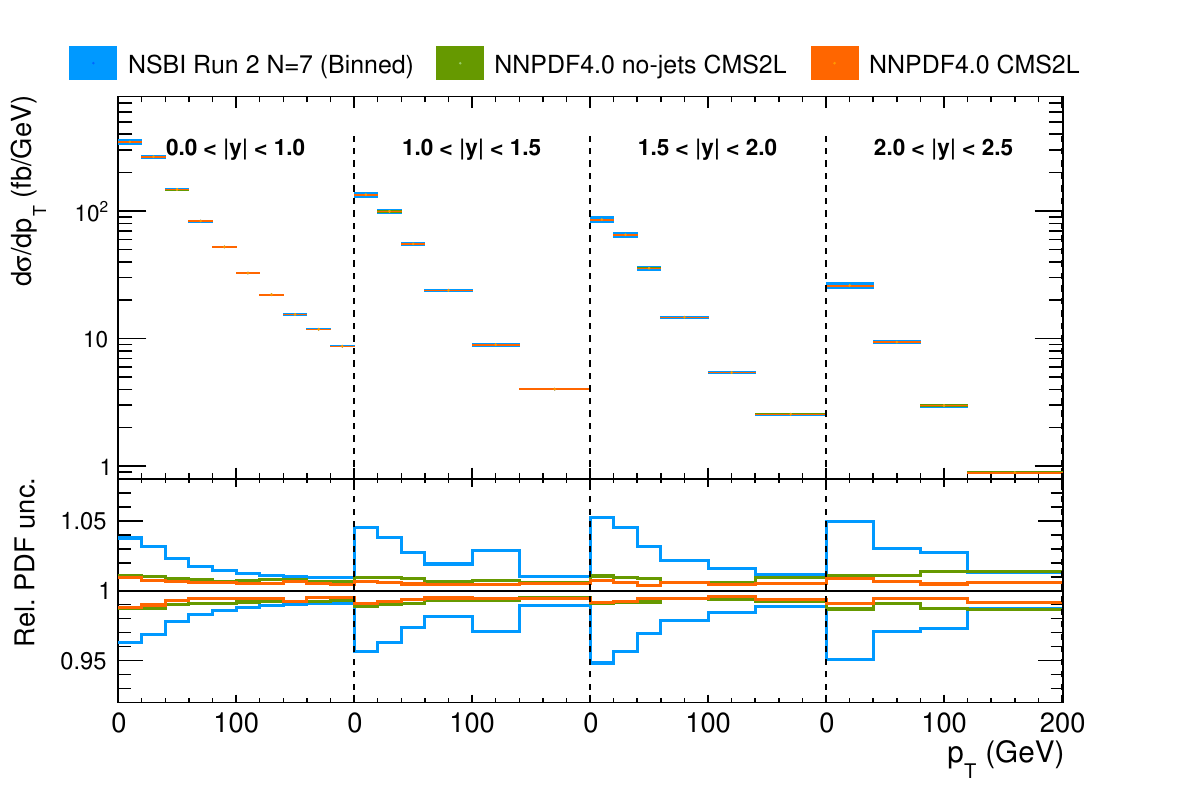}
    \caption{Same as Fig.~\ref{fig:Higgs-pheno-distribution} now comparing the predictions based on the NSBI binned baseline determination of the gluon PDF with two of the different NNPDF4.0$^{*}$ variants shown in Fig.~\ref{fig:gluon-NNPDFsynth-NSBIbinned}, see text for more details.
    \label{fig:Higgs-production-NNPDF-closure}}
\end{figure}

Fig.~\ref{fig:Higgs-production-NNPDF-closure} shows the same Higgs cross-sections as in Fig.~\ref{fig:Higgs-pheno-distribution} now with two  of the NNPDF4.0$^{*}$ variant as in Fig.~\ref{fig:gluon-NNPDFsynth-NSBIbinned}, namely the fits where one adds to a no-jet, no-top baseline first the CMS $t\bar{t}2\ell$ from Run 2 dataset and then all the jet and dijet data from NNPDF4.0.
Qualitative results are consistent with the corresponding  comparison at the PDF level:
at high-$p_T^h$ and large $|y_h|$, which maps to the gluon at large-$x$ and $Q\sim m_h$, the NSBI binned prediction is qualitatively similar in terms of PDF uncertainty to the corresponding NNPDF4.0$^{*}$ variant, while in other regions (like small-$p_T^h$, which maps to medium-$x$ instead) the former is less precise than the latter, as it should be.

\begin{figure}[t]
    \centering
\includegraphics[width=0.99\linewidth]{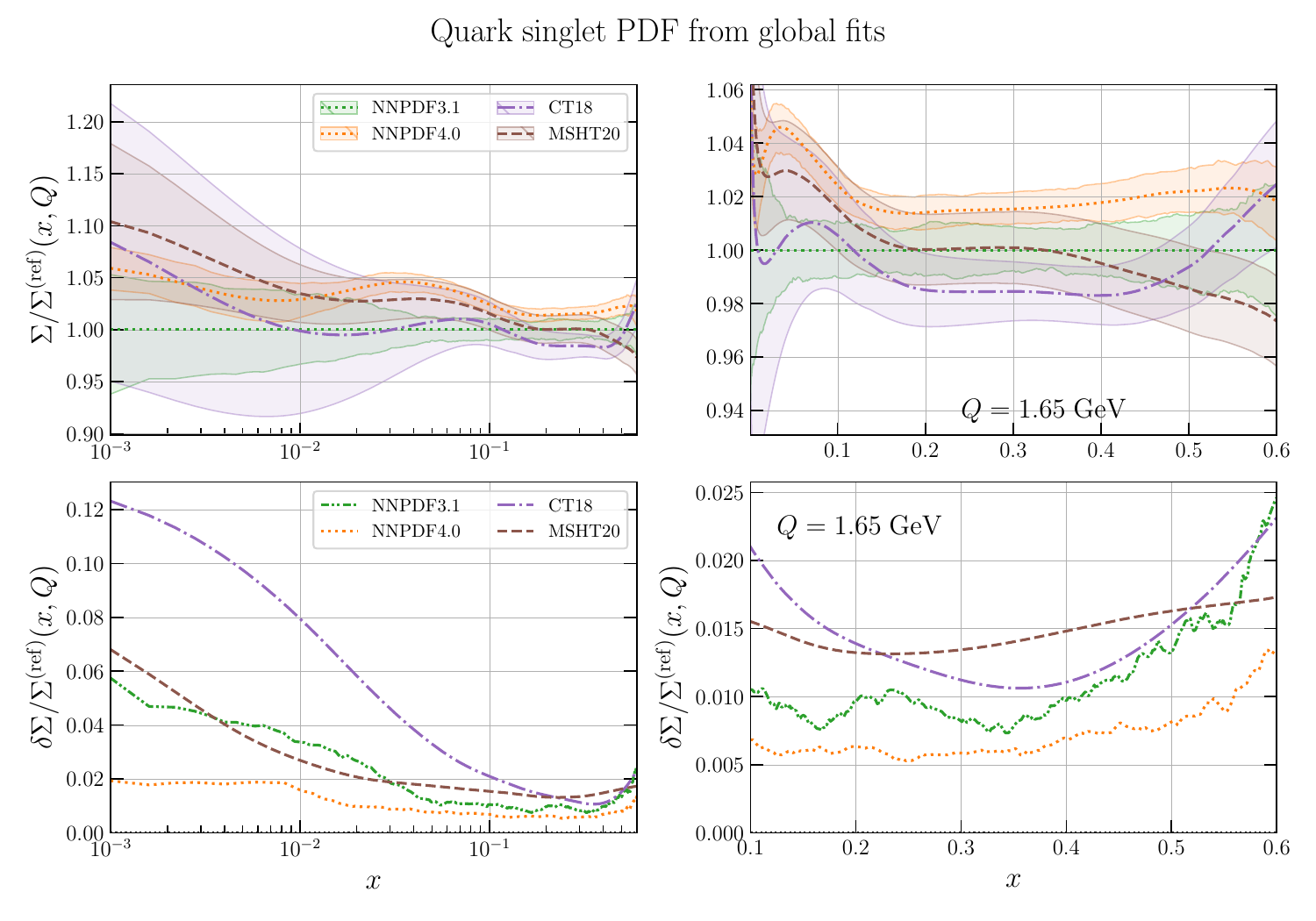}
    \caption{Same as Fig.~\ref{fig:Sec6-Fig2-NSBIg-vs-globalfits} now for the singlet PDF at the fit scale of $Q_0=1.65$ GeV.
    \label{fig:singlet-PDF-comparison}}
\end{figure}

Finally, Fig.~\ref{fig:singlet-PDF-comparison} shows the same comparison between global fits as in Fig.~\ref{fig:Sec6-Fig2-NSBIg-vs-globalfits} for the total quark singlet PDF at the fit scale of $Q_0=1.65$ GeV.
One sees that the differences between $\Sigma(x,Q_0)$ in global fits are much smaller than for the gluon PDF, specially in the region relevant for $\ttbar2\ell$ where all four global PDF fits shown are contained within a band of $\pm 2\%$.
This good agreement further justifies our strategy to fix the quark PDF to an external reference when interpreting $\ttbar2\ell$ synthetic data within our NSBI framework.

\section{The Boosted Information Tree for the full polynomial expansion}\label{app:mse_bpt}

In this appendix we construct the``Boosted Information Tree''~(BIT) algorithm used to learn the detector-level cross-section ratio introduced in Sec.~\ref{sec:nsbi_for_pdfs}. We base it on the mean squared error~(MSE) loss and subsequently extend it to a version based on the cross-entropy~(CE) loss.
This tree-based boosting algorithm builds on Refs.~\cite{Chatterjee:2021nms,Chatterjee:2022oco}, where related algorithms were developed to learn the coefficients of a polynomial expansion one at a time.
By combining these ideas with the ``Boosted Parametric Tree'' algorithm of Ref.~\cite{Schofbeck:2024zjo}, we generalize the BIT construction so that it can learn the full polynomial dependence simultaneously.
This extended variant has already been applied to the measurement of SMEFT effects in the WH and ZH processes presented in Ref.~\cite{CMS:2024ksn}.

Although the following extension of the BIT can be generalised further to arbitrary polynomial dependencies in a straightforward way, we assume here that the cross-section ratio is accurately described by a polynomial that is at most quadratic in the coefficients \(\bc\).
We write
\begin{equation}
R(\bx,\bc)=1+c_A R_A(\bx)\, ,
\label{eq:app_R_quadratic}
\end{equation}
where the multi-index \(A=\{a,ab\}\) runs over the linear coefficients \(c_a\) and the symmetric quadratic monomials \(c_a c_b\).
This assumption covers the linear PDF model as well as the case of SMEFT Wilson coefficients, so we can keep the derivation general.
Likewise, the latent event-wise PDF reweighting factor can be expanded as
\begin{equation}
\omega(\bz,\bc)=1+c_A\omega_A(\bz)\, .
\label{eq:app_omega_quadratic}
\end{equation}

We consider a nominal simulated training sample generated at \(\bc=\bzero\),
\begin{equation}
\mathcal{D}_{\rm sim}=\{\bz_i,\bx_i,w_{i,0}\}_{i=1}^{N_{\rm sim}}\, ,
\end{equation}
and a finite set of training points \(\mathcal V\subset\{\bc\}\) spanning the parameter space of the polynomial model.
We base the development of the algorithm on the MSE loss for a predictor \(\hat R(\bx,\bc)=1+c_A\hat R_A(\bx)\), which is
\begin{equation}
L_{\rm MSE}[\hat R_A]
=\sum_{\bc\in\mathcal V}\sum_{i=1}^{N_{\rm sim}}
w_{i,0}
\lp
\hat R(\bx_i,\bc)-\omega(\bz_i,\bc)
\rp^2 ,
\label{eq:app_mse_loss_start}
\end{equation}
and promote the result to the CE loss at the end of the section.
Using Eqs.~(\ref{eq:app_R_quadratic}) and~(\ref{eq:app_omega_quadratic}), the constant term cancels identically and one obtains
\begin{equation}
L_{\rm MSE}[\hat R_A]
=
\sum_{\bc\in\mathcal V}\sum_{i=1}^{N_{\rm sim}}
w_{i,0}
\lp
c_A\lp \hat R_A(\bx_i)-\omega_A(\bz_i)\rp
\rp^2 .
\end{equation}
Introducing the Gram matrix in coefficient space,
\begin{equation}
V_{AB}\equiv \sum_{\bc\in\mathcal V} c_A c_B\, ,
\label{eq:app_gram_matrix}
\end{equation}
the loss can be written as
\begin{equation}
L_{\rm MSE}[\hat R_A]
=
\sum_{i=1}^{N_{\rm sim}} w_{i,0}\,
\lp \hat R_A(\bx_i)-\omega_A(\bz_i)\rp
V_{AB}
\lp \hat R_B(\bx_i)-\omega_B(\bz_i)\rp .
\label{eq:app_mse_loss_compact}
\end{equation}
This form makes explicit that the MSE loss determines the full quadratic dependence on \(\bc\) in a single fit.
The trainable quantities are the coefficient functions \(R_A(\bx)\), while the dependence on the chosen set of training points \(\mathcal V\) enters only through the matrix \(V_{AB}\).
For the regression problem to be well defined, \(V_{AB}\) must have full rank.
This is guaranteed if the vectors in \(\mathcal V\) are chosen such that all independent linear and symmetric quadratic monomials in \(\bc\) are sampled.
A convenient choice is to consider all vectors \(\bc\) with non-negative components satisfying
\[
c_a\geq 0\, , \qquad \sum_{a=1}^N c_a \leq 2 \, .
\]
This set contains exactly
$N+\tfrac{1}{2}N(N+1)$
non-zero points, corresponding to all independent linear terms \(c_a\) and symmetric quadratic terms \(c_a c_b\).
This is precisely the number of coefficients required to determine a quadratic polynomial in \(N\) dimensions whose constant term is fixed to unity at \(\bc=\bzero\).

At the population level, the minimiser of Eq.~(\ref{eq:app_mse_loss_compact}) satisfies
\begin{equation}
\hat R_A^\ast(\bx)
=
\mathbb E\left[\omega_A(\bz)\mid \bx,\bc=\bzero\right],
\end{equation}
and therefore
\begin{equation}
\hat R^\ast(\bx,\bc)
=
1+c_A\hat R_A^\ast(\bx)
=
\mathbb E\left[\omega(\bz,\bc)\mid \bx,\bc=\bzero\right]
=
R(\bx,\bc)\, .
\end{equation}
Hence the MSE objective learns the exact detector-level cross section ratio.

\subsection{Additive boosting in coefficient space}
\label{app:mse_boosting}

A single weak learner will in general not provide a sufficiently accurate approximation of the \(\bx\)-dependence of the functions \(R_A(\bx)\).
We therefore use an additive expansion with \(B\) boosting iterations and corresponding learning rates \(0<\eta^{(b)}\leq 1\), with \(b=1,\ldots,B\).
The cumulative boosted predictor after iteration \(b\) is defined as
\begin{equation}
\hat R_{A,(b)}(\bx)=\hat R_{A,(b-1)}(\bx)+\hat R_A^{(b)}(\bx)\, ,
\qquad
\hat R_{A,(0)}(\bx)=0\, ,
\label{eq:app_boosting_update}
\end{equation}
where only $\hat R_A^{(b)}$ is iteratively fit and a fraction of $\eta^{(b)}$ of this result is accumulated~\cite{Schofbeck:2024zjo}. After \(B\) iterations,
\begin{equation}
\hat R_{A,(B)}(\bx)=\sum_{b=1}^B \eta^{(b)}\hat R_A^{(b)}(\bx)\, .
\end{equation}
Equivalently, the boosted detector-level ratio is
\begin{equation}
\hat R_{(B)}(\bx,\bc)=1+c_A\hat R_{A,(B)}(\bx)\, .
\end{equation}

At iteration \(b\), the weak learner is obtained from the current cumulative prediction \(\hat R_{A,(b-1)}\) by minimizing
\begin{equation}
L_{\rm MSE}^{(b)}[\hat R_A^{(b)}]
=
\sum_{i=1}^{N_{\rm sim}} w_{i,0}\,
\lp
\hat R_{A,(b-1)}(\bx_i)+\hat R_A^{(b)}(\bx_i)-\omega_A(\bz_i)
\rp
V_{AB}
\lp
\hat R_{B,(b-1)}(\bx_i)+\hat R_B^{(b)}(\bx_i)-\omega_B(\bz_i)
\rp .
\label{eq:app_boosting_loss}
\end{equation}
It is therefore convenient to define the residual coefficient targets
\begin{equation}
\omega_{A,i}^{(b)}\equiv \omega_A(\bz_i)-\hat R_{A,(b-1)}(\bx_i)\, ,
\label{eq:app_residual_targets}
\end{equation}
in terms of which Eq.~(\ref{eq:app_boosting_loss}) becomes
\begin{equation}
L_{\rm MSE}^{(b)}[\hat R_A^{(b)}]
=
\sum_{i=1}^{N_{\rm sim}} w_{i,0}\,
\lp
\hat R_A^{(b)}(\bx_i)-\omega_{A,i}^{(b)}
\rp
V_{AB}
\lp
\hat R_B^{(b)}(\bx_i)-\omega_{B,i}^{(b)}
\rp .
\label{eq:app_boosting_loss_residual}
\end{equation}
Thus, each boosting step amounts again to fitting the same polynomial model, now to the residuals remaining after the preceding iterations.

The boosting procedure can be summarised as follows:
\begin{itemize}
\item[(i)]
Initialise the residual targets by
\begin{equation}
\omega_{A,i}^{(1)}=\omega_A(\bz_i)\, .\label{eq:boosting-start}
\end{equation}

\item[(ii)]
At iteration \(b\), determine the weak learner \(\hat R_A^{(b)}(\bx)\) by minimizing the residual loss
\begin{equation}
L_{\rm MSE}^{(b)}[\hat R_A^{(b)}]
=
\sum_{i=1}^{N_{\rm sim}} w_{i,0}\,
\lp
\hat R_A^{(b)}(\bx_i)-\omega_{A,i}^{(b)}
\rp
V_{AB}
\lp
\hat R_B^{(b)}(\bx_i)-\omega_{B,i}^{(b)}
\rp .
\end{equation}

\item[(iii)]
Update the cumulative prediction according to
\begin{equation}
\hat R_{A,(b)}(\bx)=\hat R_{A,(b-1)}(\bx)+\eta^{(b)}\hat R_A^{(b)}(\bx)\, ,
\end{equation}
and redefine the residual targets by subtracting the learned polynomial,
\begin{equation}
\omega_{A,i}^{(b+1)}=\omega_{A,i}^{(b)}-\eta^{(b)}\hat R_A^{(b)}(\bx_i)\, .\label{eq:boosting-reweighting}
\end{equation}
\end{itemize}
In this way, boosting is carried out directly in the coefficient space labeled by \(A\).

\subsection{Tree weak learners}
\label{app:mse_tree_weak_learners}

At each boosting iteration \(b\), we represent the weak learner by a tree with piecewise-constant predictions in \(\bx\).
We denote by \(\mathcal J^{(b)}\) the corresponding phase-space partitioning, and by \(J\in\mathcal J^{(b)}\) its terminal nodes, such that
\begin{equation}
\mathcal X=\bigcup_{J\in\mathcal J^{(b)}} J\, ,
\qquad
J\cap J'=\emptyset \quad \text{for} \quad J\neq J'\, .
\end{equation}
Introducing the indicator functions
\begin{equation}
\bone_J(\bx)=
\begin{cases}
1 & \text{if } \bx\in J,\\
0 & \text{otherwise},
\end{cases}
\end{equation}
the weak learner at iteration \(b\) is written as
\begin{equation}
\hat R_A^{(b)}(\bx)
=
\sum_{J\in\mathcal J^{(b)}} \bone_J(\bx)\,\hat R_{A,J}^{(b)}\, .
\label{eq:app_tree_ansatz}
\end{equation}
Hence each terminal node predicts the full polynomial in \(\bc\) through the coefficient vector \(\hat R_{A,J}^{(b)}\).

Inserting Eq.~(\ref{eq:app_tree_ansatz}) into Eq.~(\ref{eq:app_boosting_loss_residual}), and using the disjointness of the partition, the loss decomposes into a sum over terminal nodes,
\begin{equation}
L_{\rm MSE}^{(b)}[\mathcal J^{(b)},\hat R_{A,J}^{(b)}]
=
\sum_{J\in\mathcal J^{(b)}} L_J^{(b)}[\hat R_{A,J}^{(b)}]\, ,
\end{equation}
with
\begin{equation}
L_J^{(b)}[\hat R_{A,J}^{(b)}]
=
\sum_{\{\bz_i,\bx_i,w_{i,0}\}\in\mathcal D_{\rm sim}\cap J}
w_{i,0}\,
\lp
\hat R_{A,J}^{(b)}-\omega_{A,i}^{(b)}
\rp
V_{AB}
\lp
\hat R_{B,J}^{(b)}-\omega_{B,i}^{(b)}
\rp .
\label{eq:app_node_loss}
\end{equation}
This separates the training problem into two parts. For a fixed phase-space partitioning \(\mathcal J^{(b)}\), the optimal node predictions \(\hat R_{A,J}^{(b)}\) are obtained analytically. The non-trivial task is then to determine the optimal partitioning in feature space.

To solve the node-level problem, we define the nominal node weight
\begin{equation}
\sigma_{J,0}
=
\sum_{\{\bz_i,\bx_i,w_{i,0}\}\in\mathcal D_{\rm sim}\cap J}
w_{i,0}\, ,
\end{equation}
and the residual coefficient sums
\begin{equation}
\Omega_{A,J}^{(b)}
=
\sum_{\{\bz_i,\bx_i,w_{i,0}\}\in\mathcal D_{\rm sim}\cap J}
w_{i,0}\,\omega_{A,i}^{(b)}\, .
\end{equation}
With these definitions, Eq.~(\ref{eq:app_node_loss}) can be rewritten as
\begin{equation}
L_J^{(b)}[\hat R_{A,J}^{(b)}]
=
\sigma_{J,0}\,\hat R_{A,J}^{(b)}V_{AB}\hat R_{B,J}^{(b)}
-2\,\hat R_{A,J}^{(b)}V_{AB}\Omega_{B,J}^{(b)}
+\sum_{i\in J} w_{i,0}\,\omega_{A,i}^{(b)}V_{AB}\omega_{B,i}^{(b)}\, .
\end{equation}
Differentiating with respect to \(\hat R_{C,J}^{(b)}\) yields
\begin{equation}
0
=
2\,\sigma_{J,0}\,V_{CB}\hat R_{B,J}^{(b)}
-2\,V_{CB}\Omega_{B,J}^{(b)}\, .
\end{equation}
Since the matrix \(V_{AB}\) is invertible provided the training points \(\mathcal V\) span the polynomial space, the optimal terminal-node prediction is
\begin{equation}
\hat R_{A,J}^{(b)}
=
\frac{\Omega_{A,J}^{(b)}}{\sigma_{J,0}}
=
\frac{\sum_{i\in J} w_{i,0}\,\omega_{A,i}^{(b)}}
{\sum_{i\in J} w_{i,0}}\, .
\label{eq:app_terminal_node_solution}
\end{equation}
Thus, for a fixed partitioning \(\mathcal J^{(b)}\), the node fit is analytically solvable in the coefficient space indexed by \(A\).

Inserting Eq.~(\ref{eq:app_terminal_node_solution}) back into the node loss gives
\begin{equation}
L_J^{(b)\,\ast}
=
\sum_{i\in J} w_{i,0}\,\omega_{A,i}^{(b)}V_{AB}\omega_{B,i}^{(b)}
-
\frac{1}{\sigma_{J,0}}\,
\Omega_{A,J}^{(b)}V_{AB}\Omega_{B,J}^{(b)}\, .
\end{equation}
Summing over all nodes, the optimised tree loss is
\begin{equation}
L_{\rm MSE}^{(b)\,\ast}[\mathcal J^{(b)}]
=
\sum_{J\in\mathcal J^{(b)}} L_J^{(b)\,\ast}\, .
\end{equation}
The first term is independent of the partitioning once summed over all events, and can therefore be dropped for the purpose of constructing the tree. The partitioning is thus determined by minimizing
\begin{equation}
L_{\rm MSE}^{(b)}[\mathcal J^{(b)}]
=
-
\sum_{J\in\mathcal J^{(b)}}
\frac{1}{\sigma_{J,0}}\,
\Omega_{A,J}^{(b)}V_{AB}\Omega_{B,J}^{(b)}\, ,
\label{eq:app_split_loss}
\end{equation}
or, equivalently, by maximizing the corresponding positive quantity.
This criterion can be optimised with standard tree-building algorithms such as CART~\cite{breiman1984classification} or
TAO~\cite{TAO-1,TAO-2,TAO-3,TAO-4}.

\subsection{Cross-entropy loss}\label{sec:BIT-cross-entropy}

The construction above can be generalised to the CE loss.
For a fixed parameter point \(\bc\), we consider the binary classification loss
\begin{equation}
L_{\rm CE}[f]
=
-\int \dd\sigma(\bx|\bzero)\,\log f(\bx,\bc)
-\int \dd\sigma(\bx|\bc)\,\log\!\bigl(1-f(\bx,\bc)\bigr)\, .
\end{equation}
Minimizing pointwise in \(\bx\) yields
\begin{equation}
f^\ast(\bx,\bc)
=
\frac{\dd\sigma(\bx|\bzero)}
{\dd\sigma(\bx|\bzero)+\dd\sigma(\bx|\bc)}
=
\frac{1}{1+R(\bx,\bc)}\, ,\label{eq:min-pointwise}
\end{equation}
so that the CE optimum is again in one-to-one correspondence with the detector-level cross section ratio.

Passing to the empirical loss, we first make the latent-space integration explicit.
Using the detector-level differential cross section
\[
\frac{\dd\sigma(\bx|\bc)}{\dd\bx}
=
\int \dd\bz\,
\frac{\dd\sigma(\bx,\bz|\bc)}{\dd\bx\,\dd\bz}\, ,
\]
the population-level CE loss can be written as
\[
L_{\rm CE}[f]
=
-\sum_{\bc\in\mathcal V}
\left[
\int \dd\bz\,\dd\bx\,
\frac{\dd\sigma(\bx,\bz|\bzero)}{\dd\bx\,\dd\bz}\,
\log f(\bx,\bc)
+
\int \dd\bz\,\dd\bx\,
\frac{\dd\sigma(\bx,\bz|\bc)}{\dd\bx\,\dd\bz}\,
\log\!\bigl(1-f(\bx,\bc)\bigr)
\right].
\]
Since the transfer from latent to reconstructed variables is independent of \(\bc\), the joint measure at parameter point \(\bc\) differs from the nominal one only through the event-wise PDF reweighting factor,
\[
\frac{\dd\sigma(\bx,\bz|\bc)}{\dd\bx\,\dd\bz}
=
\omega(\bz,\bc)\,
\frac{\dd\sigma(\bx,\bz|\bzero)}{\dd\bx\,\dd\bz}\, .
\]
Both terms can therefore be expressed with respect to the common nominal measure, which leads directly to the empirical loss
\begin{equation}
L_{\rm CE}[\hat f]
=
-\sum_{\bc\in\mathcal V}\sum_{i=1}^{N_{\rm sim}} w_{i,0}
\left[
\log \hat f(\bx_i,\bc)
+
\omega(\bz_i,\bc)\,
\log\!\bigl(1-\hat f(\bx_i,\bc)\bigr)
\right].
\end{equation}
The trick is now to use the analytic form of Eq.~(\ref{eq:min-pointwise}) in the ansatz for the surrogate
\begin{equation}
\hat f(\bx,\bc)=\frac{1}{1+\hat R(\bx,\bc)}\, ,
\qquad
\hat R(\bx,\bc)=1+c_A\hat R_A(\bx)\,
\end{equation}
and insert into the loss function. This yields
\begin{equation}
L_{\rm CE}[\hat R_A]
=
\sum_{\bc\in\mathcal V}\sum_{i=1}^{N_{\rm sim}} w_{i,0}
\left[
\log\!\bigl(1+\hat R(\bx_i,\bc)\bigr)
+
\omega(\bz_i,\bc)\,
\log\!\left(1+\frac{1}{\hat R(\bx_i,\bc)}\right)
\right].
\end{equation}

As in the MSE case, the cumulative predictor satisfies
\begin{equation}
\hat R_{(b-1)}(\bx,\bc)+R_{\rm residual}^{(b)}(\bx,\bc)=R(\bx,\bc)\, ,
\end{equation}
where \(R_{\rm residual}^{(b)}(\bx,\bc)\) denotes the ratio still to be learned at boosting step \(b\).
This implies that we apply the same iterative boosting procedure as in Eq.~(\ref{eq:boosting-start}--\ref{eq:boosting-reweighting}), i.e., learn \(R_{\rm residual}^{(b)}(\bx,\bc)\) from the CE loss after performing an iterative reweighting of the nominal sample.

Substituting this solution back and summing over all nodes gives the optimised loss
\begin{equation}
L_{{\rm CE}}^{(b)}
=
\sum_{\bc\in\mathcal V}\sum_{J\in\mathcal J^{(b)}}
\sigma_{J,0}
\left[
\log\!\bigl(1+\hat R_J^{(b)}(\bc)\bigr)
+
\hat R_J^{(b)}(\bc)\,
\log\!\left(1+\frac{1}{\hat R_J^{(b)}(\bc)}\right)\right],\;\;\text{with}\;\; \hat R_J^{(b)}=1+\frac{c_A\Omega_{A,J}^{(b)}}{\sigma_{J,0}}
\end{equation}
which can be minimised with CART or TAO.
The CE loss requires positive arguments of the logarithms. In the present parametrization, this amounts to demanding
\begin{equation}
1+\frac{c_A\Omega_{A,J}^{(b)}}{\sigma_{J,0}}>0
\qquad
\text{for all } \bc\in\mathcal V
\end{equation}
in every terminal node \(J\).
In practice, we enforce this by vetoing candidate node splits for which the contraction \(1+c_A\hat R_{A,J}^{(b)}\) becomes non-positive for at least one sampled training point \(\bc\in\mathcal V\) in either daughter node.
This provides a simple positivity criterion on the parameter domain actually used in the empirical CE training.
\begin{algorithm}[t]
\caption{Boosted Information Tree~(BIT) for learning a quadratic cross section ratio}
\label{alg:bit_mse}
\begin{algorithmic}
\Require nominal sample \(\mathcal D_{\rm sim}=\{\bz_i,\bx_i,w_{i,0}\}_{i=1}^{N_{\rm sim}}\), latent coefficients \(\omega_A(\bz_i)\), training points \(\bc\in\mathcal V\),\\\quad\quad\quad\;\,
boosting iterations \(B\), learning rates \(0<\eta^{(b)}\leq 1\) for \(b=1,\ldots,B\)
\Ensure \(V_{AB}=\sum_{\bc\in\mathcal V} c_A c_B\) has full rank
\State compute \(V_{AB}\gets \sum_{\bc\in\mathcal V} c_A c_B\)
\State initialise cumulative predictor \(\hat R_{A,(0)}(\bx)\gets 0\)
\State initialise residual targets \(\omega^{(1)}_{A,i}\gets \omega_A(\bz_i)\) for all events \(i\)
\For{\(b=1,\ldots,B\)}
    \State \(\mathcal J^{(b)} \gets \argmin_{\mathcal J} L_{\rm MSE}^{(b)}[\mathcal J]\) using CART or TAO (for CE use \(L_{\rm CE}^{(b)}[\mathcal J]\))
    \ForAll{\(J\in\mathcal J^{(b)}\)}
        \State \(\sigma_{J,0}\gets \sum_{\{\bz_i,\bx_i,w_{i,0}\}\in\mathcal D_{\rm sim}\cap J} w_{i,0}\)
        \State \(\Omega^{(b)}_{A,J}\gets \sum_{\{\bz_i,\bx_i,w_{i,0}\}\in\mathcal D_{\rm sim}\cap J} w_{i,0}\,\omega^{(b)}_{A,i}\)
        \State \(\hat R^{(b)}_{A,J}\gets \Omega^{(b)}_{A,J}/\sigma_{J,0}\)
    \EndFor
    \State \(\hat R_A^{(b)}(\bx)\gets \sum_{J\in\mathcal J^{(b)}} \bone_J(\bx)\,\hat R^{(b)}_{A,J}\)
    \State \(\hat R_{A,(b)}(\bx)\gets \hat R_{A,(b-1)}(\bx)+\eta^{(b)}\hat R_A^{(b)}(\bx)\)
    \State \(\omega^{(b+1)}_{A,i}\gets \omega^{(b)}_{A,i}-\eta^{(b)}\hat R_A^{(b)}(\bx_i)\) for all events \(i\)
\EndFor
\State \Return \(\hat R_A(\bx)=\sum_{b=1}^B \eta^{(b)}\sum_{J\in\mathcal J^{(b)}} \bone_J(\bx)\,\hat R^{(b)}_{A,J}\) and \(\hat R(\bx,\bc)=1+c_A\hat R_A(\bx)\)
\end{algorithmic}
\end{algorithm}
\subsection{Algorithm summary}
\label{sec:bit-summary}

In summary, the BIT algorithm for learning the full polynomial dependence consists of an iterative fit of a tree-based weak learner to the residual polynomial coefficients left over from the preceding boosting iteration.
The starting point is the nominal simulated sample \(\mathcal D_{\rm sim}\) together with the event-wise latent coefficients \(\omega_A(\bz_i)\) entering the quadratic expansion of the PDF reweighting factor.
The set of training points \(\mathcal V\) must be chosen such that the matrix \(V_{AB}\), defined in Eq.~(\ref{eq:app_gram_matrix}), has full rank.
At each boosting iteration, a weak learner is constructed by partitioning the feature space with a tree algorithm such as CART or TAO.
For a fixed partitioning, the optimal terminal-node predictions are obtained analytically from Eq.~(\ref{eq:app_terminal_node_solution}).
Overfitting is controlled by the usual tree hyperparameters, such as the maximum tree depth and the minimum number of events in each terminal node.
The resulting weak learner is multiplied by a learning rate \(\eta^{(b)}\) and used to update the residual coefficient targets, which defines the input to the following iteration.
After \(B\) boosting steps, the final result is
\begin{align}
\hat R_A(\bx)
=
\sum_{b=1}^B \eta^{(b)}
\sum_{J\in\mathcal J^{(b)}} \bone_J(\bx)\,\hat R^{(b)}_{A,J}\, ,
\qquad
\hat R(\bx,\bc)=1+c_A\hat R_A(\bx)\, ,
\end{align}
where \(\mathcal J^{(b)}\) denotes the phase-space partitioning found at boosting iteration \(b\), and \(\hat R^{(b)}_{A,J}\) are the corresponding polynomial coefficients in terminal node \(J\).
Algorithm~\ref{alg:bit_mse} provides a pseudo-code summary of these steps.

\bibliographystyle{lucas_unsrt}
\bibliography{SBI-PDF.bib}

\end{document}